\documentclass[a4paper,11pt]{article}
\pdfoutput=1

\usepackage{jheppub}
\usepackage{customprelude}

\def\IS{{\bf S}}
\def\IR{{\bf R}}
\def\IZ{{\bf Z}}
\def\IX{{\bf X}}

\def\IP{\bf P}
\def\CN{{\cal N}}

\newcommand{\Ocase}[1]{
  \subsubsection*{\centering #1}
}

\def\beq{\begin{equation}}
\def\eeq{\end{equation}}
\def\beqa{\begin{eqnarray}}
\def\eeqa{\end{eqnarray}}

\preprint{IFT-UAM/CSIC-24-126}

\title{\centering  {\huge SymTFT Fans} \\[0.3 cm] The Symmetry Theory of 4d $\cN=4$ Super Yang-Mills on spaces with boundaries }

\author{Iñaki García Etxebarria$^{\clubsuit}$,}
\author{Jesús Huertas$^{\spadesuit}$ and}
\author{Angel M. Uranga$^{\spadesuit}$}

\affiliation[\clubsuit]{Department of Mathematical Sciences,
  Durham University,\\
  Durham, DH1 3LE, United Kingdom}

\affiliation[\spadesuit]{Instituto de Física Teórica IFT-UAM/CSIC,\\
  C/ Nicolás Cabrera 13-15, Campus de Cantoblanco, 28049 Madrid, Spain}

\emailAdd{inaki.garcia-etxebarria@durham.ac.uk}
\emailAdd{j.huertas@csic.es}
\emailAdd{angel.uranga@csic.es}

\abstract{The SymTFT construction is an efficient way of studying the
  symmetries of Quantum Field Theories. In this paper we initiate the
  study of the SymTFT construction for interacting theories on spaces
  with boundaries, in the concrete example of $d=4$ $\cN=4$ Super
  Yang-Mills theory with Gaiotto-Witten boundary conditions. The
  resulting SymTFT has a number of peculiarities, most prominently a
  fan-like structure with a number of different SymTFT sectors joined
  by gapless interfaces that merge at the Gaiotto-Witten boundary
  SCFT.}

\begin{document}
\maketitle
\flushbottom

\section{Introduction}
\label{sec:intro}

In the Euclidean formulation of Lorentz-invariant Quantum Field
Theories, internal symmetries give rise to topological operators of
codimension one. During the last few years many papers, starting with
\cite{Gaiotto:2014kfa}, have convincingly shown that it is very
illuminating to think of \emph{all} the topological operators present
in any given Quantum Field Theory as symmetries, sometimes referred to
as \emph{categorical} symmetries to keep in mind both that we are
significantly broadening the traditional definition of symmetry, and
that category theory is a natural tool for talking about the
topological QFTs describing the topological subsector of the
theory. Generic topological operators in QFT are not necessarily
codimension one, and the fusion rules for topological operators are
not necessarily group-like, so the resulting algebraic structures can
be very rich, and encode very subtle information about the Quantum
Field Theory at hand. We refer the reader to any of the excellent
recent reviews
\cite{McGreevy:2022oyu,Brennan:2023mmt,Gomes:2023ahz,Shao:2023gho,Schafer-Nameki:2023jdn,Bhardwaj:2023kri,Iqbal:2024pee}
for surveys of different aspects of this rapidly developing field.

We will focus on a recent proposal for capturing the categorical
symmetries of a theory known (somewhat imprecisely in our case, for
reasons we will explain momentarily) as the SymTFT construction
\cite{Gaiotto:2014kfa,Gaiotto:2020iye,Ji:2019jhk,Apruzzi:2021nmk,Freed:2022qnc}. In
the simplest variant of this approach, the \mbox{$d$-dimensional} QFT
is realised in terms of the interval compactification of a
\mbox{$(d+1)$-dimensional} topological field theory (the
\emph{SymTFT}), with gapped boundary conditions at one end, and
gapless boundary conditions at the other end. The SymTFT construction
arises very naturally from geometric engineering in string theory and
holography, and indeed we will use holography in this work to analyse
the SymTFT construction in a regime that has not been much studied
before: $d$-dimensional theories on manifolds with boundaries. As we
will see, in this context it is still possible to understand the
symmetries of the $d$-dimensional theory on a manifold with boundary
in terms of a $(d+1)$-dimensional construction on a space with
corners, where additional gapless degrees of freedom live.

The resulting bulk is not fully topological, so it would perhaps be
better to talk about a \emph{symmetry theory} instead (as emphasised
in a related context in \cite{Apruzzi:2024htg}, for instance), but in
the cases we study in this paper the non-topological nature of the
bulk ends up being fairly mild: the $(d+1)$-dimensional theory is
topological except on non-topological $d$-dimensional interfaces. So
by a small abuse of language we will still refer to the bulk as the
SymTFT.

The specific theories we will focus on are $d=4$ $\cN=4$ $SU(N)$ SYM on a
space with boundary. The theory on spaces without boundary has a well known SymTFT, which can be obtained from the topological sector of its holographic dual \cite{Witten:1998wy}. It encodes the possible  electric and magnetic $\IZ_N$ 1-form symmetries in terms of two topological 2-form fields in a 5d SymTFT. The main focus of our work is to explore the modification of this construction due to the introduction of boundaries, and in particular analyse the fate of the 1-form symmetries via the 5d 2-form fields.

\begin{figure}[htb]
\begin{center}
\includegraphics[scale=.35]{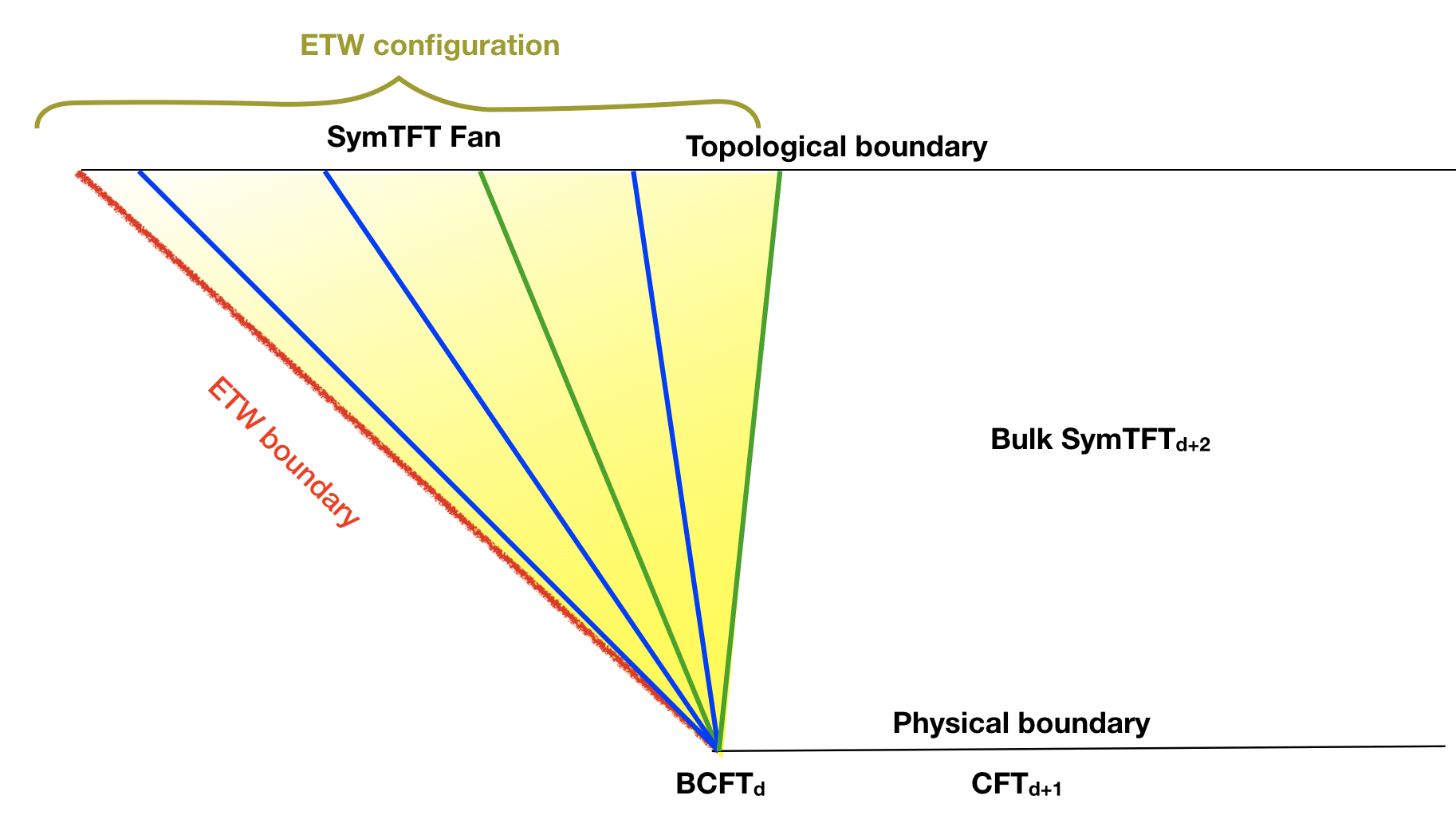}
\caption{\small Sketch of the structure of a SymTFT Fan describing a
  CFT$_{d+1}$ on a space with a boundary coupled to a BCFT$_{d}$. The
  gapless domain walls (blue and green lines) connecting different
  SymTFT wedges sticking out of the BCFT in a fan configuration.}
\label{fig:fan}
\end{center}
\end{figure}

We choose the boundary conditions preserving half
of the superconformal symmetry which have been studied by Gaiotto and
Witten in \cite{Gaiotto:2008sa,Gaiotto:2008ak}, and which we briefly
review in section~\ref{sec:gw}. Our basic tool for understanding the
SymTFT for such configurations is the holographic dual, constructed in
\cite{DHoker:2007zhm,DHoker:2007hhe,Aharony:2011yc,Assel:2011xz,Bachas:2017rch,Bachas:2018zmb},
and reviewed in section~\ref{sec:holodual}. We extract the SymTFT from
the holographic dual in section~\ref{sec:symtft}. The resulting SymTFT
is similar to the SymTree construction introduced in
\cite{Baume:2023kkf} and reviewed in appendix \ref{app:symtree}, although with a number of differences that we
describe in detail in section~\ref{sec:symtft}. A striking feature of
our setup is the various gapless domain walls connecting the different
SymTFTs are arranged in a fan configuration, with the
three-dimensional Gaiotto-Witten BCFT living at the tip of the
fan. See figure~\ref{fig:fan} for a sketch of the
SymTFT.

%%%%%%%%%%%
%%%%%%%%%%%

A number of expected features of the SymTFT can be understood from
brane physics in the original string theory setup. We develop various
aspects of the relation between these two viewpoints in
section~\ref{sec:symtft}, in particular the effect of translating
topological operators between different sectors in the SymTFT
fan. Section~\ref{sec:7branes} goes further in this direction,
studying the SymTFT interpretation of bringing 7-branes from infinity
--- an operation which is fairly natural and well understood from the
brane point of view, but leads to some puzzles (which we resolve)
when reinterpreted from the SymTFT point of view. Finally, in
\ref{sec:orientifold} we generalise the discussion in the previous
sections by allowing for orientifolds localised at the boundary.

In this work we exploit the supergravity duals of the QFT with
boundary to learn about its topological sector and extract the SymTFT,
but our results are also relevant in the reverse direction. Indeed,
gravitational solutions with End of the World (\emph{ETW}) boundary
configurations, dubbed dynamical cobordisms
\cite{Buratti:2021fiv,Angius:2022aeq,Blumenhagen:2022mqw,Blumenhagen:2023abk}\footnote{For
  related ideas, see
  \cite{Dudas:2000ff,Blumenhagen:2000dc,Dudas:2002dg,Dudas:2004nd,Hellerman:2006nx,Hellerman:2006ff,Hellerman:2007fc}
  for early references, and
  \cite{Basile:2018irz,Antonelli:2019nar,Mininno:2020sdb,Basile:2020xwi,Mourad:2021qwf,Mourad:2021roa,Basile:2021mkd,Mourad:2022loy,Angius:2022mgh,Basile:2022ypo,Angius:2023xtu,Huertas:2023syg,Mourad:2023ppi,Angius:2023uqk,Delgado:2023uqk,Angius:2024zjv,Mourad:2024dur,Mourad:2024mpg}
  for recent works.}, provide dynamical realisations of the Swampland
Cobordism Conjecture \cite{McNamara:2019rup}, generalising the concept
of bubbles of nothing \cite{Witten:1981gj} (see
\cite{Ooguri:2017njy,GarciaEtxebarria:2020xsr,Friedrich:2023tid,Blanco-Pillado:2023hog,Blanco-Pillado:2023aom,Delgado:2023uqk,Friedrich:2024aad}
for recent developments). In this spirit, the ETW configurations we
will consider were in fact described as dynamical cobordisms of
AdS$_5\times\IS^5$ in \cite{Huertas:2023syg}. Hence, our in-depth
understanding of the topological structure of the gravitational bulk
theory and its interplay with the topology of the ETW configuration
are a first step towards understanding it for more general dynamical
cobordisms, providing a valuable tool for the study of topological
properties of general cobordism defects ending spacetime.

\medskip

\emph{Note added:} The works
\cite{Copetti:2024onh,Cordova:2024iti,Copetti:2024dcz} appeared as we
were nearing completion of this paper, and also discuss various
aspects of SymTFTs for theories on spacetimes with boundaries. This topic was also covered in a number of papers that appeared simultaneously with
ours
\cite{Choi:2024tri,Das:2024qdx,Bhardwaj:2024igy,Heymann:2024vvf,Choi:2024wfm}. We
are very thankful to the authors of these papers for
agreeing to coordinate submission.

\section{Gaiotto-Witten configurations}
\label{sec:gw-holo}

In this section we will quickly review --- focusing on the ingredients
most relevant for the discussion of the SymTFT --- the boundary
conditions studied in \cite{Gaiotto:2008sa,Gaiotto:2008ak}, which
preserve half of the superconformal symmetry, namely $OSp(4|4)$, of
$d=4$ $\cN=4$ $\fsu(N)$ SYM. These boundary conditions involve a 3d
SCFTs (BCFT$_3$) on the boundary of the 4d theory. In these works the
CFT$_4$/BCFT$_3$ systems were realised as systems of $N$ semi-infinite
D3-brane ending on a set of NS5- and D5-branes (with finite segments
of D3-branes suspended between them), thus realising a quiver gauge
theory of the kind introduced in \cite{Hanany:1996ie}, which under
suitable conditions flows in the IR to an interacting 3d $\CN=4$
SCFT.

\subsection{The brane construction and its field theory}
\label{sec:gw}

We consider $N$ D3-branes along the direction 012, of semi-infinite
extent in the direction 3 (namely $x^3>0$, without loss of
generality), and at the origin in the remaining directions.  These
D3-branes end on stacks of NS5-branes, which span the directions 012
456, and at the origin in the direction 3 789, and stacks of D5-branes
spanning 012 789, and at the origin in 3 456. Actually, in order to
define the D3-brane boundary on the NS5- and D5-brane system, it is
more appropriate to consider the NS5- and D5-branes to be slightly
separated in the direction $x^3$, and to admit D3-branes suspended
between them. Once a specific set of constraints, to be reviewed
below, are satisfied, one can take the limit of coincident 5-branes,
which implements the flow to the strongly coupled IR BCFT$_3$ boundary
conditions for the CFT$_4$. Because the UV theory, defined by the
Hanany-Witten brane configuration, preserves 8 supercharges (3d
$\CN=4$), the resulting BCFT preserves maximal $OSp(4|4)$
superconformal symmetry. Note that the R-symmetry subgroup
$SO(3)\times SO(3)$ of $SO(6)$ is manifest as the rotational symmetry
in the 3-planes 456 and 789.

A key role in the brane configuration is played by the linking numbers
of the 5-branes.  These can be defined as the total monopole charge on
the 5-brane worldvolume theory \cite{Hanany:1996ie}, and can therefore
be measured at infinity of the 5-brane worldvolume. Hence they are
invariant under changes in the ordering of branes and the
corresponding brane creation effects. In practice, given an ordering
of the 5-branes, the linking number of a 5-brane is given by the net
number of D3-branes ending on the 5-brane from the right (because
D3-brane boundaries are (singular) monopoles on the 5-brane
worldvolume theory) plus the total number of 5-branes of the other
kind (i.e. NS5 vs D5) to the left of the 5-brane (because they can
lead to D3-branes ending from the right via Hanany-Witten brane
creation processes).

There are two conditions for the boundary brane configuration to
define a supersymmetric BCFT$_3$: (1) Any D5-brane on which a net
non-zero number of D3-branes ends from the right is located to the
right of all the NS5-branes. This constraint ensures that the brane
configurations have an interpretation in gauge theory, with the
NS5-branes with suspended D3-branes among them providing a quiver
gauge theory, and the D5-branes with non-positive net number of
D3-branes yielding flavours for some of the nodes. (2) For each kind
of 5-brane (NS5 or D5) the linking numbers are non-decreasing from
left to right. For D5-branes, this is automatic except for the
D5-brane to the right of all NS5-branes, and for these the condition
guarantees the absence of decoupled degrees of freedom in the limit of
coincident 5-branes; the condition for NS5-branes then follows from
S-duality.

Related to the above, linking numbers play a role in non-abelian IR enhancement of $U(1)$ global (0-form) flavour symmetries of the 3d UV theory, which are realised as symmetries on the worldvolume of the 5-branes. In the strong coupling regime, $n$ 5-branes of the same kind and with the same linking number lead to monopole operators in the field theory which enhance the symmetry to $U(n)$. Hence linking numbers provide the invariant quantities characterising the BCFT$_3$.

Let us provide an extra intuition of the interpretation of the linking number in defining the boundary conditions. Consider the case of a single kind of 5-brane, e.g. D5-branes (NS5-branes admit a similar description due to S-duality), organised in $m$ stacks, labelled by an index $b$, with multiplicities $m_b$ of D5-branes with equal linking number $L_b$. Since in this case there are no NS5-branes, the linking numbers are realised in terms of D3-branes ending on the D5-branes from the right. Each D5-brane in the $b^{\text{th}}$ stack has $L_b$ D3-branes ending on it, with boundary conditions associated to the $L_b$-dimensional irreducible representation ${\cal R}_{L_b}$ of $SU(2)$ (namely, the $SO(3)$ associated to 789), which is realised by the three (matrix-valued) complex scalars of the 4d $\CN=4$ $\fsu(N)$ SYM theory, of the form $[X^i,X^j]\sim \epsilon^{ijk} X^k$. Hence, for the complete system of D5-brane stacks, the boundary condition is associated to a (reducible, in general) representation of $SU(2)$
\[
 {\cal R}_N=\oplus_{b=1}^m m_b{\cal R}_{L_b}\, .
 \label{split}
\]
The $SU(2)$ commutation relation can be regarded as describing the
D3-branes puffing up into non-commutative funnels corresponding to the
D5-branes, whose width is related to the number of D3-branes, i.e. the
linking numbers. The ordering of linking numbers mentioned above
ensures that narrow funnels fit inside wider funnels without
intersections which would introduce additional degrees of freedom.

An efficient way to describe these systems is to focus on the 3d
$\CN=4$ theories realised on the D3-branes suspended among the
5-branes, and to describe the semi-infinite D3-branes as the gauging
of a global symmetry of the BCFT$_3$. One prototypical example is that
of the $T[SU(N)]$ theories \cite{Gaiotto:2008ak}, which are defined as
the IR fixed point of the gauge theory on a system with $N$ NS5-branes
and $N$ D5-branes, realising the gauge theory
\begin{equation}
U(1) - U(2) - \ldots - U(N-1) -[\fsu(N)]\, ,
\label{tsun}
\end{equation}
with bifundamental hypermultiplets of consecutive gauge factors, and $N$ hypermultiplets in the fundamental of the $U(N-1)$ group, acted on by the $\fsu(N)$ global symmetry. Namely, there are $N$ NS5-branes separated by intervals of (an increasing number of) suspended D3-branes, and $N$ D5-branes between the two rightmost NS5-branes, hence giving flavours to the $U(N-1)$ factor. 

The gauging of the $\fsu(N)$ symmetry is implemented by moving the D5-branes infinitely to the right, introducing $N$ semi-infinite D3-branes in crossing the last NS5-brane. We thus get a semi-infinite stack of $N$ D3-branes ending on a set of $N$ NS5-branes, all with linking number 1 (hence with an enhanced $U(n)$ flavour symmetry). A similar configuration (obtained by S-duality and exchange or 345 and 789 (i.e. 3d mirror symmetry) exists for $N$ D3-branes ending on a set of $N$ D5-branes with linking number 1. Hence this provides a boundary condition of the kind (\ref{split}), with the $N$ dimensional representation being trivial (i.e. $N$ copies of the trivial 1-dimensional representation).

Let us pause for an interesting observation nicely illustrated by this example. The symmetries of fairly general 3d $\CN=4$ theories (including those from from NS5- and D5-brane configurations) were considered in \cite{Bhardwaj:2023zix}, in particular the just discussed enhanced 0-form flavour symmetries, as well as the 1-form symmetries, so we may apply their results to our setup. In particular, considering the above $T[SU(N)]$ theories, the 1-form symmetry obtained upon gauging the flavour $SU(N)$ (meaning $\fsu(N)$ with global structure of $SU(N)$) was identified in \cite{Bhardwaj:2023zix} to be $\IZ_N$. Hence, despite the fact that the D5-branes introduce flavours in the fundamental of the 4d $SU(N)$, the electric 1-form symmetry is not fully broken, but rather there survives a diagonal combination of it with a $\IZ_N$ in the enhanced flavour $SU(N)$ on the D5-branes. Similarly, for other gaugings $SU(N)/\IZ_p$ with $N=pk$, the surviving electric 1-form symmetry is $\IZ_k$ \cite{Bhardwaj:2023zix}. 

The generalisation of the $T[SU(N)]$ theories to the general configurations of D5-branes described in (\ref{split}) is known as the $T_\rho[SU(N)]$, where $\rho$ is a partition of $N$ given by
\begin{equation}
N= \underbrace{1+\ldots+1}_{m_1\;{\rm times}}+\underbrace{2+\ldots+2}_{m_2\;{\rm times}}+\ldots 
\end{equation}
i.e. with $m_b$ parts equal to $L_b$, according to the decomposition of the representation (\ref{split}).

As a concrete example, we may consider the theory corresponding to the
trivial partition $\rho$ of $N$ into 1 part equal to $N$. We thus have
1 D5-brane with linking number $N$, namely the $N$ D3-branes end on 1
D5-brane exploiting the $N$-dimensional irreducible representation of
$SU(2)$. By extending the arguments in \cite{Bhardwaj:2023zix}, one
can check that upon gauging of $SU(N)$ the electric 1-form symmetry of
the theory is $\IZ_N$. Analogously the 1-form symmetry for
$SU(N)/\IZ_p$ gauging is $\IZ_k$ with $k=N/p$. We will recover this
pattern from the SymTFT in section \ref{sec:junction}.

There is a generalisation of the above class to the case with both
NS5- and D5-branes. The starting point is to also introduce stacks of
$n_a$ NS5-branes with linking numbers $K_a$, in addition to the
earlier stacks of D5-branes, obeying the rules provided above, to
define a 3d $\CN=4$ theory with no semi-infinite D3-branes. Then one
can rearrange the configuration by locating all the D5-branes to the
left of the NS5-branes. We end up with a set of $N$ D3-branes on a
segment, ending on D5-branes from the left (resp. NS5-branes from the
right) according to a partition $\rho$ (resp. ${\hat{\rho}}$) of $N$,
equivalently the multiplicities and linking numbers of the
5-branes. The data (ranks and fundamental flavours of all the unitary
gauge factors in the quiver) of the original gauge theories can be
obtained from the partitions. These gauge theories, under the
conditions already explained, flow to IR 3d SCFTs known as
$T_\rho^{\hat \rho}[SU(N)]$, and provide BCFT$_3$ boundary conditions
by gauging their $\fsu(N)$ global symmetry.

Let us be more explicit on this last point. In addition to the above defined linking number, it is useful to introduce an equivalent set of linking numbers ${\tilde K}_a$, ${\tilde L}_b$ as the net number of D3-branes ending on the 5-brane from the right {\em minus} the total number of 5-branes of the other kind to the {\em right} of the 5-brane. These new versions differ from the old ones just in an overall shift by the total numbers of D5- and NS5-branes, respectively. In the 3d $T_\rho^{\hat \rho}[SU(N)]$ theories, we have $N$ D3-branes ending on the NS5-branes from the right and on the D5-branes on the left, hence the total sum of NS5-brane linking numbers $\sum_a n_aK_a$ is equal to $N$, and similarly for the total sum of D5-brane linking numbers $\sum m_b{\tilde L}_b=-N$. Upon gauging the $\fsu(N)$ flavour symmetry, we have an extra set of $N$ semi-infinite D3-branes ending on the 5-brane set from the right, hence the condition that the BCFT$_3$ provides a boundary condition for the 4d $\CN=4$ $su(N)$ SYM theory is
\[
N=\sum_a n_aK_a + \sum m_b{\tilde L}_b\, ,
\label{d3-flux}
\]
These ingredients will be manifest in the gravitational dual description in the next section, and will be inherited as key ingredients in the SymTFT.

Before that, we would like to mention that the Hanany-Witten brane configurations can be used not only to provide boundary conditions for 4d $\CN=4$ $\fsu(N)$, but also more general configurations. These include configurations of NS5- and D5-branes separating two stacks of semi-infinite D3-branes (a particular case of Janus configurations \cite{Gaiotto:2008sd}), which we will exploit in the discussion of orientifold theories in section \ref{sec:orientifold}. In addition, one can also construct theories in which there are several configurations of NS5- and D5-branes separated by large distances, with sets of D3-branes suspended between consecutive configurations. These have gravitational duals including multiple AdS$_5\times\IS^5$ asymptotic regions, and will be briefly mentioned in section \ref{sec:general-holo}.

\subsection{The holographic dual: End of the World branes}
\label{sec:holodual}

In this section, we will briefly review the gravitational duals of the brane configurations described in the last section. They are given by a particular class of explicit 10d supergravity backgrounds studied in \cite{DHoker:2007zhm,DHoker:2007hhe,Aharony:2011yc,Assel:2011xz,Bachas:2017rch,Bachas:2018zmb} (see also \cite{Raamsdonk:2020tin,VanRaamsdonk:2021duo,Akhond:2021ffz,Demulder:2022aij,Karch:2022rvr,Akhond:2022oaf,Huertas:2023syg,Chaney:2024bgx} for recent applications).

%%%%%%%%%%%
\begin{figure}[htb]
\begin{center}
\includegraphics[scale=.35]{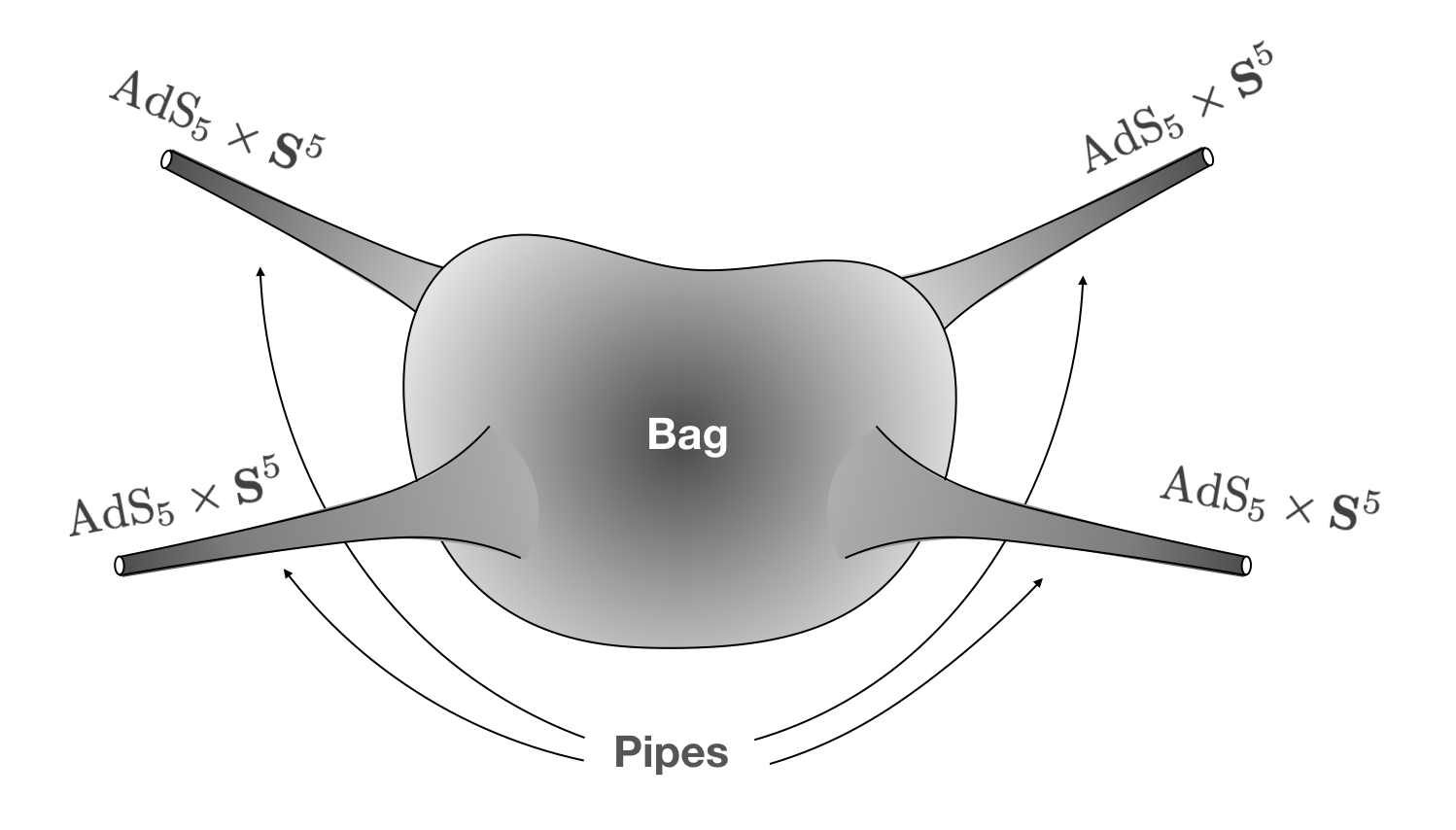}
\caption{\small The bagpipe geometries.}
\label{fig:bagpipe}
\end{center}
\end{figure}
%%%%%%%%%%%

The most general supergravity solutions preserving 16 supersymmetries
and $SO(2,3)\times SO(3)\times SO(3)$ symmetry were explicitly
constructed in \cite{DHoker:2007zhm,DHoker:2007hhe}. They are
gravitational duals of Hanany-Witten brane configurations of NS5- and
D5-branes with stacks of D3-branes as in the previous section
\cite{Aharony:2011yc}. The geometries in general have a ``bagpipe''
structure \cite{Bachas:2018zmb} (see figure \ref{fig:bagpipe}), namely
they look like AdS$_4\times \IX^6$, with $\IX^6$ a 6d compact manifold
(the ``bag''), save for a number of AdS$_5\times \IS^5$ throats (the
``pipes'') sticking out of it. We will eventually focus on the
specific case with only one AdS$_5\times \IS^5$ ending on the bag,
which provides the gravity dual of 4d $\cN=4$ $\fsu(N)$ SYM on a 4d
spacetime with boundary.

The most general supergravity solutions preserving 16 supersymmetries
and $SO(2,3)\times SO(3)\times SO(3)$ symmetry have the structure of a
fibration of AdS$_4\times \IS_1^2\times \IS_2^2$ over an oriented
Riemann surface $\Sigma$. The ansatz for the 10d metric is
\[
ds^2= f_4^2 ds^2_{AdS_4}+f_1^2 ds^2_{\IS_1^2}+f_2^2 ds^2_{\IS_2^2}+ds^2_\Sigma\, .
\label{ansatz}
\]
Here $f_1$, $f_2$, $f_3$ are functions of a complex coordinate $w$ of $\Sigma$, in terms of which 
\[
ds^2_\Sigma=4\rho^2 |dw|^2\, ,
\label{2dmetric}
\]
for some real function $\rho$. There are also non-trivial backgrounds for the NSNS and RR 2-forms and the RR 4-form; we will provide only the necessary results, referring the reader to the references for further details. 

There are closed expressions for the different functions in the above metric, describing the BPS solution. As an intermediate step, we define the real functions
\[
& W\df \partial_w h_1\partial_{\bar w}h_2+\partial_w h_2\partial_{\bar w}h_1 
\; \;,\;\; N_1\df 2h_1h_2|\partial_w h_1|^2-h_1^2 W \;\; ,\;\;N_2\df 2h_1h_2|\partial_w h_2|^2-h_2^2 W
\label{wnn}
\]
in terms of the functions $h_1,h_2$ to be specified below. The dilaton is given by
\[
e^{2\Phi}=\frac{N_2}{N_1}\, ,
\label{dilaton}
\]
and the functions are given by
\[
\rho^2=e^{-\frac 12 \Phi}\frac{\sqrt{N_2|W|}}{h_1h_2}\; ,\;\; f_1^2=2e^{\frac 12\Phi} h_1^2\sqrt{\frac{|W|}{N_1}}\; ,\;\; f_2^2=2e^{-\frac 12\Phi} h_2^2\sqrt{\frac{|W|}{N_2}}\; ,\; \;f_4^2=2e^{-\frac 12\Phi} \sqrt{\frac{N_2}{|W|}}\, .
\label{the-fs}
\]

In the following we focus on the solutions describing a single asymptotic AdS$_5\times\IS^5$ region, as befits the gravitational dual of a stack of semi-infinite D3-branes ending on a configuration of 5-branes, i.e. a 4d $\CN=4$ $\fsu(N)$ SYM theory with a boundary on which it couples to a Gaiotto-Witten BCFT$_3$.

The Riemann surface $\Sigma$ can be taken to correspond a quadrant $w=r e^{i\varphi}$, with $r\in (0,\infty)$ and $\varphi\in \left[\frac{\pi}{2},\pi\right]$. At each point of the quadrant we have ${\rm AdS}_4\times \IS_1^2\times \IS_2^2$, fibered such that $\IS^2_1$ shrinks to zero size over $\varphi=\pi$ (negative real axis) and $\IS^2_2$ shrinks to zero size over $\varphi=\pi/2$ (positive imaginary axis). 
Hence the boundaries of the quadrant $\Sigma$ are actually not boundaries of the full geometry, which closes off smoothly over those edges. The general solution includes a number of 5-brane sources, describing the NS5- and D5-branes, and spikes corresponding to asymptotic regions AdS$_5\times\IS^5$, which describe 4d $\CN=4$ SYM sectors on D3-branes. 

\begin{figure}[htb]
\begin{center}
\includegraphics[scale=.35]{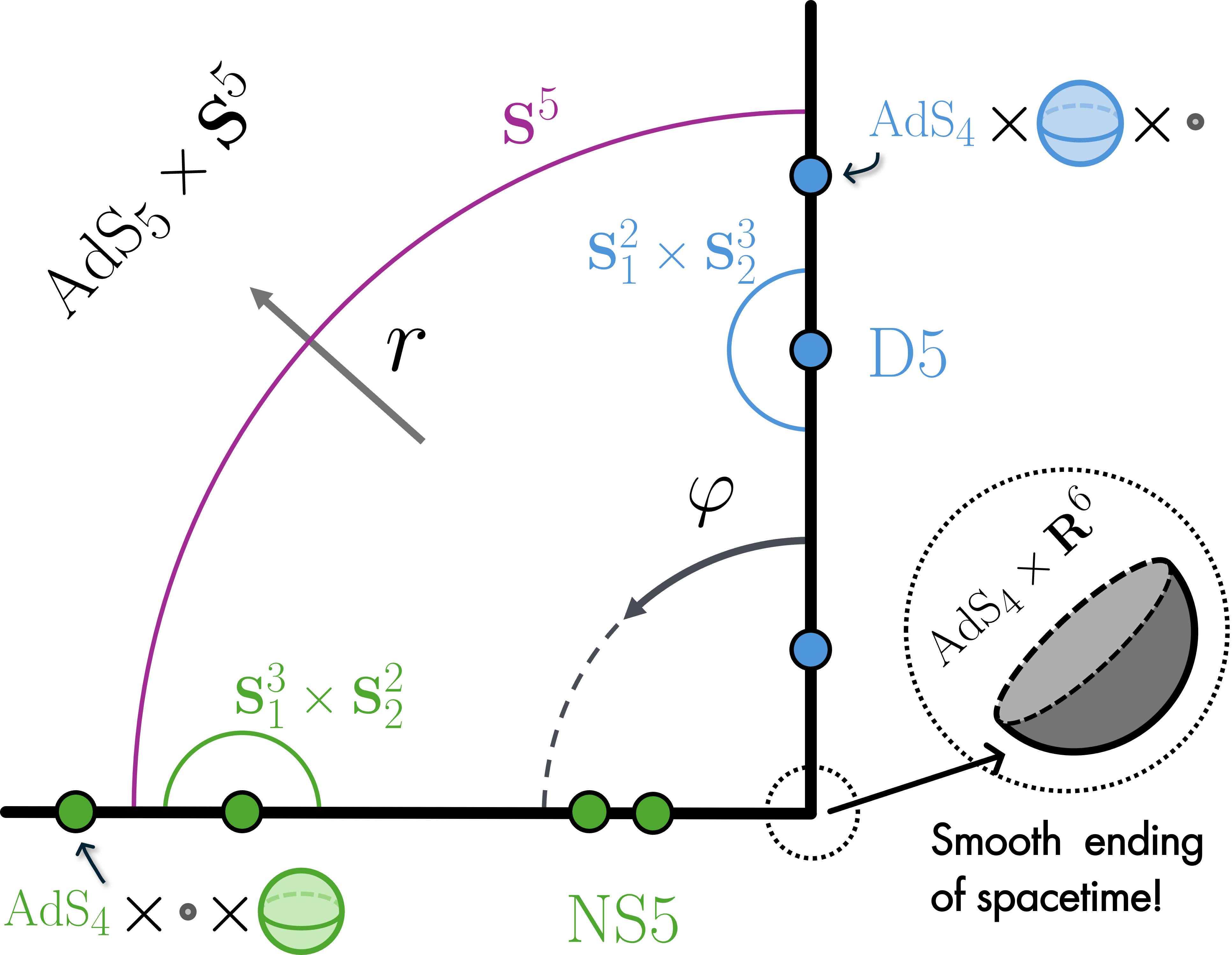}
\caption{\small Riemann surface over which the AdS$_4\times\IS^2_1\times \IS^2_2$ is fibered. The horizontal (resp. vertical) axis is the locus at which the $\IS_1^2$ (resp. $\IS_2^2$) shrinks. The green (resp. blue) dots in the horizontal (resp. vertical) axis correspond to the location of NS5-branes (resp. D5-branes). We have indicated the segments describing $\IS^5$ (purple arc) and $\IS^2\times \IS^3$'s (blue and dark green arcs) in the full fibration.}
\label{fig:quadrant}
\end{center}
\end{figure}

All these features are encoded in the quantitative expression for the functions $h_1$, $h_2$, which for this class of solutions, have the structure 
\begin{subequations}
\begin{align}
h_1&= 4{\rm Im}(w)+2\sum_{b=1}^m{\tilde d}_b \log\left(\frac{|w+il_b|^2}{|w-il_b|^2}\right)\, ,\\
h_2&=-4{\rm Re}(w)-2\sum_{a=1}^n d_a \log\left(\frac{|w+k_a|^2}{|w-k_a|^2}\right)\, .
\end{align}
\label{the-hs}
\end{subequations}
Here the supergravity solution includes 5-brane sources, which are
located at specific points in the two edges of the quadrant. Concretely the NS5-branes
are along the $\varphi=\pi$ axis, with stacks of multiplicity $n_a$ at
positions $w=-k_a$, and the D5-branes are along the $\varphi=\pi/2$
axis, with stacks of multiplicity $m_b$ at positions $w=il_b$. The
multiplicities of 5-branes are related to the parameters $d_a$,
${\tilde d}_b$ via
\[
n_a= 32\pi^2 d_a\in\IZ \quad ,\quad m_b= 32 \pi^2 {\tilde d}_b\in\IZ\, .
\label{quant1}
\]
Specifically, the supergravity solution near one of this punctures is locally of the form of a NS5-brane spanning AdS$_4\times \IS^2_2$ (respectively a D5-brane
spanning AdS$_4\times\IS^2_1$), with $n_a$ units of NSNS 3-form flux
(respectively $m_b$ units of RR 3-form flux) on the $\IS^3$
surrounding the 5-brane source. The latter is easily visualised, by
simply taking a segment in $\Sigma$ forming a half-circle around the
5-brane puncture, and fibering over it the $\IS^2$ fibre shrinking at
the endpoints, so that we get a topological $\IS^3$, see figure
\ref{fig:quadrant}. Hence, near each 5-brane puncture, the metric factorises as
AdS$_4\times \IS^2\times \IS^3$ fibered along a local radial
coordinate $\tilde{r}$ parametrising the distance to the puncture.

%%%%%%%%%%%
%%%%%%%%%%%

Asymptotically, at $r\to\infty$ far away from the 5-branes, the two
$\IS^2$'s combine with the coordinate $\varphi$ to form a $\IS^5$, so
in this limit we have AdS$_5\times\IS^5$, with $N$ units of RR 5-form
flux.  The parameter $N$ will shortly be related to other quantities
in the solution. The $\IS^5$ is manifest in figure \ref{fig:quadrant},
by fibering $\IS_1^2\times\IS_2^2$ over an arc $\varphi\in[\pi/2,\pi]$
at fixed $r>|k_a|,|l_b|$, so that each of the two $\IS^2$ shrinks to
zero size at one of the endpoints of the arc, leading to a topological
$\IS^5$.

The $\IS^5$ can be deformed to smaller values of $r$, and the 5-form
flux is preserved, except when the arc crosses one of the 5-brane
punctures. At this point, the 5-form flux can escape through the
5-brane and decreases, leaving less flux in the inner $\IS^5$ shell;
this is the gravitational manifestation that some number of D3-branes
ends on the 5-brane. Hence, the change in the 5-form flux encodes the
linking number of the 5-branes in the corresponding stack. Following
\cite{Aharony:2011yc}, in the $\IS^2\times\IS^3$ geometry around the
stacks of $n_a$ NS5-branes and of $m_b$ D5-branes, there are
non-trivial integrals
\[
\int_{\IS^2_2} C_2=K_a\quad,\quad \int_{\IS^3_1} H_3=n_a\quad ;\quad \int_{\IS_1^2} B_2={\tilde L}_b\quad,\quad \int_{\IS^3_2} F_3=m_b\, .
\label{the-fluxes}
\] 
Note that here we have used the usual linking numbers $K_a$ for the NS5-branes, but the modified linking numbers ${\tilde L}_b$, introduced in section \ref{sec:gw} for the D5-branes. We recall that the linking numbers $K_a$, $L_b$ (respectively ${\tilde K}_a$, ${\tilde L}_b$) of 5-branes are defined as the net number of D3-branes ending from the right on the 5-brane plus (respectively, minus) the number of 5-branes of opposite kind to the left (respectively, to the right) of the 5-brane.

With this proviso, the change in the flux 
\begin{equation}
{\tilde F_5}=F_5+B_2F_3-C_2H_3\, ,
\label{the-f5}
\end{equation}
with $F_5=dC_4$, upon crossing an NS5-brane (resp. D5-brane) is given by $n_aK_a$ (resp. $m_b{\tilde L}_b$), with no sum over indices. Hence the stacks gather the 5-branes with same linking number, and their $\fu(n_a)$, $\fu(m_b)$ gauge symmetry is the holographic dual of the enhanced non-abelian flavour symmetries (or their duals) for the BCFT$_3$ mentioned in \ref{sec:gw}.

An important point about (\ref{the-fluxes}) is that the integrals of NSNS and RR 2-form potentials on the $\IS^2$'s are essentially integers, related to linking numbers of the brane configuration. This may seem striking, because periods of $p$-form potentials are not topological quantities, and can be changed continuously. Hence, we clarify that the values in (\ref{the-fluxes}) correspond to a particular choice for $B_2$, $C_2$ and $C_4$, specifically leading to a vanishing $F_5=dC_4$ on the asymptotic $\IS^3\times\IS^2$'s associated to the 5-branes, see \cite{Aharony:2011yc}. Any topological deformation of the configuration, such that the periods of $B_2$, $C_2$ are no longer integer, implies a change in $C_4$ such that $F_5$ is no longer vanishing on $\IS^3\times\IS^2$, and compensates thing off, so that we recover the same flux of ${\tilde F}_5$ in (\ref{the-f5}). Thus, the invariant statement is that the 5-form flux jumps by the corresponding integer quantity upon crossing the 5-branes.

The flux over the $\IS^5$'s decreases in such crossings until we reach
the region $r<|k_a|,|l_b|$, in which there is no leftover flux, so
that the $\IS^5$ shrinks and spacetime ends in an smooth way at $r=0$,
see figure \ref{fig:quadrant}. This implies that the relation between
the asymptotic 5-form flux $N$ and the 5-brane parameters is
\begin{equation}\label{N_D3}
    N=\sum_a n_a K_a+\sum_b m_b{\tilde L}_b\, ,
\end{equation}
nicely dovetailing (\ref{d3-flux}). 

A final point we had skipped is the relation between the location of the 5-brane punctures $k_a$, $l_b$ and the linking numbers $K_a$, ${\tilde L}_b$, i.e. the 2-form fluxes (\ref{the-fluxes}). The positions of the 5-brane punctures are determined by the expressions
\begin{equation}
    K_a=32\pi \left(k_a +2\sum_b{\tilde d}_b \arctan\left(\frac {k_a}{l_b}\right)\right)\quad , \quad
    {\tilde L_b}=32\pi \left(l_b +2\sum_a d_a \arctan\left(\frac {k_a}{l_b}\right)\right)\, .\quad 
    \label{positions-linking}
\end{equation}
We refer the reader to the references for a derivation of this result and other related details.

As we had anticipated, and as emphasised in \cite{Huertas:2023syg}, the above supergravity background describes a bordism to nothing of the asymptotic AdS$_5\times\IS^5$ via an End of the World (ETW) configuration \cite{VanRaamsdonk:2021duo}. We review this perspective in what follows, also as a recap of the fundamental parameters determining the structure of the solution, and their interplay with the field theory parameters. 

Let us describe the asymptotic AdS$_5$($\times \IS^5$) region in the Poincaré patch, so the the holographic boundary (which is a half-space semi-infinite in the direction 3) is manifest. The AdS$_4$ geometry in the supergravity solution corresponds to an AdS$_4$ slicing of AdS$_5$, where the slices of the foliations correspond to radial lines sticking out of the holographic 3d boundary of AdS$_4$, see figure \ref{fig:sandwich}. The slices of this foliation are labelled by the coordinate $r=|w|$ in the Riemann surface $\Sigma$ of the supergravity solution. The asymptotic region $r\to\infty$ in the 10d supergravity solution corresponds to the radial lines closer to the 4d holographic boundary, and correspond to an AdS$_5$ ($\times\IS^5$) with $N$ units of 5-form flux. 

As one moves inward in the Riemann surface to values of $r$ corresponding to the 5-brane puncture locations, the 5d geometry hits 5-brane sources spanning AdS$_4$ slices. In the Poincar\'e picture they correspond to lines sticking out radially from the 3d boundary of the 4d holographic boundary in a fan-like structure. The 5-brane sources actually correspond to local internal geometries $\IS^2\times\IS^3$, with background fields (\ref{the-fluxes}), namely 3-form fluxes $n_a$ or $m_b$ (encoding the 5-brane multiplicity for NS5- and D5-branes, respectively), and 2-form backgrounds $K_a$, ${\tilde L}_b$. The latter determine the explicit angular locations of the 5-brane stacks in Poincaré coordinates (equivalently, the radial positions of the 5-brane punctures in $\Sigma$ in the 10d solution) via (\ref{positions-linking}). The 5-form flux over the $\IS^5$ decreases upon crossing the 5-branes by amounts $n_aK_a$ or $m_b{\tilde L}_b$, respectively. Eventually, the 5-form flux is completely peeled off and at a critical angle in  Poincaré coordinates the $\IS^5$ shrinks and spacetime ends. Hence, the solution describes and End of the World (ETW) configuration.

A final word of caution about the above 5d picture. As is familiar,
the scale of the internal geometries are comparable to those of the
corresponding AdS geometries \footnote{This is known as `no scale
  separation' in the swampland literature \cite{Lust:2019zwm} (see
  \cite{Coudarchet:2023mfs} for a review and further references).}. In
our setup this holds both for the asymptotic AdS$_5\times\IS^5$ and
the AdS$_4\times\IX_6$ ``bag'' dual of the BCFT$_3$. As usual in
holography, one may often still use a lower dimensional perspective,
by removing the internal space via a dimensional reduction, rather
than a physical effective theory below a decoupled KK scale. This is
extremely useful in cases where the dimensional reduction is a
consistent truncation, as in the case of the asymptotic
AdS$_5\times\IS^5$, where the 5d perspective has been usefully
exploited for over two decades. To some extent, as discussed in detail
in \cite{Huertas:2023syg}, this does not apply to the configuration
with the ETW configuration, due to its reduced symmetry, and the
lower-dimensional perspective has a more limited applicability. This
will however not pose any problem for our analysis, which is focused
on topological aspects related to the generalised symmetries of the
QFT and its SymTFT as obtained from the topological sector of the
holographic realisation. We turn to this study in the next section.

\section{The SymTFT}
\label{sec:symtft}

The 4d ${\cal N}=4$ $\fsu(N)$ SYM theory on spaces without boundary has a well known SymTFT, which can be obtained from the topological sector of its holographic dual \cite{Witten:1998wy}. The 5d SymTFT is a theory for two $\IZ_N$-valued 2-form fields, which encode the possible  electric and magnetic $\IZ_N$ 1-form symmetries of the 4d physical theory. In this section we explore the modification of this SymTFT due to the presence of boundaries, and in particular focus on physics of the 5d 2-form fields and its implication for the 4d 1-form symmetries.

A natural way to obtain the SymTFT in $d$-dimensional quantum field
theories with a $(d+1)$-dimensional holographic dual is by extracting
the topological sector of the dual
\cite{Witten:1998wy,Aharony:2016kai,Hofman:2017vwr,Bah:2019rgq,Bah:2020jas,Bergman:2020ifi,Bah:2020uev,DeWolfe:2020uzb,Apruzzi:2021phx,Bah:2021mzw,Bah:2021hei,Iqbal:2021tui,Bah:2021brs,Apruzzi:2022dlm,Damia:2022seq,Apruzzi:2022rei,GarciaEtxebarria:2022vzq,Bergman:2022otk,Antinucci:2022vyk,Etheredge:2023ler,Bah:2023ymy,Sela:2024okz,Heckman:2024oot,Heckman:2024obe,Bergman:2024aly}. (More
generally, it is possible to perform a similar analysis for
geometrically engineered QFTs without requiring the existence of a
tractable large $N$ gravitational dual, see for instance
\cite{DelZotto:2015isa,GarciaEtxebarria:2019caf,Morrison:2020ool,Albertini:2020mdx,Gukov:2020btk,Bhardwaj:2021pfz,Hosseini:2021ged,Braun:2021sex,Apruzzi:2021nmk,Cvetic:2022imb,DelZotto:2022joo,vanBeest:2022fss,Heckman:2022xgu,Lawrie:2023tdz,Apruzzi:2023uma,Baume:2023kkf,Yu:2023nyn,DelZotto:2024tae,GarciaEtxebarria:2024fuk,Braeger:2024jcj,Franco:2024mxa,Heckman:2024zdo,Cvetic:2024dzu}
for works in this direction.)

In our present context of the 4d $\CN=4$ $\fsu(N)$ SYM with
Gaiotto-Witten BCFT$_3$ boundary conditions, we may use this same
strategy by analysing the holographic dual reviewed in section
\ref{sec:holodual}. Indeed, in the asymptotic AdS$_5\times\IS^5$
region the recipe amounts to reducing the 10d Chern-Simons coupling of
type IIB string theory on an $\IS^5$ with $N$ units of RR 5-form flux,
which, as studied originally in \cite{Witten:1998wy} leads to the 5d
SymTFT\footnote{In this paper it will be sufficient for us to think of
  $B_2$ and $C_2$ as differential forms, as opposed to cochains in
  differential cohomology or K-theory.}
\begin{equation}
    S_{5d}=\frac{N}{2\pi}\int_{\IX_5} B_2 dC_2\, ,
    \label{bf-symtft}
\end{equation}
where we have allowed for a general 5d manifold $\IX$, and we work in
the convention where the supergravity fields $B_2$, $C_2$ are
$2\pi$-periodic. (In a number of places below we will just highlight
the parametric dependence on $N$ or similar parameters fixing the
order of the discrete symmetries.)

\begin{figure}[htb]
\begin{center}
\includegraphics[scale=.35]{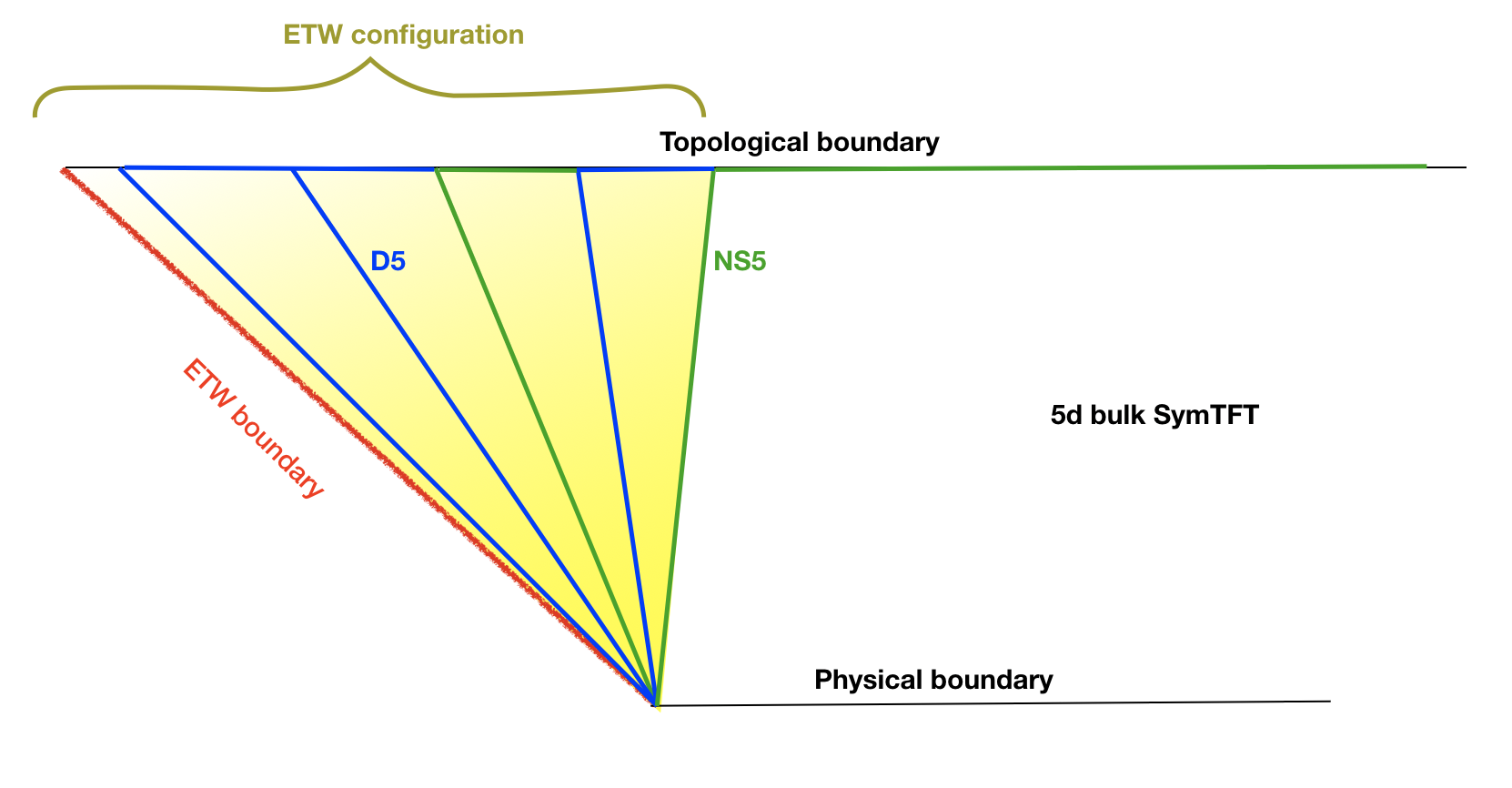}
\caption{\small Sketch of the 5d SymTFT Fan with a set of 5-branes separating different SymTFT wedge regions with different coefficient of the topological coupling (\ref{bf-symtft}), peeling off the $\IS^5$ flux, until one reaches the ETW boundary, at which spacetime ends.}
\label{fig:sandwich}
\end{center}
\end{figure}

In the case of the holographic dual of the 4d SCFT with BCFT$_3$
boundary conditions, we encounter a crucial new structure, which we
dub the SymTFT Fan, as we approach the ETW configuration. Indeed, the
holographic 5d bulk is split into several pieces by the presence of
the 5-branes, which moreover carry non-topological field theories on
their worldvolume. Each piece of the 5d bulk is still associated to a
compactification of the 10d theory on $\IS^5$, albeit with decreasing
number of units of RR 5-form flux, until the flux is eventually peeled
off and the $\IS^5$ can safely shrink, see figure \ref{fig:sandwich},
in the spirit of the cobordism to nothing explained in the previous
section. Hence, we expect that the dimensional reduction on $\IS^5$,
and the SymTFT (\ref{bf-symtft}) for different effective values of $N$
still plays and important role, but new structures are required to
describe the 5-brane theories and how they glue the SymTFTs.

%%%%%%%%%%%
%%%%%%%%%%%
 
Since our Symmetry Theory at hand has several boundaries, we fix some
terminology for them to avoid confusions, see figure
\ref{fig:sandwich}: we refer to the 4d half-space where the physical theory lives as physical boundary, as in standard SymTFTs, because it supports the local degrees of freedom of the $d=4$ $\cN=4$
theory; the same applies to the 3d space supporting the physical
BCFT$_3$, which we will also refer to as the 3d boundary of the 4d
physical theory. The topological boundary is that at which topological
boundary conditions are set for the SymTFTs. Finally, the boundary of
the 5d bulk spacetime, at which the $\IS^5$ shrinks and spacetime
ends, is referred to as ETW boundary. We also sometimes use ETW brane
or ETW configuration for the complete systems of 5-branes and ETW
boundary providing the bordism to nothing for the asymptotic 5d bulk,
in practice the AdS$_5(\times \IS^5$) region.

\subsection{The SymTFT Fan: Gluing SymTFTs via $U(1)$s}
\label{sec:gluing} 

In this section we discuss the structures underlying the gluing of the
different SymTFT wedges via the 5-brane theories into the SymTFT
Fan. This structure of the Symmetry Theory connects very nicely with
several recently introduced tools in the generalisation of the concept
of SymTFTs.

\subsubsection{Relation to SymTrees}
\label{sec:symtrees}

In this section we motivate the relation of the SymTFT Fans with the SymTree construction in \cite{Baume:2023kkf}, see appendix \ref{app:symtree}. This allows to address the key point about how the SymTFTs in the different $\IS^5$ regions of the geometry are related. 

The crucial observation is that, as explained in section \ref{sec:holodual}, in the full supergravity description, the regions near the 5-branes correspond to spikes locally corresponding to AdS$_4\times\IS^2$ times a cone over $\IS^3$, as befits the local geometry around a backreacted stack of 5-branes. For simplicity, in what follows we focus on a single stack of D5-branes (NS5-branes work similarly due to S-duality), and we denote by $n$ the number of 5-branes in the stack and $P$ its linking number.
This implies that there are $n$ units of RR 3-form flux over the $\IS^3$ (NSNS 3-form flux for NS5-branes), and there is a non-trivial background of the NSNS 2-form $B_2$ over the $\IS^2$ (RR 2-form for NS5-branes)
\begin{equation}
    \int_{\IS^2}B_2=P\, .
    \label{b2-background-5brane}
\end{equation}
%  
%%%%%%%%%%%
\begin{figure}[htb]
\begin{center}
\includegraphics[scale=.3]{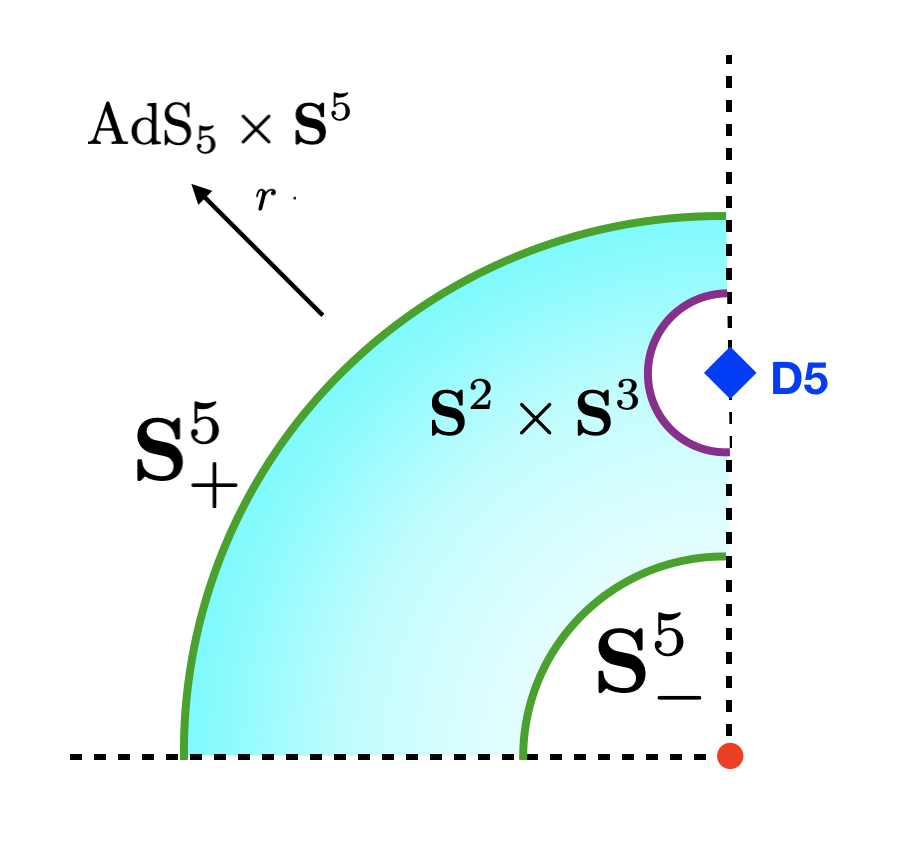}
\caption{\small Picture of the Riemann surface in the supergravity solution displaying the homology relation between the two $\IS^5$'s on two sides of a 5-brane source and the $\IS^2\times \IS^3$ around it.}
\label{fig:homology}
\end{center}
\end{figure}
%%%%%%%%%%%

This implies that when the solution approaches a 5-brane, the geometry
corresponding to the compactification on $\IS^5$ with $N$ units of RR
5-form flux splits into a region corresponding to $\IS^5$ with $N-nP$
units of RR 5-form flux (the 5d region ``across'' the 5-brane in
figure \ref{fig:sandwich}), and a region corresponding to a
compactification on $\IS^2\times\IS^3$ with $nP$ units of 5-form flux.
The latter arises from
\begin{equation}
    \int_{\IS^2\times\IS^3}{\tilde F}_5=\int_{\IS^2\times\IS^3}(F_5+B_2F_3)=nP\, ,
    \label{5form-flux-s2xs3}
\end{equation}
where we have used the fact that there are no D3-brane sources and hence no $F_5$ flux in the $\IS^2\times\IS^3$ regions. In other words, the $\IS^2\times\IS^3$ is homologically the difference of the two $\IS^5$'s in both sides of the 5-brane, see figure \ref{fig:homology}, hence the total flux of ${\tilde F}_5$ must be conserved.

%%%%%%%%%%%
\begin{figure}[htb]
\begin{center}
\includegraphics[scale=.25]{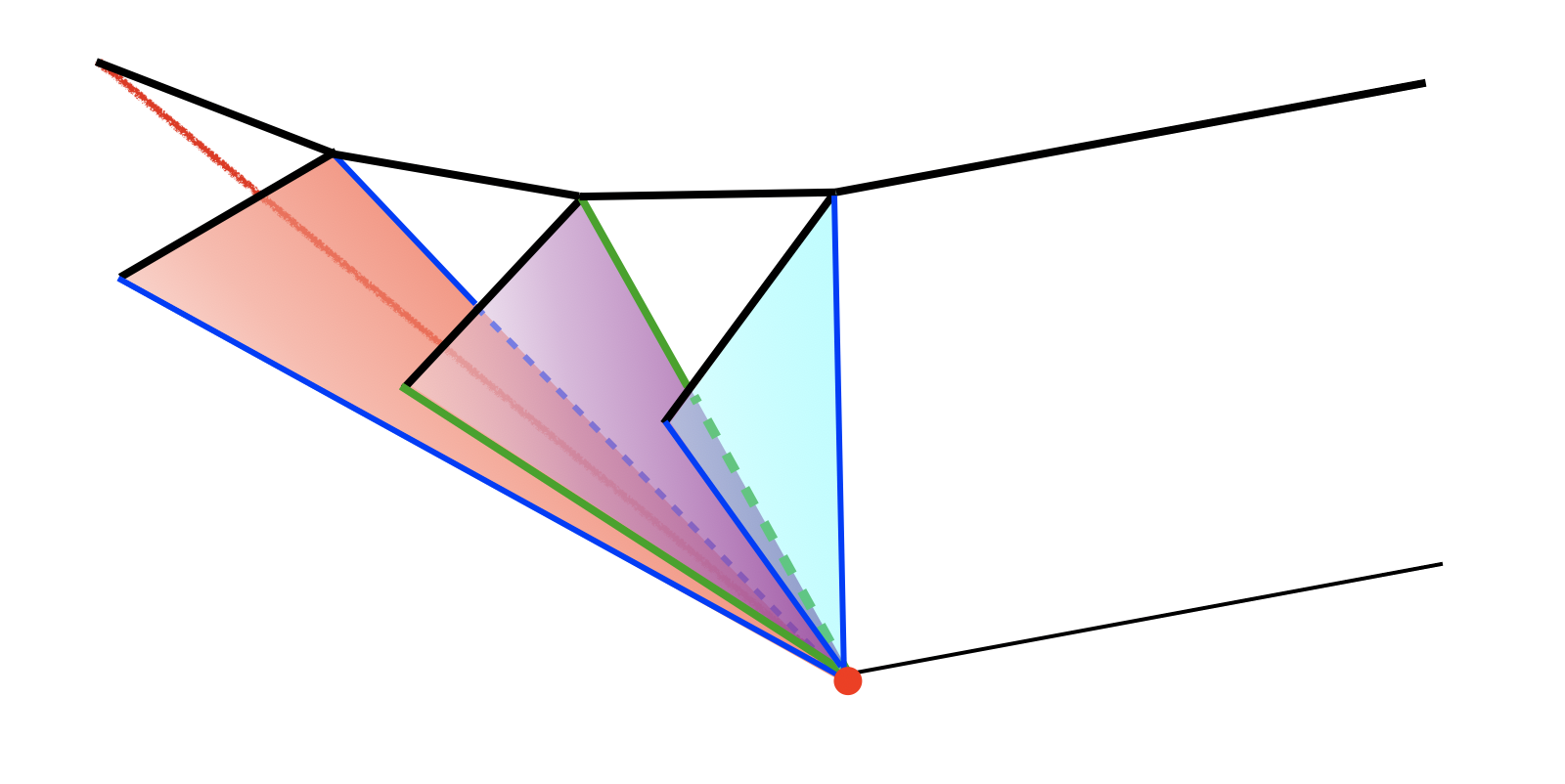}
\caption{\small Depiction of the SymTFT Fan, built as a chain a of SymTFTs of the kind (\ref{bf-symtft}) with different coefficients (shown as unpainted areas) ending in the ETW boundary (thick red line). They are separated by junction theories (green and blue lines) at which the different branch SymTFT$_{\IS^2\times\IS^3}$'s join. Each such branch SymTFT$_{\IS^2\times\IS^3}$ ends up in a physical 5-brane theory, also depicted as green or blue lines. The SymTree-like structure occurs in directions parallel to the boundaries of the Symmetry Theory. This is manifest in the topological boundary (upper thick black line); on the physical boundary, the 5-brane SymTFT branches all converge at the physical 3d BCFT providing the boundary conditions for the 4d physical theory in half-space.}
\label{fig:tree}
\end{center}
\end{figure}
%%%%%%%%%%%

The splitting of the geometry implies a splitting of the SymTFT, as follows. As suggested above, the two $\IS^5$ regions (on both sides of the 5-brane) lead to two SymTFTs of the kind (\ref{bf-symtft}) with coefficients $N$ and $M\df N-nP$, which we denote by SymTFT$_N$ and SymTFT$_{M}$, respectively. On the other hand, out of their meeting locus there sticks out another SymTFT (which we denote by SymTFT$_{\IS^2\times\IS^3}$), obtained from the reduction of the topological sector of the 10d theory on the compact $\IS^2\times\IS^3$ geometry with the above mentioned fluxes and background fields. The structure of this SymTFT$_{\IS^2\times\IS^3}$ will be discussed in section \ref{sec:5-brane-symtft}, but we anticipate it contains a coupling (\ref{bf-symtft}) with coefficient $nP$. The three SymTFTs meet at a junction, which supports non-topological degrees of freedom, which will be discussed in section \ref{sec:junction}.  The complete picture of the Symmetry Theory is hence a modification of figure \ref{fig:sandwich} in which each 5-brane is replaced by a branch SymTFT, as depicted in figure \ref{fig:tree}.

The local  (in the sense that it is away from the physical and topological boundaries of the SymTFTs) structure of several SymTFTs joining along a junction theory is identical to the SymTrees introduced in \cite{Baume:2023kkf} (see appendix \ref{app:symtree} for a quick review), in which a SymTFT can split into multiple decoupled SymTFTs at a (non-topological) junction theory. We will exploit this close relation to disclose the properties of SymTFT Fans, and their implications for the introduction of boundaries in 4d ${\cal N}=4$ $\fsu(N)$ SYM. However, it is crucial to notice that there are the following two important (and related) differences between our setup and that in \cite{Baume:2023kkf}:

\smallskip

$\bullet $ The first is that in SymTrees the splitting of the `trunk' SymTFT into the branch SymTFTs happens as one moves from the topological boundary to the physical boundary of the complete Symmetry Theory; in our present setup, the splitting of the SymTFTs happens along a direction {\em parallel} to the boundaries of the complete Symmetry Theory, namely each of the SymTFTs in the tree (as well as the junction theory) extend across the Symmetry Theory sandwich between the topological and the physical boundaries. On the topological boundary, this implies that each of the SymTFTs after the splitting has its own set of topological boundary conditions, rather than being determined by the junction theory as in \cite{Baume:2023kkf}. We will provide an appropriate interpretation for these independent (up to junction condition constraints) boundary conditions in coming sections. On the physical boundary, the parent SymTFT$_N$ ends up on the 4d SCFT on the half-space, whereas interestingly {\em all} the SymTFTs after the splitting end up on the {\em same} 3d BCFT. This is related to the next point.

\smallskip

$\bullet $ The second main difference, related to the previous one, is that in SymTrees the splitting of the SymTFT is a consequence of the existence of a energy scale below which the full theory is separated into different decoupled sectors \cite{Baume:2023kkf}. In our context, both the 4d physical theory and its 3d boundary theory enjoy superconformal symmetry, so there are no scales in the system, hence the splitting of the SymTFT is not controlled by the same criteria. The difference is clearly displayed in the holographic dual, which for SymTrees displays the splitting along the holographic direction, hence is associated to a physical energy scale, whereas in our setup the splitting occurs along a direction parallel to the holographic boundary and takes place at all scales. In fact, in the supergravity solution the amount of separation between the branches after splitting increases with the distance to the holographic boundary in a characteristic fan-like structure; this nicely dovetails the behaviour in a setup with conformal invariance, where the physical separation is related to the energy scale as dictated by dimensional analysis, due to the absence of fundamental energy scales. 

\smallskip

From the latter argument, we expect on general grounds that the fan-like structure will occur for the holographic dual of any CFT with BCFT boundary conditions. This is a particularly useful observation since explicit supergravity solutions of gravitational duals of CFTs on spacetimes with boundaries are scarce. We thus expect that our observation of the role of SymTFT Fans in 4d $\cN=4$ SYM with boundaries extends to more general classes of CFTs with boundaries, and are useful to uncover their Symmetry Theory structures, even when the supergravity solution is not known. 

For instance, we expect a similar SymTFT Fan structure for the 3d ${\cal N}=2$ BCFT boundary conditions for 4d ${\cal N}=4$ $\fsu(N)$ theory on half-space considered in \cite{Hashimoto:2014vpa,Hashimoto:2014nwa}. From the field theory perspective, this class of boundary conditions preserving a quarter of the superconformal symmetry are essentially a superposition of two kinds of Gaiotto-Witten boundary conditions, associated to sets of NS5- and D5-branes, and of differently oriented 5-branes, denoted by NS5'- and D5'-branes. Even though the supergravity solution of the near horizon limit of this brane system is not known, the Symmetry Theory can be expected to be described as a 5d SymTFT Fan with SymTFT wedges separated by 5-brane theories corresponding to NS5-, D5-, NS5'- and D5'-brane gauge sectors. This has been recently exploited in the discussion of symmetries and anomalies for 4d ${\cal N}=4$ $\fsu(N)$ theory on half-space with 3d ${\cal N}=2$ BCFT boundary conditions in \cite{Huertas:2024mvy}.

\subsubsection{The 5-brane SymTFT }
\label{sec:5-brane-symtft}

Let us now go into more detail into the SymTFT$_{\IS^2\times\IS^3}$ corresponding to the reduction of the topological sector of the 10d theory on $\IS^2\times\IS^3$. The first observation is that, in a sense that will be made more precise later in section \ref{sec:nesting}, it corresponds to the 5d SymTFT of the 4d physical theory of the stack of $n$ 5-branes compactified on $\IS^2$ (and propagating on AdS$_4$ or whatever replaces it in more general context). This has a $\fu(n)\simeq \fsu(n)\oplus \fu(1)$ gauge symmetry. Leaving the $\fu(1)$ factor for later on, we focus on the non-abelian factor, which leads to a 4d supersymmetric $\fsu(n)$ pure SYM theory. The configuration is actually closely related to the realisation in \cite{Apruzzi:2022rei} of this gauge theory as the context of the Klebanov-Strassler holographic dual \cite{Klebanov:2000hb}. In fact, this holographic dual is based on a cone over $T^{1,1}\df\IS^2\times\IS^3$ with $n$ units of RR 3-form flux over the $\IS^3$ and $nP$ units of 5-form flux over $\IS^2\times\IS^3$. Therefore the derivation of the SymTFT of the 5-brane theory essentially follows \cite{Apruzzi:2022rei}. We will not need its detailed structure and simply note that it includes a coupling (\ref{bf-symtft}) with coefficient $nP$ due to (\ref{5form-flux-s2xs3}); we refer the reader to appendix A in \cite{Apruzzi:2022rei} for additional details.

One difference with the simple SymTFT construction is that the 5-brane SymTFT in our setup does not have the standard ``sandwich'' structure, with parallel physical and topological boundaries separated by a gapped bulk. Actually, the boundary structure of this 5d SymTFT sector belongs to a novel generalization of the concept of SymTFT which has only recently been introduced \cite{Cvetic:2024dzu}, and will be explained in section \ref{sec:nesting}. In any event, there is still a topological boundary, at which topological boundary conditions are fixed for the 5d topological fields, in particular the 2-forms. These determine the symmetry structure of the 5-brane theory, the flavour theory of the BCFT$_3$. These properties, e.g. the global structure of the flavour symmetry group, do no have a direct effect on the 3d theory as a QFT. However, they are relevant in setups in which such symmetries are gauged, such as brane realizations of the theory or the holographic setup. Related to this, they are interesting from a more abstract pure QFT perspective as determining the possible gauging of flavour symmetries of the BCFT$_3$. This is however outside the main scope of our work.

\subsubsection{The junction theory }
\label{sec:junction}

In addition to the $\fsu(n)$ theory, there is the $\fu(1)$ degree of freedom. As explained for SymTrees in \cite{Baume:2023kkf}, the $\fu(1)$ is actually not localised on the 5-brane physical theory, but rather belongs to the junction theory. This actually plays  a crucial role in the construction, since the $\fu(1)$ allows the different SymTFTs, which contain $\IZ_N$-valued 2-form fields for {\em different} values of $N$, to be coupled consistently, as we describe below. The price to pay is that this introduces a non-topological sector in the 5d theory, which takes it beyond the pure SymTFT realm.

Let us now describe the structure of this $\fu(1)$ junction theory. Actually, from our earlier discussion about the general structure of the splitting of SymTFTs and the matching of the $B_2,C_2$ sectors with action (\ref{bf-symtft}) with coefficients $N$, $M$, $nP$, the local structure of the junction is identical to that in the SymTree corresponding to an adjoint Higgsing $\fsu(N)\to \fs\, [\, \fu(N_1)\oplus \fu(N_2)\,]$, with $N_1=nP$ and $N_2=M$ in our case. Following \cite{Baume:2023kkf}, the $\fu(1)$ dynamics can be isolated as the decomposition $\fsu(N)\to \fsu(N_1)\oplus \fsu(N_2)\oplus \fu(1)$ inherited from $SU(N)\to [SU(N_1)\times SU(N_2) \times U(1)]/\IZ_L$ with $L={\rm lcm} (N_1,N_2)$ the least common multiple of $N_1, N_2$.

Let us provide a simple heuristic argument that there is a $\fu(1)$ theory supported at the junction theory describing the gluing of two $\IS^5$'s and a $\IS^2\times\IS^3$ geometry (which 5-form fluxes related by their homology relation). The argument is in the spirit of the SymTree structure in \cite{Baume:2023kkf} for a Higgsing, but, as explained, the local structure of the SymTrees applies analogously to our case. Consider a stack of $m$ D3-branes located at a conifold singularity, and split it into two stacks of $m_1$, $m_2$ D3-branes (with $m_1+m_2=m$) located at slightly different positions away from the singular point. We may take the distance between the two points much smaller than the (already small) distance from them to the singular point. This corresponds to performing a Higgsing of the $\fsu(m)\oplus \fsu(m)$ Klebanov-Witten theory \cite{Klebanov:1998hh} to the diagonal $\CN=4$ $\fsu(m)$ SYM theory, which is subsequently Higgsed down to two decoupled $\CN=4$ $\fsu(m_1)$ and $\fsu(m_2)$ SYM theories. At the level of the Symmetry Theory, the above system is described by a SymTree, where one branch corresponds to a SymTFT$_{\IS^2\times\IS^3}$ (i.e. reduction over the base space of the conifold singularity $T^{1,1}\df \IS^2\times\IS^3$), which turns into a SymTFT$_m$ (i.e. reduction on the $\IS^5$ around the joint stack of $m$ D3-branes) to reproduce the Higgsing to the diagonal $SU(m)$, which then splits in a junction into a SymTFT$_{m_1}$ and SymTFT$_{m_2}$ (corresponding to the reductions on the $\IS^5$'s around the individual stacks of $m_1$ and $m_2$ D3-branes). Overall, we have a  SymTree with asymptotic branches given by a SymTFT$_{\IS^2\times\IS^3}$ and the two $\IS^5$ theories, SymTFT$_{m_1}$ and SymTFT$_{m_2}$. The junction theory is obtained by smashing together the transition SymTFT$_{\IS^2\times\IS^3}\;\to$ SymTFT$_m$ and the junction of the SymTFT$_m$, SymTFT$_{m_1}$ and SymTFT$_{m2}$. The latter junction was studied explicitly in \cite{Baume:2023kkf}, where it was shown that it supports a $\fu(1)$ gauge theory, arising from the group theory analysis of the Higgsing 
\begin{equation}
\fsu(m)\to \fs[\fu(m_1)\oplus \fu(m_2)]=\fsu(m_1)\oplus \fsu(m_2)\oplus \fu(1)_L\, ,
\label{higgsing}
\end{equation}
with $L={\rm lcm}\,(m_1,m_2)$. On the other hand, the former transition, associated to the Higgsing $\fsu(m)\oplus \fsu(m)\to \fsu(m)$ does not support any $\fu(1)$'s, because the Higgsing by the bifundamental matter lowers the rank; hence we do not expect additional non-trivial degrees of freedom. The complete junction theory is therefore a 4d $\fu(1)$ gauge theory identical to that associated to the Higgsing (\ref{higgsing}). 

The above process generalizes to a multi-branch SymTree implementing the partition of $N$ (\ref{N_D3}), with a junction per 5-brane stack, with each junction described analogous to that in an adjoint Higgssing. More precisely, the complete analogous Higgsing is
\begin{equation}
\fsu(N)\, \to\,   \mathfrak{s}[\,\oplus_a {K_a }\,\fu(n_a)\oplus_b {\tilde L}_b\,\fu(m_b)\,\big]\, \to\,   \mathfrak{s}[\,\oplus_a \fu(n_a) \oplus_b \fu(m_b)]\quad\quad
\label{general-decomp}
\end{equation}
where the last Higgsing is to the diagonal subalgebra. One particular case is when the partition has a single term $N=nk$, corresponding to a single stack of $n$ 5-branes with linking number $K$. In this case there is no $\fu(1)$ degree of freedom in the decomposition $\fsu(N)\to \fsu(n)$, hence one should expect no non-trivial junction theory. This is reproduced in the SymTree because in the branch theory across the 5-brane there is no 5-form flux, hence the SymTFT is trivial. The trunk theory simply becomes the theory on the non-trivial branch, without any junction required. Interestingly, this always happens in our SymTFT Fans when one reaches the last 5-brane before the ETW boundary. Hence the number of non-trivial junctions is given by the number of 5-brane stacks minus one, in agreement with the number of $\fu(1)$ terms in the decomposition (\ref{general-decomp}).

Let us return to the simple trivalent SymTree for one of the junctions describing the crossing of the 5-brane in the Symmetry Theory of the 4d $\CN=4$ $\fsu(N)$ SYM on a half-space coupled to BCFT$_3$ degrees of freedom. We may even follow the analysis in \cite{Baume:2023kkf} to write down the junction conditions between the background fields of the 1-form symmetries in the different SymTFTs. In our case, for a SymTree corresponding to a D5-brane SymTFT$_{\IS^2\times\IS^3}$ branch, focusing on the coupling of the junction theory to the RR 2-form background field $C_2$, at the location of the junction we have
\begin{equation}
\left.\frac{m_1}{g}C_2^{(m_1)}\right|_{\rm jnct.}=\left.\frac{m_2}{g}C_2^{(m_2)}\right|_{\rm jnct.}=\left.\frac{m}{g}C_2^{(m)}\right|_{\rm jnct.}\, .
\label{junction-condition}
\end{equation}
Here the relation is in $\IZ_g$, with $g={\rm gcd}(m_1,m_2)$, and in
our case\footnote{Note that in our system, we have $m_2+m=m_1$,
  i.e. $m=m_1-m_2$ as opposed to the above Higgsing, in which
  $m=m_1+m_2$. The sign flip is however irrelevant for the formal
  arguments to derive the presence of the $\fu(1)$ in the junction
  theory and the derivation of the junction conditions. It is formally
  equivalent to considering that $m_2$ describes a stack of
  anti-D3-branes.} $m_1=N$, $m_2=nP$, $m=M$. The superindex in the
fields $C_2$ corresponds to its value in each of the SymTFTs,
evaluated at the location of the junction. There is a similar
gluing condition for $B_2$.

An interesting implication of the above junction conditions is that there is a local 1-form symmetry $\IZ_g$. In fact, this reproduces the electric 1-form symmetries studied in particular cases in section \ref{sec:gw}. For instance, for the boundary conditions corresponding to the $T[SU(N)]$ theory, the Symmetry Theory contains one stack of $N$ D5-branes, each with linking number 1, joining the SymTFT$_N$ of the asymptotic 5d theory, and a trivial SymTFT corresponding to the region with no 5-form flux near the ETW boundary. The jump in 5-form flux is hence from $N$ units directly to zero, and the junction conditions above lead to an unbroken $\IZ_N$ symmetry, matching the expected electric 1-form symmetry. Actually, the specific symmetry realised in the physical theory is determined by the choice of boundary conditions; this can be exploited to reproduce the electric 1-form symmetries for other gaugings $SU(N)/\IZ_p$, although we leave the details as an exercise for the interested reader.

Similarly, for the configuration corresponding to a single D5-brane with linking number $N$, the junction conditions indicate an unbroken $\IZ_N$, in agreement with the field theory result in section \ref{sec:gw}. We will provide a different argument for the $\IZ_g$ symmetry in the general case in section \ref{sec:action-baryon}.

Since the junction theory stretches out to the topological boundary, the junction conditions above also apply to the boundary values for the 2-form fields. This implies that the independent boundary conditions of the SymTFT$_N$ and SymTFT$_M$ on both sides of a 5-brane determine, via the junction conditions, the boundary conditions of the SymTFT$_{\IS^2\times\IS^3}$ of the 5-brane theory. As explained in section \ref{sec:5-brane-symtft}, the latter hence determine the symmetries realized by the 5-brane theory, such as the global structure of the BCFT$_3$ flavour symmetry group. 

In the following section we provide a rederivation of the results in this section from a different perspective, which exploits the worldvolume fields and couplings in explicit D5-branes.

\subsubsection{Retraction of the 5-brane SymTFT }
\label{sec:retraction}

An interesting operation introduced in \cite{Baume:2023kkf} for
SymTrees is the retraction of one or a subset of the branches, which
in their context corresponded to smashing together the physical
boundary of those branches with the junction theory. In our present
setup, we can also consider the retraction of branches in the SymTree,
and in particular it is interesting to consider the retraction of the
branches corresponding to the 5-branes. This now corresponds to simply
collapsing the corresponding SymTFT$_{\IS^2\times\IS^3}$, bringing the
$\fsu(n)$ physical theory on top of the junction theory. In terms of
figure \ref{fig:tree}, we collapse the blue/green lines of each
physical 5-brane theory with the blue/green line of its corresponding
junction theory. The result of this retraction should precisely
correspond to figure \ref{fig:sandwich}, with a fan of SymTFTs of the
kind (\ref{bf-symtft}) separated by the non-topological theories on
the worldvolume of explicit 5-branes.

In fact it is possible to recover the main properties derived from the geometric realisation of the SymTree junction by using the properties of explicit 5-branes in the system, as follows. Since the interesting structure is related to the center of mass $U(1)$, we simply take a single D5-brane, i.e. $n=1$ (general $n$ simply introduces some multiplicities in the coefficients; also NS5-branes lead to similar results up to S-duality). Let us consider the topological couplings arising from the D5-brane worldvolume, in particular
\begin{equation}
   \int_{D5} (\, C_4 {\cal F}_{D5}+ C_2 {\cal F}_{D5}{\cal F}_{D5}\,)\, .
   \label{d5-coupling-general}
\end{equation}
where 
\begin{equation}
    {\cal F}_{D5}=B_2 + F_{D5}\, .
\end{equation}
Here $B_2$ is the pullback of the bulk $B_2$ on the D5-brane worldvolume, and $F_{D5}$ is the  field strength for the D5-brane worldvolume $U(1)$ 1-form gauge connection $F_{D5}=dA_{D5}$. In (\ref{d5-coupling-general}) and similar forthcoming expressions we will skip numerical factors in the coefficients, and just highlight the parametric dependence on the flux or brane multiplicities.

We now use the linking number is given by the total worldvolume monopole charge, namely
\begin{equation}
    \int_{\IS^2}  {\cal F}_{D5}=P\, .
    \label{monopole-charge}
\end{equation}
This is the D5-brane probe version of the supergravity background (\ref{b2-background-5brane}), because the NSNS 2-form $B_2$ in the 5-brane backreacted geometry encodes the 2-form $B_2$ in the probe approximation together with the 5-brane worldvolume gauge field $A_{D5}$ (which is just a trivialisation of the latter). Hence we get the 4d couplings
\begin{equation}
   P\int_{4d} (\, C_4 + C_2 {\cal F}_{D5}\,)\, .
   \label{d5-coupling-4d}
\end{equation}
The first term in (\ref{d5-coupling-4d}) describes the coupling of the D5-brane to the RR 4-form, and implies that the RR 5-form flux on $\IS^5$ jumps by $P$ units as one crosses the D5-brane. Regarding the second term, the contribution of $B_2$ in ${\cal F}_{D5}$ leads to a coupling
\begin{equation}
P\int_{4d} C_2 \wedge B_2\, .
\label{BF-level-jump}
\end{equation}
This nicely reproduces the topological coupling of the 5-brane branch SymTFT$_{\IS^2\times\IS^3}$ in the former SymTree before the retraction. It accounts for the jump in the coefficient of the trunk SymTFT (\ref{bf-symtft}) by $P$ as one crosses the 5-brane,  nicely dovetailing the change in RR 5-form flux discussed in the previous paragraph.

Finally, the remaining term has the structure
\begin{equation}
P\int_{4d} C_2 \wedge F_{D5}\, .
\label{junction-coupling1}
\end{equation}
This should be interpreted as the coupling of the $U(1)$ in the junction theory with the background field $C_2$, as anticipated in the previous section. Note that in the case $n=1$, the $\fsu(n)$ piece of the 5-brane theory is trivial, so the 5-brane theory is just the $\fu(1)$ junction theory.

Note that for a D5-brane, the junction theory couples to the background field $C_2$, ultimately related to the 1-form symmetry of the holographic dual 4d SCFT. Clearly NS5-branes lead to similar couplings for $B_2$, the background field ultimately related to the dual 1-form symmetry of the holographic dual 4d SCFT. In later sections this will become manifest in the behaviour of diverse charged operators upon crossing the 5-branes.

\subsubsection{The boundary of the 5-brane SymTFT and its corner theories}
\label{sec:nesting}

We would now like to remark on an interesting point. We have emphasised that the supergravity solution for the ETW configuration with the 5-brane sources in section \ref{sec:holodual} is the gravity dual of the 3d Gaiotto-Witten BCFT$_3$. It is thus reasonable to expect that at the topological level the structure of the Symmetry Theory in the ETW configuration, with the SymTree-like structure in figure \ref{fig:tree} corresponds to the Symmetry Theory of the 3d BCFT. 

This however leads to a seemingly striking feature: The 3d BCFT is actually coupled to a set of non-topological 4d gauge theories, i.e. the physical 4d theory on the 5-brane worldvolumes. Each of the latter provides the physical boundary of a 5d SymTFT$_{\IS^2\times\IS^3}$, for which the 3d BCFT lies at a corner (the intersection of the 4d 5-brane physical theory, and the 4d junction theory of the SymTree). In fact the 3d theory actually lies at the corner of several such 5d structures, as we saw in Figure \ref{fig:tree}.

%%%%%%%%%%%
\begin{figure}[htb]
\begin{center}
\includegraphics[scale=.35]{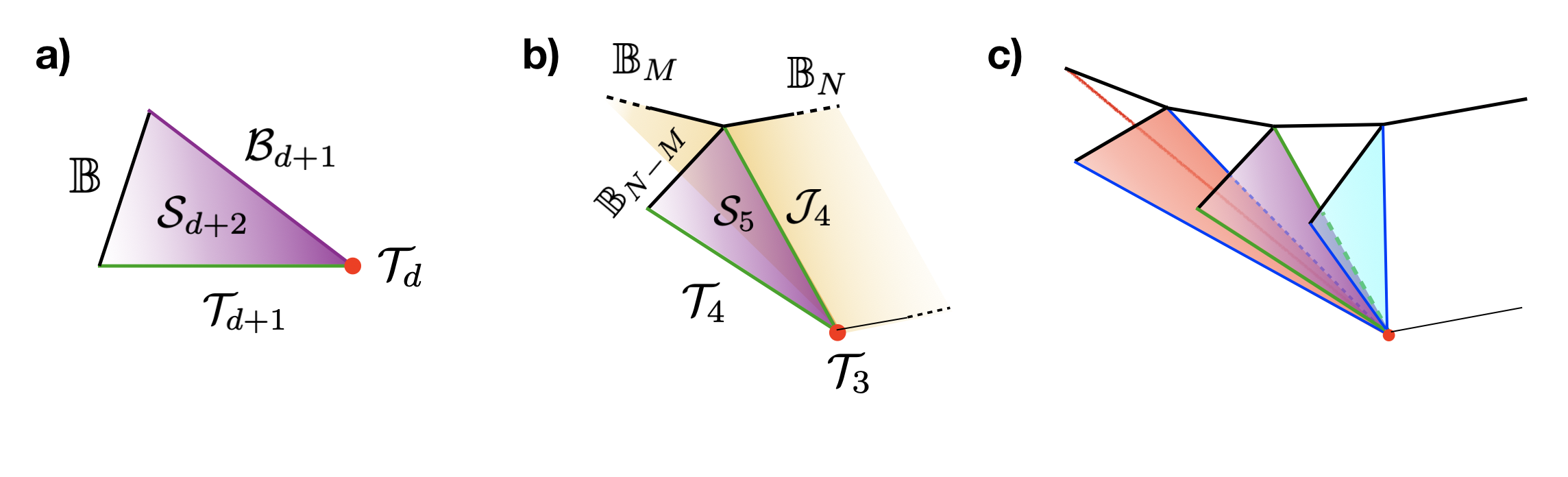}
\caption{\small a) General structure of a corner theory \cite{Cvetic:2024dzu}. The physical $d$-dimensional theory ${\cal T}_d$ is couples to a $(d+1)$-dimensional physical theory ${\cal T}_{d+1}$ describing its flavour structure, with global boundary conditions ${\cal B}_{d+1}$. Both are coupled to a $(d+2)$-dimensional gapped SymTFT ${\cal S}_{d+2}$, with gapped boundary conditions ${\mathbb B}$ at a further boundary. b) The variant of this structure arising for the 5-brane theories in the SymTFT Fan. The BCFT$_3$ is a corner theory of the 5d SymTFT$_{\IS^2\times\IS^3}$ (here denoted by ${\cal S}_5$), which is the symmetry theory of the 5-brane physical theory (here denoted by ${\cal T}_4$). In this case, the topological boundary ${\cal B}_4$ is actually replaced by a junction theory (denoted by ${\cal J}_4$), coupled to the 5d SymTFT$_N$ and SymTFT$_M$ theories. The topological boundary conditions ${\mathbb B}$ (here denoted by ${\mathbb B}_{N-M}$) are correlated to those of the latter SymTFTs (denoted by ${\mathbb B}_{N}$ and ${\mathbb B}_{M}$) by the junction conditions. c) The actual SymTFT Fan, c.f. figure \ref{fig:tree}, includes one copy of the structure in figure (b) per 5-brane theory, making the BCFT$_3$ a corner theory of several 5d SymTFTs.}
\label{fig:nesting}
\end{center}
\end{figure}
%%%%%%%%%%%

Although this structure may seem exotic, it is actually a variant the recent generalised structure of corner theories introduced in \cite{Cvetic:2024dzu}. A typical presentation of a corner theory is shown in figure \ref{fig:nesting}a. This structure provides the Symmetry Theory for a $d$-dimensional physical theory ${\cal T}_d$ whose flavour symmetries are described in terms of a $(d+1)$-dimensional physical theory ${\cal T}_{d+1}$ sticking out of ${\cal T}_d$. The configuration in figure \ref{fig:nesting}a the Symmetry Theory of the combined system. It is given by $(d+2)$-dimensional gapped SymTFT ${\cal S}_{d+2}$ of the $(d+1)$-dimensional flavour physical theory, bounded by the $(d+1)$-dimensional gapped  boundary ${\cal B}_{d+1}$, and the additional gapped boundary ${\mathbb B}$. In other words, the $d$-dimensional physical theory ${\cal T}_d$ sits at a corner of the $(d+2)$ SymTFT ${\cal S}_{d+2}$, at the intersection of the physical theory ${\cal T}_{d+1}$ and the topological boundary ${\cal B}_{d+1}$.

The variant of this corner theory construction arising in our case is shown in figure \ref{fig:nesting}b. The BCFT$_3$ (denoted by ${\cal T}_3$ in the figure, to maintain the general notation) is sitting at the corner of the 5d SymTFT$_{\IS^2\times \IS^3}$ (denoted by ${\cal S}_5$), at the intersection of a 4d physical theory  of the 5-branes (denoted by ${\cal T}_4$) and a 4d boundary, which is now not at topological one, but is rather given by the junction theory (denoted by ${\cal J}_4$, of the SymTFT$_{\IS^2\times \IS^3}$ with the $\IS^5$ SymTFT$_N$ and SymTFT$_M$. The junction reaches the former topological boundary ${\mathbb B}$, leading to a junction of topological boundaries, denoted by ${\mathbb B}_N$ and ${\mathbb B}_M$ for the SymTFT$_N$ and SymTFT$_M$ and ${\mathbb B}_{N-M}$ for ${\cal S}_{d+2}$, as explained in section \ref{sec:junction}.

In fact, the situation is even more complex because there is one such corner structure per 5-brane, so the BCFT$_3$ is actually a 3d corner theory of a fan of several such 5d SymTFT's, as shown in figure \ref{fig:nesting}b, c.f. figure \ref{fig:tree}.

\subsection{Topological operators and crossing the SymTFT Fan}
\label{sec:actions1}

We have discussed the behaviour of the different topological fields in the SymTFT Fan of the 4d $\CN=4$ $\fsu(N)$ theory on a spacetime with a boundary coupled to a Gaiotto-Witten BCFT$_3$. In this section we discuss the behaviour of various topological operators, in particular as they move across 5-branes between different SymTFT wedges. 

\subsubsection{Generalities}
\label{sec:generalities-operators}

Since our SymTFT Fans have a local SymTree structure, studying the motion of operators across 5-branes in our setup is similar to studying the motion of topological operators among the trunk and the branch SymTFTs in a SymTree in \cite{Baume:2023kkf}; we review this construction in appendix \ref{app:symtree}. The main differences arise from the different orientation of the SymTree with respect to the topological and physical boundaries of the SymTFTs, and in the corresponding QFT interpretation. 

%%%%%%%%%%%
\begin{figure}[htb]
\begin{center}
\includegraphics[scale=.4]{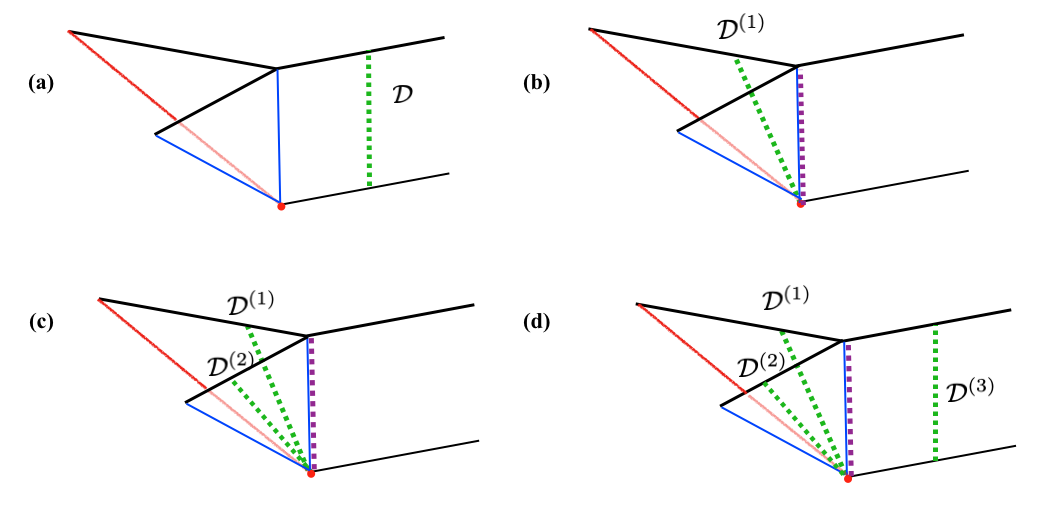}
\caption{\small Sketch of various charged defect topological operators in a SymTFT Fan. Figure (a) provides the  initial configuration of a charged topological defect in the trunk SymTFT of the 4d bulk CFT. Pushing it onto the ETW boundary, it crosses the junctions and can turn into various possible configurations, describing charged defects of the BCFT, with some illustrative examples depicted in figures (b), (c). The purple line on the junction indicates the appearance of some junction operator. Figure (d) is a general configuration with line operators for both the CFT and BCFT, obtained by fusion of previous ones.}
\label{fig:symtft-fan-defects}
\end{center}
\end{figure}
%%%%%%%%%%%

For instance, let us focus on charged defects in our setup. They correspond to topological operators stretching between the two boundaries of the 5d Symmetry Theory. In the asymptotic region corresponding to the SymTFT$_N$ of the 4d $\CN=4$ $\fsu(N)$ SYM theory they correspond to the familiar line operators. The effect of the spacetime boundary and the BCFT boundary conditions can be described by pushing the line operator to the ETW region. The operator will thus cross the different junction theories in the SymTFT Fan, and can be ultimately turned into a set of line operators of the different SymTFT$_{\IS^2\times\IS^3}$'s of the 5-branes. This latter description corresponds to an operator of the BCFT$_3$, and in particular allows to read how it transforms under the non-abelian flavour symmetries, as we describe in section \ref{sec:lineops}.

Clearly, the crossing of the SymTFT line operators across junctions is locally described in the same way as in SymTrees, see section \ref{app:crossing}. The main difference is that the crossings of charged defects in the SymTFT Fan do not correspond to that of charged defects in SymTrees, but are rather analogous to those of symmetry generators, cf. figure \ref{fig:symtree-generators}. For convenience, we adapt the latter figure to our present purposes and display in figure \ref{fig:symtft-fan-defects} some of the  charged defects crossings in the SymTFT Fan which appear in the explicit discussion to be carried out in section \ref{sec:lineops}.

\begin{figure}[htb]
\begin{center}
\includegraphics[scale=.4]{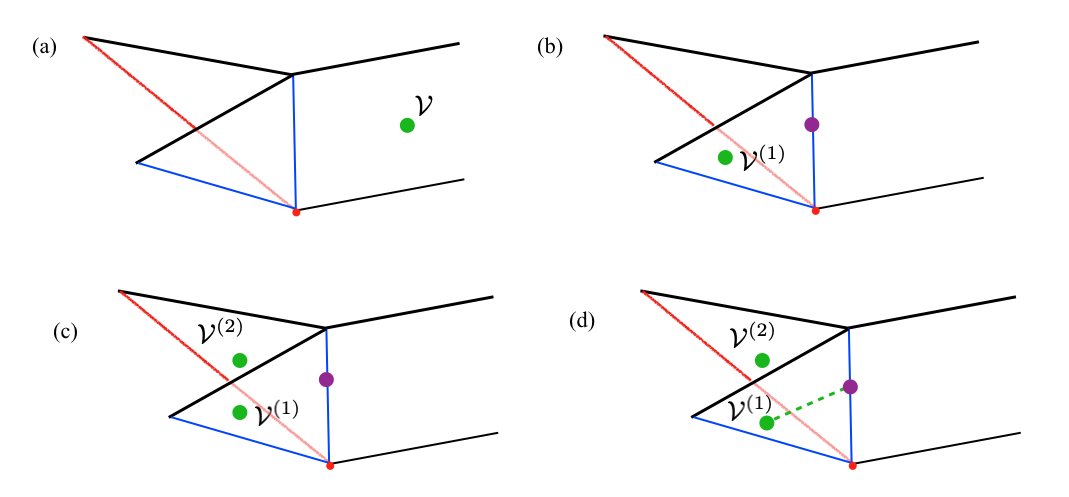}
\caption{\small Sketch of various topological operators for symmetry generators in a SymTFT Fan. Figure (a) provides the  initial configuration of a symmetry generator in the trunk SymTFT of the 4d bulk CFT. Pushing it onto the ETW boundary, it crosses the junctions and can turn into various possible configurations, describing symmetry generators of the BCFT, with some illustrative examples depicted in figures (b), (c), (d). The purple line on the junction indicates the appearance of some junction operator. Figure (d) illustrates the possibility that stretched operators may be created in the crossing with the junction.}
\label{fig:symtft-fan-generators}
\end{center}
\end{figure}

A similar story occurs for symmetry generators, which are topological operators linking the above charged defects. Starting from a symmetry generator in the asymptotic SymTFT$_N$ of the 4d CFT, one can push it to the ETW boundary, crossing it across the junction theories, to turn it into symmetry generators for the SymTFT$_{\IS^2\times\IS^3}$'s of the 5-brane theories. In figure \ref{fig:symtft-fan-generators}, we provide some illustrative examples of the results of these crossing, in particular those to arise in the explicit discussions in sections \ref{sec:lineops}, \ref{sec:action-baryon} and \ref{sec:action-dual-baryon}.

%%%%%%%%%%%
%%%%%%%%%%%

In section \ref{sec:symtrees} we had introduced the complementary representation of  SymTFT Fans by retracting the SymTFT branches corresponding to the 5-branes, as a set of wedges of SymTFTs (\ref{bf-symtft}) with different coefficients, separated by explicit D5- and NS5-branes carrying induced D3-brane charge to ensure conservation of the 5-form flux. 
It is easy to follow this retraction in the configurations in figures \ref{fig:symtft-fan-defects}, \ref{fig:symtft-fan-generators}. We simply smash together the two blue lines corresponding to the junction theory and the physical boundary of the 5-brane branch SymTFT$_{\IS^2\times\IS^3}$, and stack the possibly present operators of the latter on top of the possibly present defects of the junction theory. In the discussion of topological operators below we will find that the same crossing transitions admits different brane interpretations in terms of the full SymTree configuration or its retraction. This will be discussed in the explicit examples below.

We now turn to the study of explicit examples of topological operators and their behaviour with respect to the junctions in the SymTFT Fans of 4d $\CN$ $\fsu(N)$ SYM with Gaiotto Witten BCFT$_3$ boundary conditions.

\subsubsection{Line operators, F1s and D1s}
\label{sec:lineops}

In this section we consider line operators, which are essential for the understanding of the electric and magnetic 1-form symmetries of the 4d $\CN=4$ $\fsu(N)$ SYM theory and its behaviour under the introduction of the BCFT$_3$ boundary conditions.
The different operators can be realized using F1 and D1 strings in the 5d theory, with general $(p,q)$-string operators arising by fusion. As is familiar since \cite{Witten:1998wy}, in the asymptotic SymTFT$_N$ of the CFT$_4$ the basic set of topological operators are realised in string theory in terms of F1 or D1 strings stretching from the topological to the physical boundary (representing Wilson and 't Hooft line operators, respectively), and D1 or F1 string worldsheets linking them (representing the symmetry generating operators, respectively). 

As explained in the previous section, when moving these objects to the ETW region they cross the junction theories and enter other SymTFT$_M$ wedges. This naively leads to the following puzzle. Consider for instance the F1 stretched between the topological and physical boundaries in the SymTFT$_N$, and act on it with a D1 linking it, so it transforms by a phase $e^{2\pi i/N}$. We can consider the same kind of configuration, but in the SymTFT$_M$, in which case the charged operator transforms by a phase $e^{2\pi i /M}$. The puzzle arises because seemingly both the F1 and the D1 can be moved from the SymTFT$_N$ to the SymTFT$_M$ without any discontinuity, i.e. without crossing the 5-brane, by simply avoiding it in the internal geometry. Yet the phase transformation seems to jump discontinuously in the transition. 

The solution to the puzzle is essentially the one realized in SymTrees in \cite{Baume:2023kkf}, reviewed in appendix \ref{app:symtree}, which is realized in our setup as follows. For simplicity we focus the discussion on Wilson line operators, realized in terms of F1 strings stretching between the topological and physical boundaries, and we choose topological boundary conditions compatible with their existence, i.e. global $SU(N)$ form of the 4d gauge group, and similarly for the other branches of the SymTFT topological boundaries. Consider first the case of a Wilson line operator of the $SU(N)$ theory crossing a stack of $n$ D5-branes with linking number 1, so that the junction separates the SymTFT$_N$ associated to the $SU(N)$ theory, the SymTFT$_{\IS^2\times\IS^3}$ of the 5-brane $SU(n)$ theory, and the SymTFT$_M$, with $M=N-n$. Locally the behaviour of the line operators in this crossing is identical to that of operators in a SymTree for the adjoint Higgsing $SU(N)\to SU(N-n)\times SU(n)\times U(1)$, as already mentioned in section \ref{sec:junction}. Hence, the behaviour of Wilson line operators under this crossing, in a way compatible with the action of discrete symmetries, can be obtained by adapting the discussion for adjoint Higgsing SymTrees, reviewed in appendix \ref{app:adjoint}. In the following, we carry out this discussion in a general case of a complete SymTFT Fan with multiple junction crossings.

Let us carry out this explicit constructions, in a fairly general and illustrative case of the $T_\rho[SU(N)]$ theories. As introduced in section \ref{sec:gw-holo}, these correspond to having the $N$ D3-branes ending on sets of $m_b$ D5-branes with linking numbers ${\tilde L}_b$, associated to a partition $\rho$ of $N=\sum_b m_b{\tilde L}_b$.  We again focus on Wilson line operators and consider topological boundary conditions compatible with them in all necessary SymTFTs. In general, charged defects of the asymptotic $SU(N)$ theory will turn into charged defects of the BCFT$_3$, which are moreover charged under the flavour symmetry $\prod_b U(m_b)$. The SymTFT Fan interpretation is that when the F1 string line operator in the asymptotic SymTFT$_N$ is pushed to the ETW region, it succesively crosses the D5-branes and can become a F1 string line operator of the different $U(m_b)$ theories. In SymTree terms, this is associated to the Higgsing
\begin{equation}
SU(N)\, \to\,   S[\,\otimes_b U(m_b{\tilde L}_b)\,]\, \to\, S\big[\,\otimes_b U(m_b)^{{\tilde L}_b}\,\big]\, \to\,   S[\,\otimes_b U(m_b)\,]
\label{higgsing-bis}
\end{equation}
The Higgsing is presented in several steps to make the structure more manifest. The first step maintains the rank and corresponds to a simple adjoint Higgsing, while the second and third take into account that the $m_b{\tilde L}_b$ D3-brane charge in the $b^{th}$ stack of D5-branes collapses onto the diagonal combination $U(m_b)$. As explained in section \ref{sec:junction} the $U(1)$'s in the above $U(m_b)$ groups correspond to those of the junction theories. 

Following appendix \ref{app:adjoint},\footnote{As explained in section \ref{sec:generalities-operators} the meaning of Wilson line operators in SymTrees and SymTFT Fans is different, due to the different relative orientation with respect to the SymTFT boundaries. However, since they are identical locally in the SymTFT bulk, the computations work in a similar way.}, we can now consider Wilson line operators of the $SU(m_b)$ theories, dress them with $U(1)_b$ operators, and form combinations which transform properly under the discrete $\IZ_N$ symmetry of the asymptotic $SU(N)$ theory. These are indeed the BCFT$_3$ operators realized by pushing the asymptotic F1 string line operator to the ETW region. The construction must be slightly generalized to account for the additional steps in the Higgsing (\ref{higgsing-bis}), but essentially is as follows. Consider an $SU(N)$ Wilson line operator, for simplicity in the fundamental representation. Consider now the $SU(m_b)$ Wilson line operator ${\cal W}_b^{\rm naive}$ in the fundamental representation. This transforms properly under the electric $\IZ_{m_b}$ 1-form symmetry of $SU(m_b)$, but not under the asymptotic $\IZ_N$. This can be repaired by dressing it with a $U(1)_b$ Wilson line operator to get the analogue of (\ref{dressed-wls-tree})
\beqa
{\cal W}_b={\cal W}_b^{\rm naive} \exp \left( -i\frac{1}{m_b}\int a_b\right)
\label{dressed-wls-fan}
\eeqa
where $a_b$ is the $U(1)_b$ gauge connection. The decomposition of the fundamental of $SU(N)$ under (\ref{higgsing-bis}) motivated to introduce a weighted product of such operators for all the SymTree branch theories, analogous to (\ref{dressed-wl-sum})
\beqa
{\mathbb W}_{\fund}=\otimes_b \,{\cal W}_b^{{\tilde L}_b}
\eeqa
transforms properly under $\IZ_N$. This is therefore the BCFT$_3$ operator obtained when the Wilson line of the asymptotic $SU(N)$ theory is pushed to the boundary of spacetime. The generalization for Wilson lines in general representations of $SU(N)$ can then be simply obtained by fusion. It is easy to realize that the above process corresponds to pushing the operator in figure \ref{fig:symtft-fan-defects}a to the ETW boundary, resulting in operators of the form in e.g. figures \ref{fig:symtft-fan-defects}b, \ref{fig:symtft-fan-defects}c, depending on the choices of representations.

Clearly, a similar exercise may be carried out for symmetry generator topological operators. Also, the discussion can be generalized to other more involved situation, for instance  ETW configurations including NS5-branes, generalizing the discussion of F1 string operators to D1 strings or general $(p,q)$ strings, or junctions thereof, and allowing for general choices of topological boundary conditions for the diverse SymTFTs. We refrain from entering this detailed account, hoping that the above explicit discussion suffices to illustrate the main features of the application of SymTFT Fans to relate operators in the CFT$_4$ and the BCFT$_3$.

Let us conclude with a comment of the above argument from the string theory microscopic viewpoint. From the string theory perspective, both the F1 and the D1 charges are initially $\IZ$-valued. But the presence of e.g. $N$ units
of 5-form flux in the compact space implies there are physical
processes which violate their charge in $N$ units. These are the
baryonic vertex given by D5-brane wrapped on the 5d compact space,
which emits $N$ F1 strings \cite{Witten:1998xy}, and its S-dual
wrapped NS5-brane, which emits $N$ D1 strings.  This allows to recast
the 5d action (\ref{bf-symtft}) in terms of $\IZ_N$-valued fields in a
5d topological $\IZ_N$ theory. Our statements mean that these discrete
fields for the different SymTFTs are to be identified but up to an
overall normalisation, which in fact corresponds to the junction
conditions (\ref{junction-condition}). We now turn to the discussion of these interesting operators in the next section.

\subsubsection{The baryon vertex}
\label{sec:action-baryon}

The change in the order of the discrete symmetry between the SymTFTs is hence encoded in the existence of the baryon vertex and the dual baryon vertex, and how they behave when moved across the 5-branes, as we study in this section.

In short, as we will see in explicit examples below,  in the full SymTree configuration the crossing transitions can be understood in terms of geometric relations and Freed-Witten consistency conditions \footnote{As usual we use this term to refer to incompatibilities of branes and fluxes in general for non-torsion classes \cite{Maldacena:2001xj}, i.e. beyond the original Freed-Witten study of the torsion case \cite{Freed:1999vc}.}. On the other hand, in the Symmetry Theory after the retraction, the crossing transitions can be understood in terms of Hanany-Witten brane creation effects \cite{Hanany:1996ie} involving the explicit 5-branes. This is expected given the general relation between Freed-Witten consistency conditions and Hanany-Witten effects (see \cite{Berasaluce-Gonzalez:2012awn} for discussion).

For simplicity, we focus on the setup of a single D5-brane stack, as has been implicitly done in the previous pictures, with a single D5-brane with linking number $P$. Hence, we consider a SymTree with a trivalent junction in which the SymTFT$_N$ (corresponding to reduction on $\IS^5$ with $N$ units of 5-form flux) splits into a SymTFT$_M$ (with $M$ units of flux on $\IS^5$) and SymTFT$_{\IS^2\times\IS^3}$ with $N-M=P$ units of 5-form flux (from 1 unit of RR 3-form flux over $\IS^3$ and a NSNS 2-form period $P$ on the $\IS^2$). Equivalently, after the retraction, we have the SymTFT$_N$ and SymTFT$_M$ separated by one explicit D5-brane wrapped on a maximal $\IS^2\subset \IS^5$ and carrying $P$ units of induced D3-brane charge due to the non-trivial integral (\ref{monopole-charge}) of ${\cal F}_{D5}$ over $\IS^2$. 

Consider the baryonic vertex of the SymTFT$_N$, namely a D5-brane wrapped on the $\IS^5$ and emitting $N$ fundamental strings \cite{Witten:1998xy} ending in the physical boundary corresponding to the 4d QFT on the half-space. As explained above, although the baryonic vertex corresponds to the identity operator in the SymTFT$_N$, it has a non-trivial meaning in the microscopic string theory realisation, as encoding the actual reduction of $\IZ$-valued charges to $\IZ_N$.

%%%%%%%%%%%
\begin{figure}[htb]
\begin{center}
\includegraphics[scale=.35]{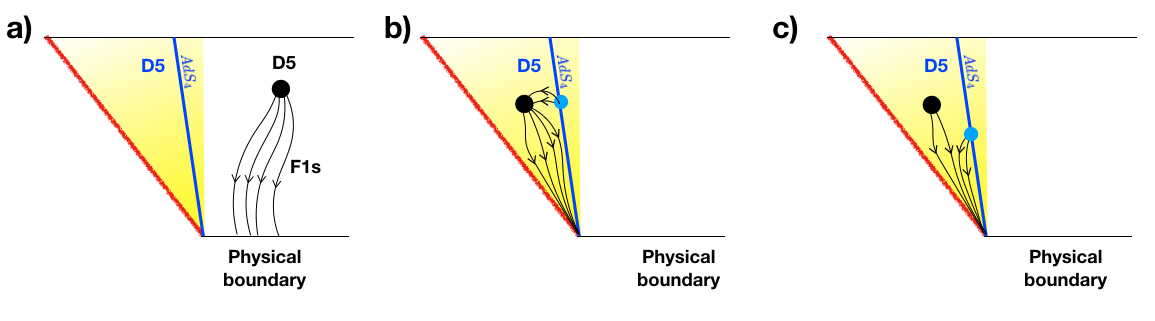}
\caption{\small (a) The SymTFT Fan with a 4d D5-brane (blue) and a baryon D5-brane (black) with its $N$ fundamental strings. (b) Crossing the baryonic D5-brane across the 4d D5-brane leads to the creation of $P$ fundamental strings between them. (c) Recombining $P$ fundamental strings, the D5-brane baryon now emits $M$ fundamental strings. In figures (b) and (c) we have depicted the $P$ fundamental strings stretching out from the same point in the D5-brane, and indicated it with a light blue dot, indicating a localised operator on the 4d D5-brane. }
\label{fig:transition-baryon}
\end{center}
\end{figure}
%%%%%%%%%%%

Let us consider the picture of the retracted SymTree, and consider moving this baryonic D5-brane (and its $N$ attached F1s) across the 4d subspace spanned by the D5-brane wrapped on $\IS^2$ (which we refer to as 4d D5-branes to avoid confusion). In the process of crossing to the SymTFT$_{M}$, there is a Hanany-Witten effect between the baryonic D5-brane and the $P$ units of induced D3-brane charge on the 4d D5-branes, resulting in the creation of $P$ fundamental strings stretching between them. One can now recombine them with $P$ of the original $N$ F1 strings and detach them from the baryonic D5-brane, leaving a D5-brane on $\IS^5$ with just $M$ F1 strings, see figure \ref{fig:transition-baryon}. This is precisely the baryonic vertex appropriate to represent the identity in the in the SymTFT$_{M}$. 

Note however that we still have the $P$ detached strings, which now
stretch from the 4d D5-brane to the physical boundary of the
BCFT$_3$. In holographic language, they correspond to $P$ Wilson
line operators, and correspond to the dynamical quarks localised at
the 3d boundary of the 4d CFT. In the language of the Symmetry Theory,
they define an operator of the junction theory left over after the
crossing, as explained above.

Let us rederive the above crossing of the baryonic operator but in terms of the full SymTree configuration, rather than its retraction. In this picture, the baryonic operator in the SymTFT$_N$, namely the D5-brane on the $\IS^5$ with $N$ units of 5-form flux (hence emitting $N$ F1 strings), is homologically equivalent to the baryonic operator in the SymTFT$_{M}$, i.e. a D5-brane on the $\IS^5$ with $M$ units of flux (hence emitting $M$ F1 strings) and a D5-brane on the $\IS^2\times\IS^3$ of the SymTFT$_{\IS^2\times\IS^3}$ emitting $P$ F1 strings due to the 5-form flux on this branch. It is easy to check that the latter are due to the induced D3-brane charge, and connect directly with the  Hanany-Witten picture above: the D5-brane on $\IS^2\times\IS^3$ has $P$ units of induced D3-brane charge due to the NSNS 2-form background (\ref{b2-background-5brane}); the $P$ induced D3-branes wrap on the $\IS^3$, which has 1 unit of RR 3-form flux (encoding the 4d D5-brane charge), hence have a Freed-Witten anomaly and must emit $P$ F1 strings. The picture of emission of F1 strings due to 5-form flux or RR 3-form flux on the induced D3-branes is just a reflection of the fact that the 5-form flux in this branch arises from the Chern-Simons couplings, c.f. (\ref{5form-flux-s2xs3}). 

%%%%%%%%%%%
\begin{figure}[htb]
\begin{center}
\includegraphics[scale=.4]{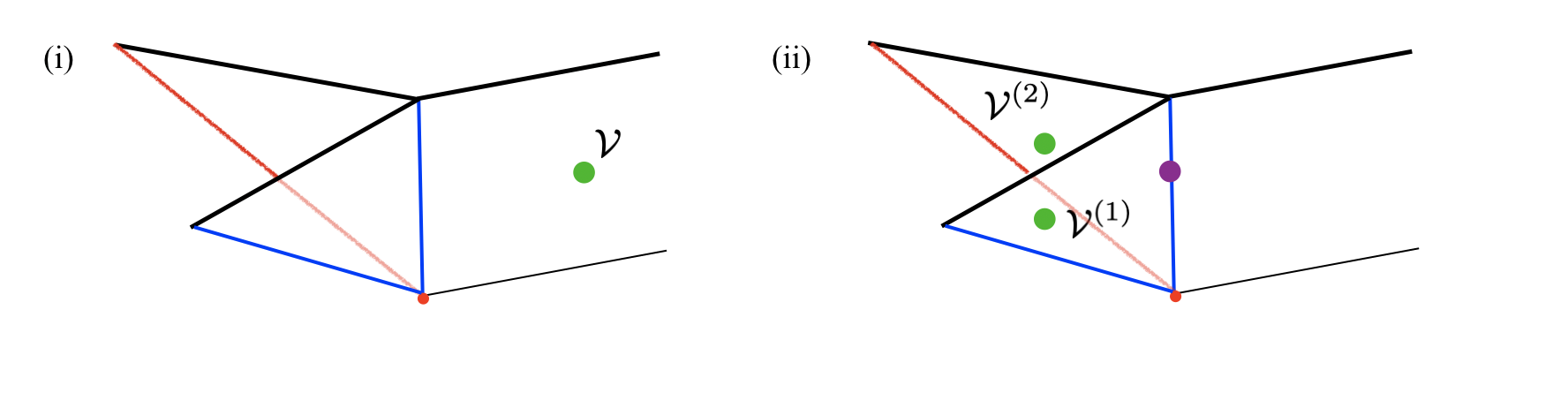}
\caption{\small Transition of operators in the SymTFT Fan, corresponding to the the baryonic operator of the SymTFT$_N$ (i) across the junction, leading to a baryonic operators of the SymTFT$_M$ and SymTFT$_{\IS^2\times\IS^3}$ as well as a junction theory operator (ii).}
\label{fig:d5-crossing}
\end{center}
\end{figure}
%%%%%%%%%%%

We have thus rederived the crossing process in a simple geometric
fashion, and recovered the appearance of the operator on the SymTFT of
the 5-brane theory, see figure \ref{fig:d5-crossing} (analogous to the
transition (a)$\to$ (c) in figure \ref{fig:symtft-fan-generators}). We note that, although the geometric picture of the
crossing would seem to suggest that there is no junction operator
created in the process, the $\fu(1)$ degree of freedom is delocalised,
hence we depict the presence of a junction operator, as discussed with
the explicit 4d D5-brane in the Hanany-Witten description. We also
emphasise that, although the operators we are discussing correspond to
the identity operator in each of the SymTFTs, there is non-trivial
information about how they are related upon crossing the
junction. This is the reflection of the junction conditions of the
underlying fields.

We finish by emphasising that the baryon vertices of the different SymTFTs can be used to derive the local $\IZ_g$ 1-form symmetry unbroken by the junction conditions. Near the junction of the SymTFT$_N$, the SymTFT$_M$ and the SymTFT$_{\IS^2\times\IS^3}$, we can consider processes annihilating F1 strings in sets of $N$, or in sets of $M$ (or equivalently, of $N-M$) or integer linear combinations thereof. So by Bezout's theorem, F1 strings are conserved mod $g$ with $g={\rm gcd}(N,M)$. For instance, in the particular case of the SymTFT of the $T[SU(N)]$ boundary conditions, we have one stack of $n$ D5-branes with linking number 1, so $M=0$, and hence we get a $\IZ_N$ 1-form symmetry, as discussed in the field theory setup in section \ref{sec:gw}, and in section \ref{sec:junction} from the viewpoint of the junction conditions.

\subsubsection{The dual baryon vertex}
\label{sec:action-dual-baryon}

Let us now consider the dual baryon vertex, which is given by an NS5-brane wrapped on the $\IS^5$, emitting $N$ D1 strings, ending on the boundary of the physical 4d theory on the half-space. In the absence of boundary for the 4d QFT, S-duality of 4d ${\CN}=4$ $\fsu(N)$ SYM relates this to the baryonic vertex in the previous section; however, in the presence of boundaries for the 4d QFT, the Symmetry Theory contains explicit 4d D5- or NS5-branes (or their SymTFTs in the SymTree picture), which are not invariant under S-duality. Hence the dual baryon operator must the studied on its own.

We start with a dual baryon vertex of the SymTFT$_N$, given by one NS5-brane wrapped on the $\IS^5$ and emitting $N$ D1 strings due to the 5-form flux. Let us first consider the picture of the retracted SymTree, namely with an explicit 4d D5-brane with $P$ units of induced D3-brane charge, separating the SymTFT$_N$ and SymTFT$_{M}$ regions. If we now move the NS5-brane across the 4d D5-brane, there is a Hanany-Witten effect between the 4d D5-brane and the NS5-brane, leading to the creation of one D3-brane wrapped on the $\IS^2$ and stretched between the NS5- and the D5-brane (for a stack of a number $n$ of 4d D5-branes, there would be $n$ D3-branes created). In addition, there is a non-trivial Hanany-Witten effect of the NS5-brane with the $P$ induced D3-brane charge on the 4d D5-brane, which creates $P$ D1 strings stretched between them. Note that they can be equivalently regarded as $P$ units of D1-brane charge induced on the previous D3-brane due to the non-trivial NSNS 2-form background on the $\IS^2$; for clarity we prefer to keep the discussion of the created D3- and D1-branes separate.

As in the discussion above for the F1 strings of the baryonic D5-brane, one can recombine the newly created $P$ D1 strings with $P$ of the D1 strings emitted by the NS5-branes, to obtain an NS5-brane emitting $M$ D1 strings, precisely as required by the change in the 5-form flux on the $\IS^5$. In the process we are also left with $P$ D1 strings stretching from the D5-brane to the boundary corresponding to the physical theory of the BCFT$_3$.

%%%%%%%%%%%
\begin{figure}[htb]
\begin{center}
\includegraphics[scale=.35]{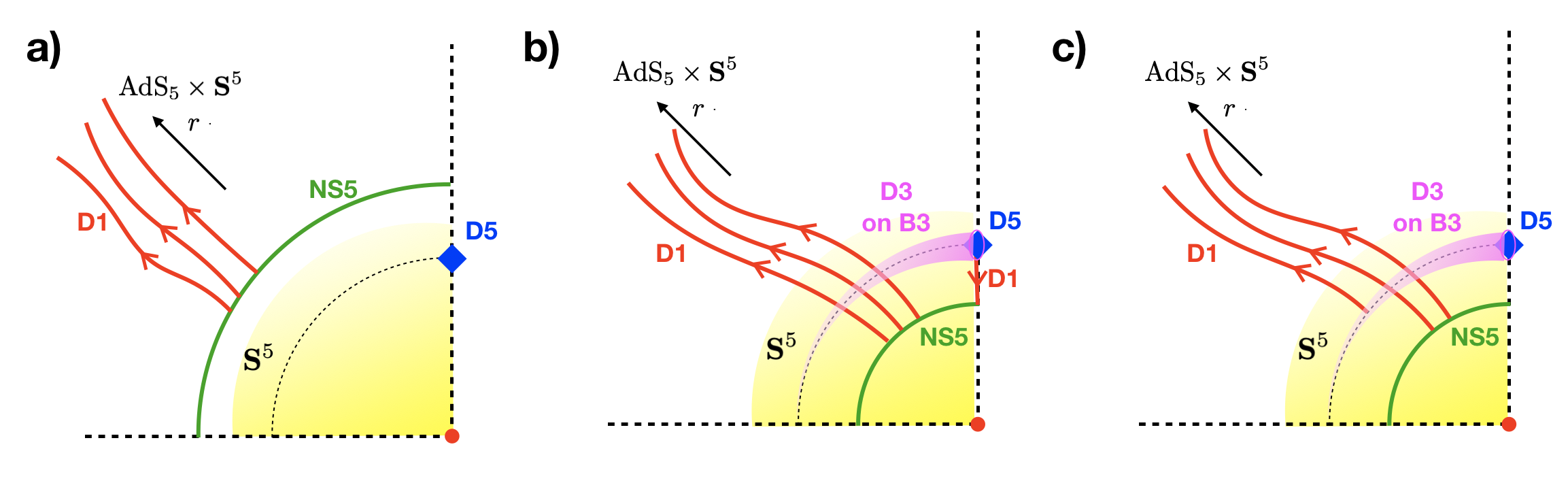}
\caption{\small (a) The gravity dual with a 4d D5-brane (blue) and a dual baryon NS5-brane (in green) with its $N$ D1 strings (red), in the picture of the fibration over the quadrant Riemann surface. (b) Crossing the  NS5-brane across the 4d D5-brane leads to the creation of $P$ D1 strings between them, and a D3-brane (in magenta) which ends up wrapping a hemisphere $B_3$ whose boundary is the $\IS^2$ on the D5-brane. (c) After recombination, the NS5-brane emits $M$ D1 strings, while the D3-brane on $B_3$ emits $P$ D1 strings.}
\label{fig:half-d3}
\end{center}
\end{figure}
%%%%%%%%%%%

In addition, we have the single D3-brane, wrapped on the $\IS^2$ and stretched between the NS5-brane and the 4d D5-brane. An important point is now that the $\IS^2$ is topologically trivial on the $\IS^5$, but not on the worldvolume of the D5-brane (as it wraps precisely this $\IS^2$). This means that the boundary of the D3-brane on the NS5-brane is actually trivial, and the D3-brane can unwind and snap away. On the other hand, the boundary of the D3-brane on the D5-brane is non-trivial, so the D3-brane cannot unwind completely. The result is that the D3-brane wraps a 3-chain $B_3$ whose boundary is the $\IS^2$ wrapped by the D5-brane, $\partial B_3=\IS^2$. A minimal-volume representative of this 3-chain is a half-$\IS^3$ in the $\IS^5$ at the location of the 4d D5-brane. In the picture of the geometry as an $\IS^2\times \IS^2$ fibration over a Riemann surface (the quadrant), the process and the structure of the 3-chain is shown in figure \ref{fig:half-d3}. In analogy with the previous section, we expect the $P$ D1 strings and the D3-brane stretching out should correspond to an operator in the D5-brane theory. This can be made more clear using the full SymTree picture, as we do next.

Consider the crossing of the NS5-brane in the full SymTree, i.e. growing back the SymTFT$_{\IS^2\times\IS^3}$ associated to the 5-brane theory. We start with the dual baryon vertex of the SymTFT$_N$, namely the NS5-brane wrapped on the $\IS^5$ with $N$ units of 5-form flux and emitting $N$ D1 strings. To move it across the junction, we simply use the homology relation among the $\IS^5$'s and $\IS^2\times\IS^3$. We obtain an NS5-brane wrapped on the $\IS^5$ with $M$ units of 5-form flux, emitting $M$ D1 strings, and an NS5-brane wrapped on $\IS^2\times\IS^3$, with $P$ units of 5-form flux, emitting $P$ D1 strings. The former is simply the dual baryon vertex of the SymTFT$_{M}$, as expected. The NS5-brane on $\IS^2\times\IS^3$, in addition to the Freed-Witten anomaly enforcing the emission of the $P$ D1 strings, has a Freed-Witten anomaly due to the single unit of RR 3-form flux on $\IS^3$, which enforces the emission of a D3-brane wrapped on $\IS^2$ (as above, we keep their discussion of the emitted D3- and D1-branes separate). The situation is very similar to the above Hanany-Witten derivation of the crossing, but with an interesting novelty. The $\IS^2$ wrapped by the emitted D3-brane is {\em non-trivial} in the $\IS^2\times\IS^3$ geometry of the SymTFT$_{\IS^2\times\IS^3}$, although it is trivial in the global geometry. Namely, the D3-brane can unwind but only in the region of the junction, in which the $\IS^2\times\IS^3$ geometry blends with the $\IS^5$ in which the $\IS^2$ is trivial. This means that the 3-chain $B_3$ wrapped by the D3-brane stretches from the 4d D5-brane on the physical boundary of the SymTFT$_{\IS^2\times\IS^3}$ to the junction theory, and ends there.

%%%%%%%%%%%
\begin{figure}[htb]
\begin{center}
\includegraphics[scale=.4]{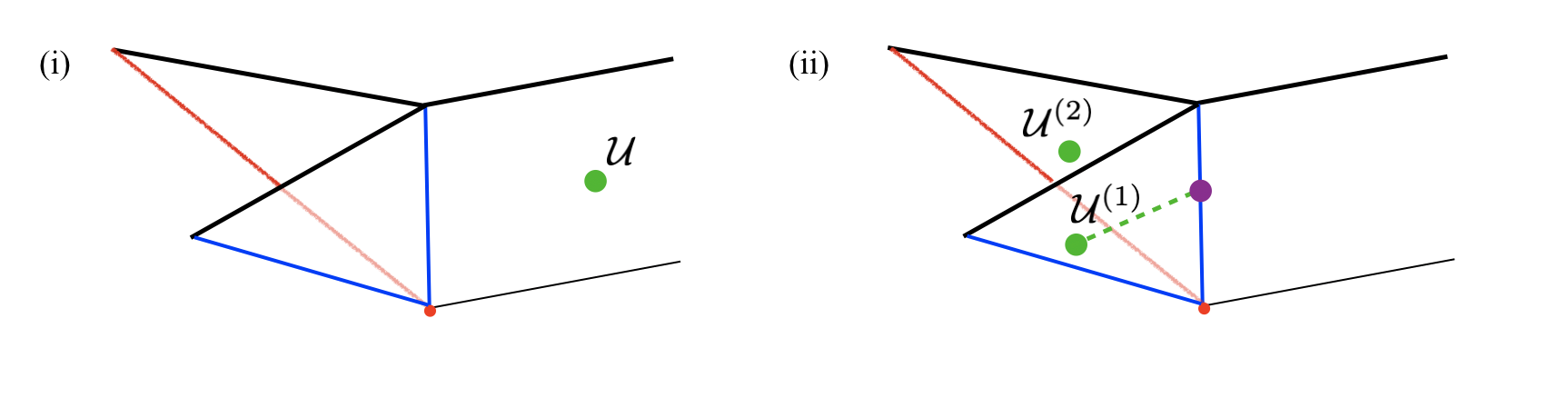}
\caption{\small Transition of operators in the SymTFT Fan, c.f. figures \ref{fig:symtft-fan-generators}a, \ref{fig:symtft-fan-generators}d, corresponding to the the magnetic baryonic operator of the SymTFT$_N$ (i) across the junction, giving a dual baryonic operator of the SymTFT$_M$ and a (non-genuine) dual baryon operator of the SymTFT$_{\IS^2\times\IS^3}$ (plus a junction theory operator).}
\label{fig:ns5-crossing}
\end{center}
\end{figure}
%%%%%%%%%%%

In formal terms, the dual baryon operator in the
SymTFT$_{\IS^2\times\IS^3}$ emitting $P$ D1 strings is non-genuine, so
it must be dressed by a operator which stretches to the junction
theory, as sketched in figure \ref{fig:ns5-crossing} (c.f. transition (a)$\to$ (d) in figure \ref{fig:symtft-fan-generators}).

It is satisfying to recover the by now familiar phenomenon that in
SymTFTs arising from holography subtle effects involving topological
operators in the SymTFT description follow easily from long-understood
string theory physics. We will encounter similar phenomena in even
richer setups in the next sections.

\section{Inclusion of 7-branes}
\label{sec:7branes}

In the previous section we have emphasised that the Symmetry Theory of our system has the local structure of a SymTree, albeit with a different location, as compared with \cite{Baume:2023kkf}, of the topological and physical boundaries. This implies that the junction theory reaches the ``topological'' boundary, and requires the introduction of physical, non-topological, boundary conditions there. In a string theory context, a natural choice is to use string theory objects to provide boundary conditions. In this section we explore the introduction of 7-branes on which the 4d 5-branes end.

The introduction of 7-branes provides a natural choice of boundary conditions for the 5-branes, making the flavour symmetry of the BCFT more explicit. The objectives of this section are to explain the properties of such configurations from the string theory perspective, and to clarify the implications of the $SL(2,\IZ)$ monodromy for the structure of the SymTFT Fan. We will explain the key role of  7-branes as duality defects \cite{Heckman:2022xgu} in the understanding of the monodromy effects on diverse  operators of the SymTFT, in particular line operators and the baryon and dual baryon operators.

\subsection{Generalities}
\label{sec:generalities-7branes}

We will phrase our discussion in terms of the retracted SymTree picture. Namely, the SymTFT Fan is described as a set of SymTFT wedges with topological action (\ref{bf-symtft}) with different levels, separated by explicit NS5- and D5-branes, which are ending on the 7-branes, see Figure \ref{fig:7branes-symtft}. The system of 5- and 7-branes carry gauge theories associated to the BCFT flavour symmetries. The presence of the explicit 5-branes facilitates the discussion of their ending on 7-branes. The description of the system in the full SymTree picture is possible, but more involved, as we briefly explain in section \ref{sec:7branes-3branes}.

\subsubsection{The 7-brane boundary conditions}
\label{sec:7branes-bc}

As in previous sections, we focus on the case of 4d D5-branes separating the SymTFT$_N$ and SymTFT$_M$ (by S-duality, similar consideration follow for NS5-branes ending on the S-dual 7-branes). We also focus on the case of a single D5-brane, with $P=N-M$ units of induced D3-brane charge. As is familiar (already from \cite{Hanany:1996ie} in a T-dual setup), D5-branes can end on NS5-branes or on D7-branes. The two choices differ in what boundary condition they impose on the D5-brane worldvolume gauge fields. The choice of D7-branes is natural in our setup since it allows to nicely translate the flavour symmetries realised on the D5-branes to the D7-branes worldvolume theory.

Hence, we consider D7-branes wrapped on $\IS^5$, and spanning a 3d Minkowski subspace on the AdS$_4$ slice, initially at a location pushed onto the topological boundary, so that they provide boundary conditions for the D5-branes ending on them. Since the D7-branes are located at a boundary, there are no closed paths in the Symmetry Theory which go around the D7-branes, and the axion $C_0$ can be defined globally, i.e. there is no need to specify a branch cut (equivalently, it is sticking out outwards, away from the Symmetry Theory).

It is interesting to notice that the D7-branes just introduced are of
the kind considered in \cite{Heckman:2022xgu}, in the sense that they are object of real codimension 2 in the 5d Symmetry Theory. As explained there (in the context of QFTs without boundaries), general $(p,q)$ 7-branes can be located in the interior of the 5d bulk as real codimension 2 defects, endowed with a branch cut emerging from them, which implement the $SL(2,\IZ)$ S-duality monodromy corresponding to the 7-brane $(p,q)$ label. The orientation of the branch cut determines the field theory interpretation of the corresponding object. The case interesting for us is when the branch cut ends on the topological boundary of the 5d Symmetry Theory, in which case the branch cut produces an $SL(2,\IZ)$ monodromy on the topological boundary conditions. Upon collapsing the 5d sandwich, the 5d codimension 2 defect becomes a codimension 1 duality interface defect in the 4d field theory.

In our present setup we have a version of this for 4d theories with boundaries, but the description in terms of the 5d bulk theory is otherwise similar. Going back to the simplest example of D7-branes as endpoints of D5-branes in the SymTFT Fan, we can follow \cite{Heckman:2022xgu} and consider moving the D7-branes into the 5d bulk of the Symmetry
Theory. In doing that, it is necessary to introduce the branch cut
implementing the $SL(2,\IZ)$ monodromy of the 7-brane. From our 
discussion above, it is clear it stretches from the location of the 7-brane
in the bulk to the topological boundary, as anticipated above. Hence, the endpoint of the
branch cut at the topological boundary still encodes the relevant
information about the boundary condition of the D5-brane (or junction)
theory. From \cite{Heckman:2022xgu}, this choice of branch cut
orientation corresponds to the 7-branes creating a duality
interface. Namely, the topological boundary conditions of the
SymTFT$_N$ and SymTFT$_M$ are related by a non-trivial $SL(2,\IZ)$
duality transformation due to crossing of the branch cut. As
emphasized in \cite{Heckman:2022xgu}, although the particular shape of
a D7-brane branch cut is not physically observable, its endpoint at
the topological boundary is, and leads to observable effects.

%%%%%%%%%%%
\begin{figure}[htb]
\begin{center}
\includegraphics[scale=.35]{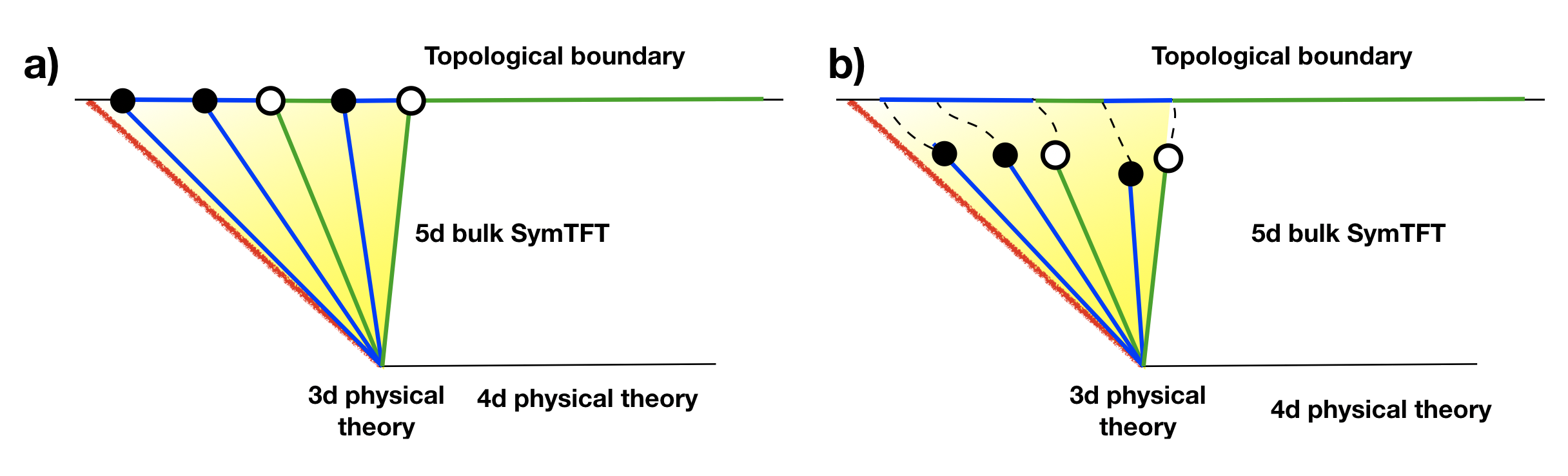}
\caption{\small Sketch of the 5d SymTFT Fan with 5-branes ending on 7-branes. Figure a) shows the situation with the 7-branes on the topological boundary, whereas figure b) shows the situation where the 7-branes are pushed into the bulk (and have branch cuts stretching out to the topological boundary).}
\label{fig:7branes-symtft}
\end{center}
\end{figure}
%%%%%%%%%%%

Hence the topological boundary of the SymTFT Fan is a set of gapped boundary conditions for the different SymTFTs, separated by 7-branes on which the 5-branes are ending, see figure \ref{fig:7branes-symtft}a. If the 7-branes are pushed into the bulk, to make their effects more manifest, the SymTFT gapped boundary conditions are separated by the endpoints of the 7-brane branch cuts, see figure \ref{fig:7branes-symtft}b.

Since we are focusing on the case of a single D5-brane, it is clear that it suffices to consider a single D7-brane on which it ends. We would like however to make a comment regarding the general case of a stack of $n>1$ D5-branes. Recall that the worldvolume gauge fields on the D5-brane have Neumann boundary condition on the physical boundary, so as to couple to the currents of the $\fu(n)$ enhanced global symmetry of the BCFT$_3$, as required by the holographic interpretation. On the other hand, they have Dirichlet boundary conditions on the D7-brane. This implies that there is a version of the s-rule in \cite{Hanany:1996ie}, which requires that each D7-brane can be the endpoint of at most one D5-brane, so that one needs $n$ D7-branes. A complementary heuristic explanation is that the $\fsu(n)$ global symmetry of the physical theory is  morally realised as the worldvolume $\fsu(n)$ gauge symmetry on a stack of $n$ D7-branes (more precisely, on the bound system of the stack of $n$ D5-branes ending on $n$ D7-branes)\footnote{This is similar to the realisation of flavour symmetries in Hanany-Witten setups when the flavour branes are moved off the interval between the NS5-branes, a process which creates new D-branes ending on the flavour branes.}. 

\subsubsection{The 5-form flux and the need for explicit D3-branes}
\label{sec:7branes-3branes}

We return to the setup with a single D5-brane separating the SymTFT$_N$ and the SymTFT$_M$ with $P=N-M$, and ending on a D7-brane in the bulk of the Symmetry Theory, with a branch cut stretching to the topological boundary.

%%%%%%%%%%%
\begin{figure}[htb]
\begin{center}
\includegraphics[scale=.35]{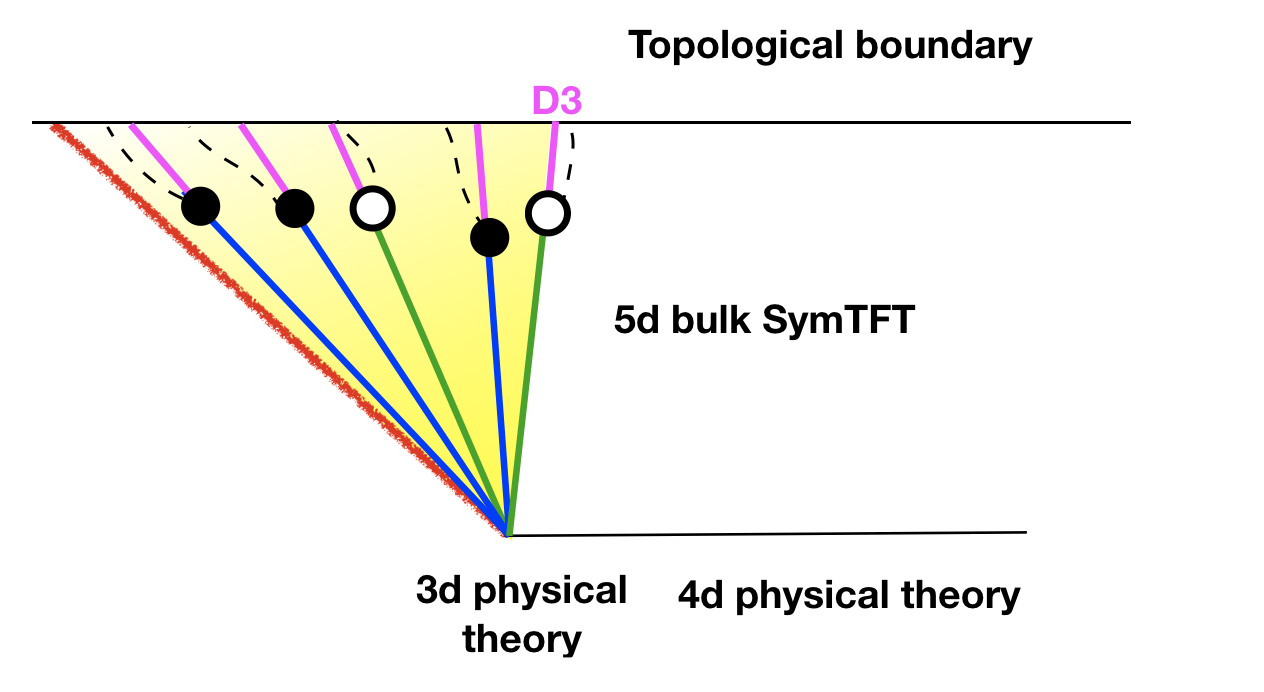}
\caption{\small The 5-branes ending on 7-branes have to be complemented with a stack of $P$ D3-branes stretching from the 7-brane to the topological boundary of the SymTFT, so that the jump in $P$ units of 5-form flux is reproduced no matter where one crosses across the D5/D7/D3 configuration.}
\label{fig:the-d3s}
\end{center}
\end{figure}
%%%%%%%%%%%

As explained earlier, the jump in flux across the D5-brane, implicit in the separation between the SymTFT$_N$ and the SymTFT$_M$, is accounted for by the $P$ units of induced D3-brane charge. Since the D5-brane ends on the D7-brane, which is localised in the 5d bulk, it would now be possible to cross from the SymTFT$_N$ to the SymTFT$_M$ without encountering any discontinuity, other than crossing the D7-brane branch cut. But the D7-brane branch cut simply implements the $SL(2,\IZ)$ transformation, which does not act on the 5-form flux. Hence, we are forced to introduce an explicit stack of $P$ D3-branes stretching from the D7-brane to the topological  boundary of the SymTFT, see figure \ref{fig:the-d3s}. 

The D3-brane charge is actually not ending on the 7-brane, but rather
it continues along the corresponding 5-brane in the form of induced
D3-brane charge. This is as it should be, since D3-branes cannot end
on 7-branes. This is also clear when looking at the 4d D3-brane
worldvolume coupling
\begin{equation}
P\int_{4d} C_2 B_2   \, ,
\end{equation}
which is the continuation of the D5-brane coupling (\ref{BF-level-jump}) and explains the jump 
in the coefficient of the 5d topological coupling $B_2F_3$ from the SymTFT$_N$ to the SymTFT$_M$. Hence the structure of the SymTFT Fan is maintained even when the 5-branes end on the 7-branes.

%%%%%%%%%%%
\begin{figure}[htb]
\begin{center}
\includegraphics[scale=.32]{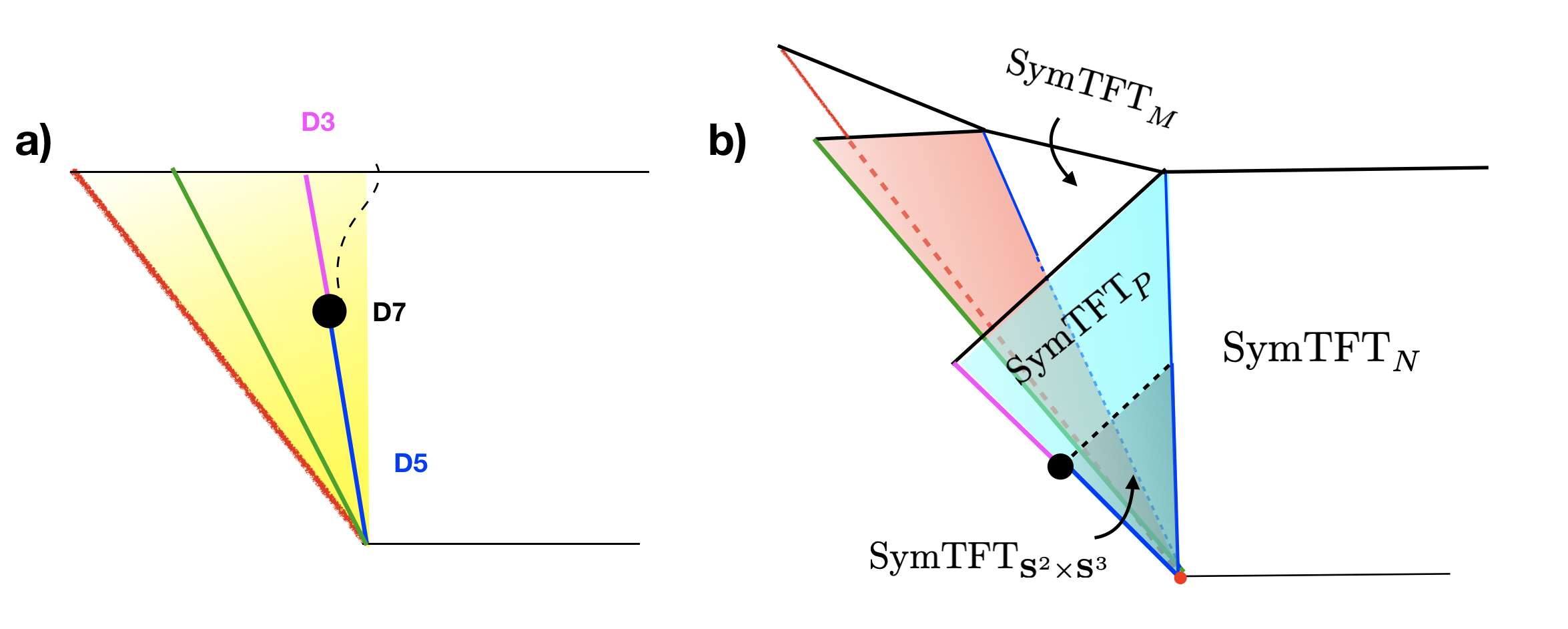}
\caption{\small a) Sketch of a SymTFT Fan with one D5-brane ending on a D7-brane, with an explicit D3-brane stack beyond it. b) Sketch of the SymTFT Fan with a SymTree whose junction joins the SymTFT$_N$ and the SymTFT$_M$ with either the SymTFT$_{\IS^2\times\IS^3}$ of the D5/D3-brane system, or the SymTFT$_P$ of the D3-brane stack. The latter two are separated by a junction theory (dashed black line), corresponding to the 4d Symmetry Theory of the D7-brane 3d theory.}
\label{fig:nest-in-tree}
\end{center}
\end{figure}
%%%%%%%%%%%

Although we will not use it, for completeness we would like to remark on the description of the system in the full SymTree picture, shown in figure \ref{fig:nest-in-tree}b. Namely, we replace the explicit 5-brane by a SymTFT$_{\IS^2\times\IS^3}$ branch, the 7-brane by a 4d SymTFT, and the $P$ explicit D3-branes by a SymTFT$_P$. The non-abelian sectors of the 5-/7-brane and the D3-brane theories remain as physical boundaries of this 5d Symmetry Theory branch; on the other hand, the would-be topological boundary is actually a $U(1)$ junction theory joining it to the SymTFT$_N$ and SymTFT$_M$, already present in the retracted picture discussed above. Finally notice that these two boundaries of the 5d Symmetry Theory branch join at the 3d BCFT, which is thus a corner theory in the sense of \cite{Cvetic:2024dzu}, see section \ref{sec:nesting}.

We conclude by noting that the effect of the D3-brane charge was crucial for the Hanany-Witten effect for the baryonic D5- and NS5-branes across the 4d D5-brane in sections \ref{sec:action-baryon} and \ref{sec:action-dual-baryon}. In the configuration where 4d D5-brane ends on a D7-brane, the extra $P$ explicit D3-branes from the D7-brane to the topological boundary guarantee that the same effect takes place if the baryon vertex is moved from the SymTFT$_N$ to the SymTFT$_M$ across the D7-brane branch cut.

In the following section we discuss in more detail the effects related to moving objects between the SymTFTs across the branch cut. As expected from \cite{Heckman:2022xgu}, this implements the $SL(2,\IZ)$ monodromies corresponding to a duality interface in the theory.

\subsection{The $SL(2,\IZ)$ monodromy}
\label{sec:monodromy}

In this section we discuss the effects experienced in the Symmetry
Theory and its topological operators as one crosses from the
SymTFT$_N$ to the SymTFT$_N$ across the D7-brane branch cut (and the
$P$ explicit D3-branes). This is a particular case of the phenomena
discussed in \cite{Heckman:2022xgu}, and implements the $SL(2,\IZ)$
monodromies corresponding to a duality interface in the theory.

 As above, we focus on the simple setup of a single D5-brane ending on a single D7-brane. In our discussions below, the effect of the D3-brane charge (either explicit or induced on the 4d D5-brane), which has already been mentioned, will we included implicitly, except for a few explicit mentions.

\subsubsection{Stacking of TQFT}
\label{sec:across-branch}

We now derive the effect of the D7-brane branch cut on the Symmetry Theory. Recall that the physical effect of the D7-brane branch cut in string theory is that the 10d RR axion shifts $C_0\to C_0+1$. From the perspective of the holographic dual 4d $\CN=4$ $\fsu(N)$ SYM theory, this corresponds to a shift of the $\theta$ angle. There is a mixed anomaly between the shift of the $\theta$ angle and the electric 1-form symmetry, encoded in the 5d TFT
\[
S=2\pi i \frac{N-1}{N}\int_{Y} \frac{d\theta}{2\pi}\cup \frac{{\cal P}({\rm B})}2\, ,
\label{anomaly-theta}
\]
where ${\rm B}$ is the background coupling to the electric 1-form symmetry and ${\cal P}$ is the Pontryagin square.
For the geometries of our interest (5d spaces foliated with 4d slices given by spin manifolds over which $\theta$ is constant) ${{\cal P}({\rm B})}$ is even, so the only relevant piece is the $1/N$ term. We will also be sensitive only to the de Rham version of ${\cal P}\sim {\rm B}^2$.

We now follow the derivation in \cite{Bergman:2022otk} of this 5d TQFT from the holographic setup. We use the well-defined 3-form field strength
\[
{\tilde F}_3= dC_2-C_0dB_2\, ,
\label{tildef3}
\]
to rewrite the 5d topological coupling (\ref{bf-symtft}) as
\[
N\int_{5d} B_2F_3=N\int_{5d}B_2 {\tilde F}_3-N\int_{5d} dC_0 B_2 B_2 \, .
\] 
The last term reproduces (with the usual redefinitions $B_2\sim {\rm B}/N$) the $1/N$ term in (\ref{anomaly-theta}).

Hence, if we take an interval $I$ across the D7-brane branch cut (so that $C_0\to C_0+1$ along it),  and integrate the 5d coupling, its variation is given by
\[
\Delta \int_{5d} \left(- NdC_0 B_2 B_2 \right) =
-N\int_{4d} B_2B_2\, .
\label{jump-pontryagin}
\]
This is precisely the kind of variation required, corresponding to the stacking of a TQFT in the SymTFT (i.e. of a counterterm in the dual QFT), as befits the shift of the $\theta$ angle. One may worry that, on the two sides of the D5/D7 system, the 5-form flux changes by $P$ units, and this requires the coefficient of the above coupling to jump as $N\to M$. Actually, this is solved because of the additional crossing of the $P$ explicit D3-branes, which have a worldvolume coupling
\[
P\int_{4d}C_0 B_2B_2\, .
\]
This amounts to $P$ units of a boundary contribution of the coupling $\int_{5d} dC_0 B_2B_2$, so that on the SymTFT$_M$ side the coefficient is effectively reduced from $M$ to $N$.

In short, the effect of the D7-brane branch cut is the stacking of a TQFT, according to the effect of the $SL(2,\IZ)$ as a shift of the $\theta$ angle corresponding to the duality interface.

\subsubsection{Line operators and $(p,q)$ string webs}
\label{sec:string-webs}

\begin{figure}[htb]
\begin{center}
\includegraphics[scale=.25]{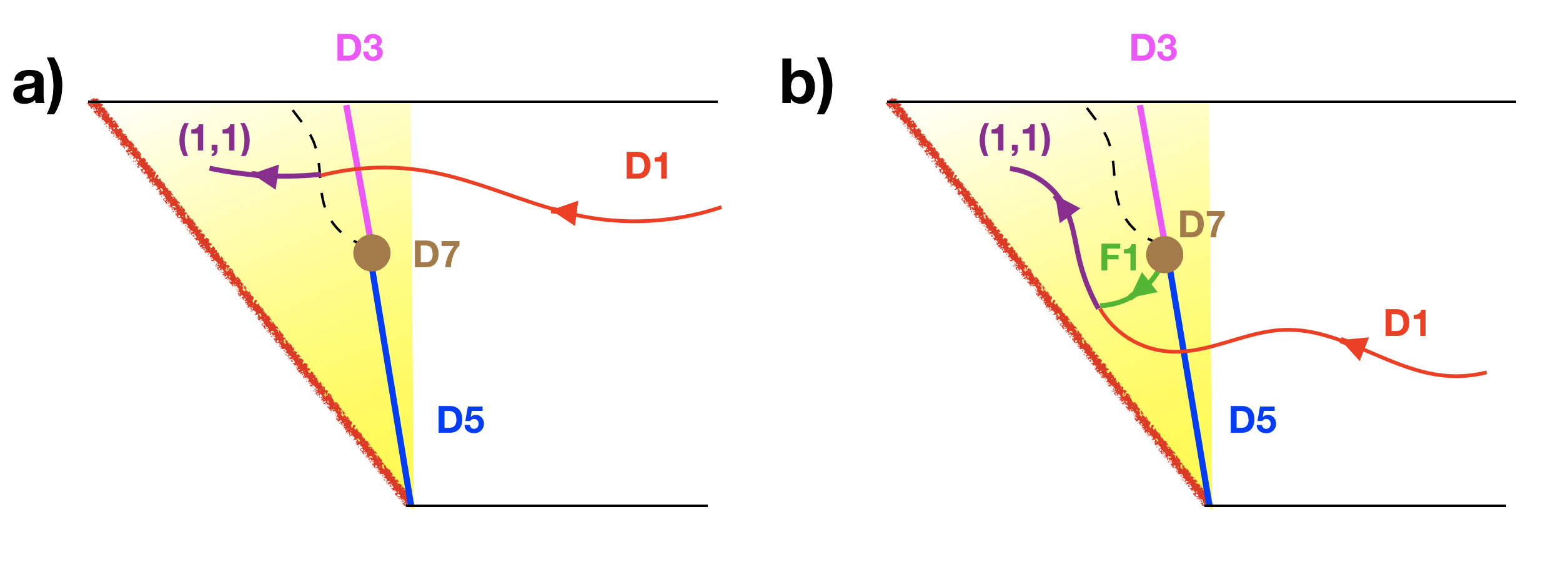}
\caption{\small a) Transformation of the D1 string when it crosses the branch cut of a bulk D7-brane. b) When moved across the D7-brane, the D1 string turns into a string web due to a string creation process.}
\label{fig:crossing-D1}
\end{center}
\end{figure}

Let us consider the fate of line operators described in terms of F1 and D1 strings in the SymTFT$_N$ as they cross the branch cut. As explored in section \ref{sec:lineops}, this can be efficiently discussed by considering F1 or D1 strings which stretch between the SymTFT$_N$ and the SymTFT$_M$ across the junction theory. The crossing over the $P$ explicit D3-branes leads to the effects already discussed in section \ref{sec:lineops}, which account for the change in the order of the discrete valuedness of the background fields, and hence of the topological operators\footnote{Incidentally, note that in this context, the D3-branes separating the SymTFT$_N$ and the SymTFT$_M$ is the retracted version of a SymTree corresponding precisely to adjoint Higgsing. Namely, the SymTFT corresponding to the D3-branes is a SymTFT$_P$ associated to reduction on $\IS^5$ with $P$ units of 5-form flux.}. Hence we only need to consider their transformation across the branch cut of the 7-brane, i.e. the $SL(2,\IZ)$ transformation $\begin{pmatrix} 1 & 1 \\ 0 & 1 \end{pmatrix}$. The F1 is a $(1,0)$ string, so it is invariant under this action. On the other hand, the D1 is a $(0,1)$ string, and hence turns into a $(1,1)$ string, as shown in figure \ref{fig:crossing-D1}a. For charged line operators, this is just the manifestation of the transformation of monopoles into dyons upon a shift of the $\theta$ angle.

%%%%%%%%%%%
%%%%%%%%%%%

It is interesting to consider the above transformation of the D1 strings when they are moved across the D7-brane, so that they cross over the D5-brane, see figure \ref{fig:crossing-D1}b. In this crossing there is a Hanany-Witten effect creating an F1 string stretching from the D7-brane to the D1 string, and turning the latter into a $(1,1)$ string due to charge conservation in the resulting string junction. Hence we recover the same overall result, even though the D1 string has not crossed over the D7-brane branch cut. 

Hence we recover the interpretation of D7-branes (with branch cuts stretched to the topological boundary) as duality interfaces in \cite{Heckman:2022xgu}, via their action on F1 and D1 strings.

\subsubsection{Baryon and dual baryon vertices}
\label{sec:7branes-baryons}

In the discussion in the previous section we have briefly mentioned the jump in the order of topological operators when they cross from the SymTFT$_N$ to the SymTFT$_M$. In order to get a more direct insight we quickly derive the transformation of baryon and dual baryon vertices as they move across the D7-brane branch cut (and the $P$ D3-branes), extending the discussion in sections \ref{sec:action-baryon} and \ref{sec:action-dual-baryon}.

We start with the baryon vertex in the SymTFT$_N$ region, namely a D5-brane wrapped on the $\IS^5$ and emitting $N$ F1 strings. Both the baryonic D5-brane and the F1 strings are left invariant when moved across the D7-brane branch cut. On the other hand, when moved across the $P$ D3-branes, there are $P$ newly created F1 strings. They combine with $P$ of the original F1 strings emitted by the baryonic D5-brane, leaving the latter with $M$ F1 strings. This corresponds to the baryon vertex of the SymTFT$_M$ region. In addition, there remain $P$ F1 strings being emitted by the D3-branes. In the full SymTree picture, these would correspond to the baryon vertex of the SymTFT$_P$ region, as well as an operator of the $\fu(1)$ junction theory. The result is as in section \ref{sec:action-baryon}, with the explicit D3-branes here playing the role of the induced D3-brane charge there.

%%%%%%%%%%%
\begin{figure}[htb]
\begin{center}
\includegraphics[scale=.35]{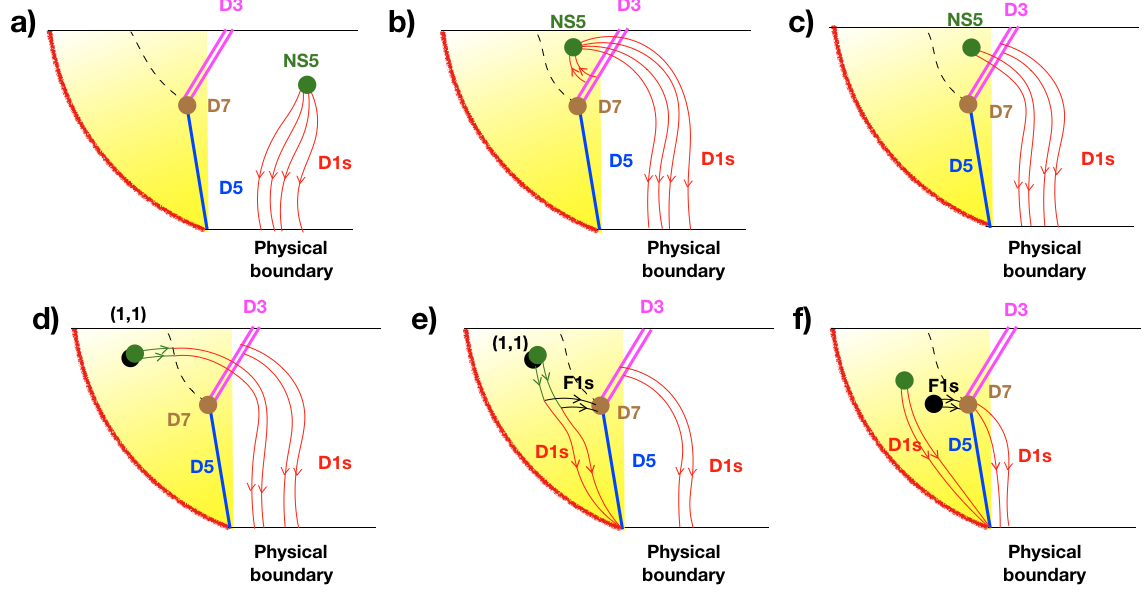}
\caption{\small Step by step crossing of the NS5-brane across the explicit D3-branes and the D7-brane branch cut. See the main text for the explanations. We have curved the ETW boundary (red line) for clarity of the pictures.}
\label{fig:mother-of-the-crossings}
\end{center}
\end{figure}
%%%%%%%%%%%

Consider now the dual baryon vertex in the SymTFT$_N$ region, namely an NS5-brane wrapped on the $\IS^5$ and emitting $N$ D1s, see figure \ref{fig:mother-of-the-crossings}a. When moved across the $P$ D3-branes, there is a Hanany-Witten creation of $P$ suspended D1 strings, see figure \ref{fig:mother-of-the-crossings}b. Upon recombining, we have an NS5-brane emitting $M$ D1 strings, and $P$ D1 strings stretching from the $P$ D3-branes, see figure \ref{fig:mother-of-the-crossings}c. These $P$ D1s will be pretty much spectators in what follows, so we do not mention them explicitly, and focus on the NS5-brane. We now move the NS5-brane across the D7-brane branch cut, dragging its $M$ D1 strings with it. Upon crossing, the NS5-brane turns into a $(1,1)$ 5-brane, which emits $M$ $(1,1)$ strings, which turn into D1s when they cross the branch cut, see figure \ref{fig:mother-of-the-crossings}d. This reproduces the expected effect of the $SL(2,\IZ)$ monodromy on the dual baryonic vertex.

It is interesting to continue manipulating the above object in order to relate it to a simple dual baryonic vertex of the SymTFT$_M$. To do so, we move across the 4d D5-brane the $M$ D1 strings, which necessarily cross the D7-brane without possibly avoiding it. This crossing produces $P$ F1 strings which combine with the $(1,1)$ and D1 strings forming a $(p,q)$ string web, see figure \ref{fig:mother-of-the-crossings}e, in analogy with section \ref{sec:string-webs}. At the topological level, we can split the $(1,1)$ 5-brane, and the string web including its $M$ attached $(1,1)$ strings, as 1 NS5-brane emitting $M$ D1 strings, and 1 D5-brane emitting $M$ F1 strings which end on the D7-brane, see figure \ref{fig:mother-of-the-crossings}f. 

The NS5-brane emitting $M$ D1 strings is just the dual baryon vertex of the SymTFT$_M$. On the other hand, the baryonic D5-brane emitting $M$ F1s ending on the D7-brane can be pushed onto the latter and regarded as a defect associated to the junction theory.

We hope these examples suffice to illustrate the behaviour of different objects in the configuration, and their interplay with the $SL(2,\IZ)$ duality interface introduced by the D7-branes.

\section{Orientifold and Orbifold Boundaries}
\label{sec:orientifold}

\subsection{Generalities on orientifold and orbifold 5-planes}
\label{sec:orientifold-generalities}

In this section we discuss a generalisation of the above boundary
configuration by allowing for the introduction of orientifold planes,
and orbifolds related to them. We will focus on orientifold/orbifolds
which are localised at the boundary in the direction 3. Namely, we are
not considering for instance the well-studied case of O3-planes
parallel to D3-branes, leading to $\cN=4$ $\fso(n)$ or $\fsp(n)$
theories (see
\cite{Witten:1998xy,Aharony:2016kai,Bergman:2022otk,Etheredge:2023ler}
for discussions of the holographic duals of such systems). Moreover,
we are also interested in configurations that preserve the same
symmetries and supersymmetries preserved by D5- and NS5-brane boundary
conditions, so as to stay in the set of boundary conditions studied in
\cite{Gaiotto:2008sa, Gaiotto:2008ak}. This essentially restricts us
to consider the introduction of O5-planes in the same directions as
the D5-brane, namely 012 789, or orbifold 5-planes in the same
directions as the NS5-branes, namely 012 456.

Orientifold 5-planes (O5-planes for short) are defined by quotienting
by the orientifold action $\Omega R$, where $\Omega$ is worldsheet
parity and $R$ is a geometric $\IZ_2$ involution acting as a
coordinate flip in the directions 3456,
$(x^3,x^4,x^5,x^6)\to (-x^3,-x^4,-x^5,-x^6)$. As discussed, this
preserves the same supersymmetries as a D5-brane along 012789. The 6d
plane fixed under $R$ is an O5-plane, and there are several discrete
choices for its properties (in analogy with O3-planes in
\cite{Witten:1998xy}), classified by discrete RR and NSNS backgrounds
on the $\IR\IP_3$ geometry around it \cite{Hanany:2000fq}. The
O5$^-$-plane and O5$^+$-plane differ in the value of the NSNS 2-form
on an $\IR\IP_2\subset \IR\IP_3$, carry RR charge $\mp 2$ (as measured
in D5-units in the double cover), and project the symmetry $\fu(n)$ on
a stack of D5-branes on top of them down to $\fso(n)$ or $\fsp(n)$,
respectively.
There are variants of these, dubbed ${\widetilde{\rm O5}}^\pm$, which
arise when the non-trivial $\IZ_2$ RR background $C_0$ is turned on at
the O5-plane location. The ${\widetilde{\rm O5}}^-$ has RR charge $-1$
and can be described as the O5$^-$ with one stuck D5-brane on top, so
it leads to a gauge symmetry $\fso(n+1)$ when $n$ additional D5-branes
are located on top of it. The ${\widetilde{\rm O5}}^+$ is an exotic
version of the O5$^+$-plane, has RR charge $+2$ and also leads to a
group $\fsp(n)$ when $n$ D5-branes (counting before orientifolding,
here $n$ must be even) are located on top of it.

Let us briefly discuss orbifold 5-planes. They are obtained by
quotienting by the orbifold action $R'(-1)^{F_L}$, where $R'$ is a
geometric $\IZ_2$ involution flipping the coordinates
$(x^3,x^7,x^8,x^9)\to(-x^3,-x^7,-x^8,-x^9)$. As discussed, this
preserves the same supersymmetries as an NS5-brane along 012456. The
6d plane fixed under $R'$ is the orbifold 5-plane, and there are also
several variants \cite{Hanany:2000fq}. Interestingly, the orbifold
5-planes are related by S-duality to O5-planes \cite{Sen:1996na}; in
particular the perturbative one, which has a localised twisted sector
spectrum given by the 6d $(1,1)$ $U(1)$ vector multiplet, is S-dual to
the O5$^-$ with 2 D5-branes on top, which realises the same localised
field content as an $\fso(2)$ vector multiplet in the open string
sector. Because of this kind of S-duality relations, we will focus our
discussion on the introduction of O5-planes, with additional NS5- and
D5-branes, and will not consider orbifold 5-planes any further.

\subsection{Brane configurations}
\label{sec:orientifold-hw}

In order to describe Hanany-Witten brane configurations for systems of D3-branes ending on a set of NS5- and D5-branes in the presence of O5-planes, it is useful to consider the configuration in the covering space. There we have a $\IZ_2$ invariant system, with one stack of $N$ semi-infinite D3-branes in $x^3\to \infty$ and its image stack in $x^3\to -\infty$, ending on a ($\IZ_2$-invariant) `middle' configuration of NS5-, D5-branes, and D3-branes suspended among them, with the O5-plane sitting at $x^3=0$. 

Thus the configuration is morally a back-to-back double copy of the kind of systems considered in previous sections. One key difference is that at the ETW boundary, which is given by the O5-plane (and possible additional 5-branes on top of it), the number of D3-branes need not vanish. This will be nicely reproduced in the gravity dual picture, as discussed later on. In general the number of D3-branes at the O5-plane location is different from the asymptotic value $N$, so we denote it by $n$ in the following.

Given the above, most of the discussion regarding NS5- and D5-branes away from the O5-plane behave locally just like the configurations of NS5- and D5-branes in the previous sections. The only position at which the presence of the orientifold quotient is felt {\em locally} is precisely the O5-plane. Hence, we next describe different possible configurations of 5-branes on top of the O5-plane. Each of them can then be used to construct a general class of 3d theories defining boundary conditions, by simply sprinkling additional NS5- and D5-branes and their $\IZ_2$ images. We hence turn to study the different configurations of 5-branes on top of the O5-plane.

\subsubsection{Boundary conditions for O5-planes with no NS5-branes}
\label{sec:bc-o5-no-NS5}

The simplest possibility is that the O5-plane does not have additional
NS5-branes on top. This possibility corresponds to the boundary
conditions reducing the gauge symmetry considered in
\cite{Gaiotto:2008sa,Gaiotto:2008ak}. Hence, we locally have a stack
of $n$ D3-branes in the double cover, intersected by an O5-plane, and
we can then use standard orientifold rules to read out the local
breaking of the $\fsu(n)$ symmetry due to the orientifold
projection. In this respect, recall that the orientifold projection of
the O5-plane on the D3-branes is of the opposite kind (namely $\fso$
vs $\fsp$) as compared with the action on D5-branes, because the
corresponding mixed open string sector has 4 DN directions
\cite{Pradisi:1988xd,Witten:1995gx,Gimon:1996rq}.

For instance, the boundary conditions defined by an O5$^+$-plane reduce the symmetry $\fsu(n)$ to $\fso(n)$, which corresponds to Class I in \cite{Gaiotto:2008ak}. In more detail, splitting the 4d $\CN=4$ vector multiplet in terms of a local 3d $\CN=4$ vector multiplet and a 3d $\CN=4$ hypermultiplet in the adjoint of $\fsu(n)$, the $\fso(n)$ part of the vector multiplet is even under the O5$^+$-plane action (and the remaining generators are odd), and the $\symm$ part (plus a singlet) of the hypermultiplet is even (while the $\asymm$ part is odd). This provides the definition of Dirichlet or Neumann boundary conditions for the different fields.

Similarly, the O5$^-$-plane boundary condition project the $\fsu(n)$
gauge symmetry down to $\fsp(n)$ (recall that $n$ is necessarily even
for this orientifold projection), which corresponds to Class II in
\cite{Gaiotto:2008ak}. In more detail, in the covering space the
$\fsp(n)$ part of the 3d $\CN=4$ vector multiplet is even (and the
rest is odd), while the $\asymm$ part of the hypermultiplet is even
(and the $\symm+{\bf 1}$ is odd).

For the ${\widetilde{\rm O5}}^-$-plane, the configuration is
equivalent to the O5$^-$-plane with one stuck D5-brane on top. Hence,
to the above boundary condition we must add one localised
half-hypermultiplet flavour (in the fundamental of the local $\fsp(n)$
symmetry and charged under the $\IZ_2\simeq O(1)$ on the stuck
D5-brane) arising from the D3-D5 open string sector. Finally, the
${\widetilde{\rm O5}}^+$-plane behaves similarly to the O5$^+$-plane,
with minor modifications due to the non-trivial value of the RR axion
$C_0=1/2$ on top.

Clearly, the above configurations can be enriched by addition additional D5-brane pairs on top of the O5-plane. At the level of the gauge theory, the main effect is the addition of extra localised flavours of the D3-brane gauge theory. We will not discuss these possibilities explicitly, but they are implicitly included in our analysis. For instance, as already mentioned, by allowing for a pair of D5-branes on top of an O5$^-$-plane we reproduce the physics of the S-dual orbifold 5-planes. In fact, such orbifolds provide a realisation  of Class III in \cite{Gaiotto:2008ak}; in terms of the O5$^-$-plane with two D5-branes stack on them, one splits the stack of D3-branes as $n=p+q$, with the two stacks of $p$, $q$ D3-branes ending on the two possible stuck D5-branes (equivalently, carrying opposite charges under the D5-brane $\fso(2)\simeq \fu(1)$ worldvolume symmetry).

In addition, recall that the configurations can be completed by adding general configurations of NS5- and D5-branes obeying the rules of the configurations in previous sections (and their $\IZ_2$ images), to define general classes of quiver gauge theories providing  Gaiotto-Witten BCFT$_3$ orientifold boundary conditions.

\subsubsection{Boundary conditions for O5-planes with one stuck NS5-brane}
\label{sec:bc-o5-yes-NS5}

A second possible local configuration corresponds to including one stuck NS5-brane on top of the O5-plane\footnote{This is possible for the O5$^\pm$-planes and the ${\widetilde{\rm O5}}^-$-plane, but not for the ${\widetilde{\rm O5}}^+$-plane, because it requires $C_0=1/2$, and the orientifold image of an NS5-brane is not just an NS5-brane, but a $(1,1)$ 5-brane.} (this was introduced in brane configurations in \cite{Landsteiner:1997ei}, see \cite{Giveon:1998sr} for review). 
In the covering space, we have two $\IZ_2$ image D3-brane stacks ending on the NS5-brane from opposite sides in $x^3$, and the O5-plane crossing through their intersection. The D3-brane gauge symmetry in the covering space is $\fsu(n)\oplus \fsu(n)$ and it projects down to $\fsu(n)$ after the orientifold. Hence, interestingly, in the quotient the $\fsu(n)$ gauge symmetry is not reduced by the boundary conditions. We again can consider the different possible kinds of O5-plane in turn. They differ in how the orientifold acts on the 3d $\CN=4$ hypermultiplet in the bifundamental $(\fund,\ov\fund)$ of the parent theory.

For the O5$^+$-plane, the bifundamental matter is projected down to a 2-index symmetric representation $\symm$ (plus a singlet) of the $\fsu(n)$ symmetry. The simplest way to check this is to realise that giving a vev to this field corresponds to recombining the half D3-branes and moving the recombined infinite D3-brane away from the NS5-brane and along the O5-plane (i.e. the directions 789); for fields in the $\symm$, the vev breaks the symmetry down to $\fso(n)$, in agreement with the  expected symmetry on D3-branes intersecting an O5$^+$ with no NS5-brane at the intersection, see section \ref{sec:bc-o5-no-NS5}.

Similarly, for the O5$^-$-plane the bifundamental matter is projected down to a $\asymm$. For the ${\widetilde{\rm O5}}^-$ (equivalently an O5$^-$-plane with one stuck D5-brane), in addition to the $\asymm$ we get an additional full hypermultiplet flavour in the fundamental. We get a full, rather than a half, hypermultiplet because in this case the D3-branes are split in two halves by the NS5-branes, and the orientifold swaps the the D5-D3 open strings for the two halves, in contrast with the previous section, where there was a single D3-D5 open string sector which was mapped to itself under the quotient, enforcing the orientifold projection down to a half-hypermultiplet. Of course this matches the group theoretical fact that the fundamental of $\fsu(n)$, even when charged under the D5-brane $O(1)\simeq \IZ_2$, is not a pseudoreal representation, so it is not possible to have half-hypermultiplets. For the ${\widetilde{\rm O5}}^+$ we get a variant of the result of the O5$^+$, namely matter in the $\symm$ plus a singlet.

In conclusion, the brane configurations are therefore very similar to those considered in previous sections, with the only novelty of the new object in the game. We now turn to a quick discussion of their role in the 1-form symmetries of the system, and then turn to the discussion of the supergravity dual, and the SymTFT picture it produces.

\subsubsection{The 1-form symmetries}
\label{sec:orientifold-1form-sym}

In this section we quickly carry out the discussion of the impact of the O5-plane orientifold quotient in the 1-form symmetries of the system. Recall that the configurations with one stuck NS5-brane in section \ref{sec:bc-o5-yes-NS5} or without it in section \ref{sec:bc-o5-no-NS5} are related by a Higgs mechanism by giving vevs to the localised hypermultiplets in the former. Thus, since 1-form symmetries are insensitive to vevs for local operators, the results are essentially identical and  will thus depend mainly on the kind of O5-plane in the configuration. We can then study the 1-form symmetries in either of the two pictures, or carry out the discussion simultaneously.

\Ocase{The $O5^-$-plane configurations and 1-form symmetries}

Let us start discussing the configurations with an O5$^-$-plane. In
the configuration with no stuck NS5-brane, we have seen in section
\ref{sec:bc-o5-no-NS5} that the $n$ D3-branes intersecting the
O5$^-$-plane have boundary condition projecting the vector multiplets
down to $\fsp(n)$ and the hypermultiplets down to the
$\asymm$. Considering the possible global structures of an $\fsp(n)$
theory\footnote{This can be made more precise by considering the
  semi-infinite D3-branes to end on some faraway NS5-branes, so that
  it describes a 3d genuine $\fsp(n)$ theory. We will implicitly work
  similarly in the discussion of 1-form symmetry in forthcoming
  examples.} \cite{Aharony:2013hda}, we can have the $\Sp(n)$ or
$(\Sp(n)/\IZ_2)_\pm$ theories. The former admits Wilson lines in the
fundamental representation, while the latter do not (and admit
magnetic or dyonic lines, respectively for the $\pm$ choices). In
short, we have potential $\IZ_2$ electric or magnetic (or the dyonic
diagonal combination) 1-form symmetries, which are jointly described
by $\IZ_2$-valued 2-form background fields $C_2$, $B_2$. We will
recover this picture in the SymTFT arising from the holographic dual
in section \ref{sec:orientifold-symtft}.

We recover the same result for the configuration with one stuck NS5-brane, in which we have an $\fsu(n)$ theory with a localised hypermultiplet in the $\asymm$, recall section \ref{sec:bc-o5-yes-NS5}. For simplicity, we start with the case of even $n$. The presence of the dynamical hypermultiplet implies that the potentially present electric 1-form symmetry of the $\fsu(n)$ theory is broken to at most $\IZ_2$. Therefore the global structure of the gauge theory is either $SU(n)$ with a $\IZ_2$ electric 1-form symmetry, admitting $\IZ_2$ valued Wilson lines, or $(SU(n)/\IZ_{\frac n2})_{0,1}$ in two variants, admitting magnetic or dyonic Wilson lines. This matches the picture discussed above.

With one stuck NS5-brane, it is possible to have odd $n$. This is compatible with the Higgsing realised by moving the NS5-brane off the D3-brane stack because one of the D3-branes is dragged along with the NS5-brane; equivalently, the Higgsing with a hypermultiplet in the $\asymm$ for odd $n$ necessarily has one vev entry equal to zero. Hence, only in the limit of infinite separation we really connect with the case with no stuck NS5-brane and even $n$.
For finite separation, the presence of this extra D3-brane may modify the structure of symmetries in the theory. In fact, it is possible to tensor the dynamical matter in the $\asymm$ to build dynamical objects in the fundamental representation, so that no leftover 1-form symmetry is present.

\Ocase{The $O5^+$-plane configurations and the non-BPS spinor}

Let us now consider the  configurations with the O5$^+$-plane. In principle one could expect this case to be fairly similar to the above, but there are two main (and related) differences. The first is that the orientifold projection on a set of $n$ D3-branes intersecting the O5$^+$-plane leads to an $\fso(n)$ symmetry, whose pattern of possible global structures is more involved than for $\fsp(n)$, due to the role of line operators in the spinor representations. The second is that, as we discuss next, on top of matter in two-index tensor representations, the O5$^+$-plane supports new dynamical degrees of freedom, precisely in the spinor representation. In the following we discuss the appearance of this spinor states.

Building on the characterisation in \cite{Witten:1998cd} of D-brane charges via K-theory, \cite{Gukov:1999yn} proposed the 
classification of D-brane charges localised on O$p$-planes based on equivariant K-theory on $\IR^{p+1}\times (\IR^{9-p}/\Omega {\cal I}_{9-p})$, where ${\cal I}_{9-p}$ flips the coordinates or $\IR^{9-p}$. In particular, localised $(d+1)$-dimensional branes on the $(p+1)$-dimensional O$p^{\pm}$-planes are classified by the real or symplectic K-theory groups $KR(\IR^{9-p}, \IR^{p-d})$ and $KH(\IR^{9-p}, \IR^{p-d})$. These are easily  obtained by the isomorphisms
\begin{equation}
KR(\IR^{9-p}, \IR^{p-d})\simeq KO(\IS^{2p-d-1})\quad ,\quad KH(\IR^{9-p}, \IR^{p-d})\simeq KO(\IS^{2p-d+3})\, .
\end{equation}
In particular, upon using mod-8 Bott periodicity of $KO$ groups, for an O5$^+$-plane the relevant K-theory groups are  $KO(\IS^{5-d})$. The non-trivial groups are
\begin{equation}
    KO(\IS^0)=KO(\IS^4)=\IZ\quad ,\quad KO(\IS^1)= KO(\IS^2)=\IZ_2\, ,
\end{equation}
which describe the BPS D5- and D1-brane, and a $\IZ_2$-charged non-BPS 4-brane and a 3-brane. 

We are interested in the 3-brane, for which we now provide a
microscopic construction. To describe a $\IZ_2$ charged 3-brane in the
directions 0 789 (hence along the O5$^+$-plane), we consider a
D7-brane along 03 456 789 and at the origin in the directions 12, and
its orientifold image, namely an anti-D7-brane (which we denote by
${\ov{\rm D7}}$) spanning 03 456 789 and also at the origin in the
directions 12. Away from the O5$^+$-plane they can both annihilate via
tachyon condensation, but the tachyon is odd under the projection and
must vanish at the O5$^+$-plane location, leading to a localised
3-brane charge. A simple way to check the above picture is to
(formally) perform T-duality along the directions 3456, so that the
O5$^+$-plane maps to an O9$^+$-plane, and the D7-${\ov{\rm D7}}$ pair
maps into a D3-${\ov{\rm D3}}$ pair, whose tachyon is projected out,
so it provides a stable non-BPS state. This non-BPS D3-brane in the
non-supersymmetric $\Sp(32)$ theory in \cite{Sugimoto:1999tx} was
identified in \cite{Witten:1998cd} with the class $KSp(\IS^6)=\IZ_2$
and built explicitly with precisely the above microscopic
construction.

Let us now introduce a stack of $n$ D3-branes intersecting the O5$^+$-plane. Using the above microscopic construction, it is easy to see that the open strings between the D7-${\ov{\rm D7}}$ and the $n$ D3-branes have 8 DN directions, and lead to fermion zero modes in the fundamental representation, so the resulting state transforms in the spinor representation of the $\fso(n)$ in the $n$ D3-branes. We also observe that for even $n$ one in fact gets spinors of both chiralities; this is in contrast with other instances of non-BPS spinor states (such as the type D0-brane) for which a worldvolume $\IZ_2$ gauge symmetry enforces a projection eliminating one of the chiralities \cite{Sen:1998tt,Witten:1998cd}.

One may worry that the state is higher-dimensional, as it additionally extends in the directions 456 789, and would see to play no role in the discussion of symmetries of the 4d theory. However the relevant point is that it has a 1d intersection with the D3-branes, so it effectively defines a line operator in the spinor representation for most purposes\footnote{In the holographic dual in section \ref{sec:orientifold-gravity-dual} and its SymTFT sector in section \ref{sec:orientifold-symtft} this will be more manifest, as the additional dimensions are actually compact in the near horizon geometry of the D3-branes.}. In particular, the global structure of the symmetry group must be compatible with the existence of objects transforming in the spinor representation. In addition, one may consider configurations where the non-BPS brane is compactified in the extra directions 789, and form the analogue of  dynamical finite energy brane-antibrane pairs, which can be nucleated and break line operators in the spinor representation. It is in this sense that we refer to them as spinor states.

The presence of these spinor states will be key in our identification below of the 1-form symmetries of this system when regarded as a boundary configuration of the 4d theory. Incidentally, we point out that such spinor states are absent for the O5$^-$ configurations, so that our identification of the 1-form symmetry structure in the earlier discussion is not modified.

\Ocase{The $O5^+$-plane configurations and 1-form symmetries: odd $n$}

Consider now the configurations with the O5$^+$-plane, for which $n$ may be even or odd. We start with the case of $n$ odd and look first at the case with no stuck NS5-brane.  As shown in section \ref{sec:bc-o5-no-NS5}, the $n$ D3-branes intersecting the O5$^-$-plane have boundary condition projecting the vector multiplets down to $\fso(n)$ and the hypermultiplets down to $\symm$. For odd $n$, the center of $\fso(n)$ is $\IZ_2$, with the $\IZ_2$ charge carried by the spinor representation. The possible global structures are $Spin(n)$, admitting spinor Wilson line operators, and $SO(n)_\pm$, without them but admitting monopole or dyon line operators, respectively. Because of the existence of the non-BPS spinor discussed above, we are led to a $Spin(n)$ global structure. This would seemingly lead to a $\IZ_2$ electric 1-form symmetry, but would be problematic, because the bulk $\fsu(n)$ has no $\IZ_2$ subgroup of its center $\IZ_n$ for odd $n$. Happily, the fact that the spinor is {\em dynamical} (i.e. can be used to break line operators in the spinor representation) saves the day, because it breaks the $\IZ_2$ 1-form symmetry. Hence, there is no left-over 1-form symmetry in this case.

If there is one stuck NS5-brane, recalling section \ref{sec:bc-o5-yes-NS5}, the boundary conditions preserve an $\fsu(n)$ symmetry but there is a localised hypermultiplet in the $\symm$. Hence, it would seem that there is a potential $\IZ_2$ electric 1-form symmetry, with the $\IZ_2$ charge carried by  line operators in the fundamental representation. This would however be at odds with the Higgsing to the $\fso(n)$ theory, where the $\IZ_2$ 1-form symmetry is ultimately broken by the spinor state. However, since the latter is a stable state, it must survive in the unHiggsing to the $\fsu(n)$ theory and be present even when the NS5-brane sits on top of the O5-plane. In this situation, although the presence of the NS5-brane prevents a fully microscopic description, it is possible to guess its $\fsu(n)$ quantum numbers. In particular, two of them can be combined into a fundamental representation of $\fsu(n)$ (just as two spinors can combine into a vector in the $\fso(n)$ subgroup), hence the $\IZ_2$ 1-form symmetry is broken, matching the above result.

\Ocase{The $O5^+$-plane configurations and 1-form symmetries:
  even $n$}

We now quickly consider the case of even $n$, starting with the configuration with no stuck NS5-brane. As shown in section \ref{sec:bc-o5-no-NS5}, the $n$ D3-branes intersecting the O5$^+$-plane have boundary condition projecting the vector multiplets down to $\fso(n)$ and the hypermultiplets down to $\symm$. The center ${\cal C}$ of $\fso(n)$ is $\IZ_4$ for $n=4k+2$ or $\IZ_2\times\IZ_2$ for $n=4k$, so the global structures are either $Spin(n)$ or $Spin(n)/H$ with $H$ a non-trivial subgroup of ${\cal C}$. The former is the only structure admitting spinors of both chiralities, hence is the appropriate global structure in our case. The $\IZ_4$ or $\IZ_2\times \IZ_2$ electric 1-form symmetry is however broken by the {\em dynamical} presence of the spinors themselves, so no non-trivial 1-form symmetry remains.

In the configuration with a stuck NS5-brane, as discussed in section \ref{sec:bc-o5-yes-NS5}, we have $\fsu(n)$ boundary conditions with a localised hypermultiplet in the $\symm$. As in the case of odd $n$, the would-be present $\IZ_2$ 1-form symmetry is broken by the $\IZ_2$-charged non-BPS states, so no 1-form symmetry is preserved in this case.

\Ocase{The ${\widetilde{\rm O5}}^\pm$-plane configurations}

Let us now consider the configurations with the ${\widetilde{\rm O5}}^-$-plane, which can be regarded as the previous ones with the addition of one stuck D5-brane on top of the O5-plane. Clearly, the present of the extra fundamental flavours from the D3-D5 open string sector, the electric 1-form symmetry is completely broken. For the case of the ${\widetilde{\rm O5}}^+$-plane, we have a situation similar to the O5$^+$-plane, and we will not discuss it further.

\subsection{Holographic dual}
\label{sec:orientifold-gravity-dual}

In this section we describe the gravitational dual of the configurations of 4d $\CN=4$ $\fsu(N)$ SYM on half-space with orientifold boundary conditions of the kind considered in section \ref{sec:orientifold-hw}. The discussion of the corresponding supergravity solutions require a slight generalisation beyond those in section \ref{sec:holodual}, as we explain next.

\subsubsection{Supergravity solutions with multiple AdS$_5\times\IS^5$ asymptotic regions}
\label{sec:general-holo}

The supergravity solutions considered in \cite{DHoker:2007zhm,DHoker:2007hhe} (see also \cite{Aharony:2011yc,Assel:2011xz,Bachas:2017rch,Bachas:2018zmb} for further works) are actually the most general compatible with $SO(2,3)\times SO(3)\times SO(3)$ symmetry and 16 supersymmetries. In fact, they describe the near horizon solution of stacks of D3-branes with general Hanany-Witten configurations of NS5- and D5-branes. These include configurations defining BCFT$_3$ boundary conditions for 4d $\CN=4$ $\fsu(N)$ SYM, but also configurations in which several 4d $\CN=4$ $\fsu(N_i)$ SYM sectors are separated by configurations of NS5- and D5-branes (with extra D3-branes suspended among them). We quickly review their basic structure, emphasising only the points necessary for the generalisation we need, and refer the reader to the literature for further details.

The corresponding gravity duals are given by
AdS$_4\times \IS^2_1\times\IS^2_2$ fibered over a Riemann surface,
which can be conveniently described as a disk. Its boundary is divided
into segments of two kinds, in which either $\IS^2_1$ or $\IS^2_2$
shrink. Segments of different kinds are separated by punctures,
which in general can correspond to asymptotic AdS$_5\times\IS_5$
regions, where the $\IS^5$ is realised by fibering
$\IS_1^2\times\IS_2^2$ over an arc in the Riemann surface with one
endpoint in each of the two boundary segments, as we discussed in the
example in figure \ref{fig:quadrant}. We denote by $N_i$ the RR 5-form
flux on the $\IS^5$ in the $i^{th}$ puncture. In case this flux is
zero, the geometry at this point is actually smooth, and describes the
shrinking of $\IS^5$ to zero size, as in the case of the origin in
figure \ref{fig:quadrant}.

Within each segment, there may be additional punctures, which describe NS5- or D5-branes, according to the kind of segment on which it is located. These punctures describe local regions with geometry AdS$_4\times\IS^2\times\IS^3$, where the $\IS^2\times\IS^3$ is obtained by fibering $\IS_1^2\times\IS_2^2$ over an arc in the Riemann surface with the two endpoints on both sides of the puncture in the boundary segment, as we discussed in the example in figure \ref{fig:quadrant}. There is an NSNS or RR 3-form flux over $\IS^3$ encoding the 5-brane charge, and an integral of the 2-form field (of the opposite kind) over $\IS^2$, encoding the 5-brane linking number (or worldvolume monopole charge). We note that the argument below (\ref{the-f5}) applies also here, essentially relating the integer value of the 2-form field background to the integer value of the RR 5-form ${\tilde F}_5$ flux on $\IS^2\times\IS^3$.

A particular class is that of solutions with only one asymptotic AdS$_5\times\IS^5$ region, which we have studied in earlier sections. In this case, in our discussion the disk of the Riemann surface was distorted into the quadrant in figure \ref{fig:quadrant}, with the boundary corresponding to the horizontal and vertical semi-infinite lines, plus the point at infinity. The two axes correspond to segments of the two kinds, and they are separated by a puncture at the origin, with no 5-form flux on the $\IS^5$, hence describing a smooth point, and by a puncture at the point at infinity, with $N$ units of 5-form flux, describing the asymptotic AdS$_5\times\IS^5$ dual to 4d $\fsu(N)$ SYM theory on half-space. Along each segment/axis, there are 5-brane punctures which describe the BCFT$_3$ and which have been the focus of most of our study.

The previous paragraphs show that most of the ingredients of the supergravity solution in the general case are essentially already present in the case of an ETW configuration with a single AdS$_5\times\IS^5$ asymptotic region. Hence, most of our discussion of the SymTFT structure of the topological sector of the gravitational background extend to the general case. 
The general structure is that of a collection of SymTrees, analogous to those introduced in \cite{Baume:2023kkf} to discuss compactifications with several sectors associated to isolated singular points. In our case, we have a set of SymTFT$_{N_i}$'s associated to the AdS$_5\times\IS^5$ asymptotic regions, which are separated by $U(1)$ junction theories connecting them with the different SymTFT$_{\IS^2\times\IS^3}$'s associated to the 5-branes. We also have a retracted SymTree picture in which the SymTFT$_{\IS^2\times\IS^3}$'s are collapsed onto the junction theories to yield the explicit 5-brane probe theories.  The coefficients of the topological couplings (\ref{bf-symtft}) jump across the junctions, as dictated by the induced D3-brane charge on the 5-branes in the retracted SymTree picture, or equivalently by the coefficient of the sector (\ref{bf-symtft}) of the corresponding SymTFT$_{\IS^2\times\IS^3}$ in the full SymTree picture. We will not consider this general  setup any further, but simply adapt it to the class of solutions necessary to describe the gravity dual of the orientifold boundary conditions in section \ref{sec:orientifold-hw}, and to extract their Symmetry Theories. 

\subsubsection{Orientifold ETW configurations}

The general class of solutions in the previous section includes the gravitational duals of the orientifolded brane configurations in section \ref{sec:orientifold-hw}, when described in the double cover, as we explain in this section. 

The appropriate class of solutions is that with two identical asymptotic AdS$_5\times\IS^5$ regions, with the same number $N$ of RR 5-form flux units (and same value of other fields, e.g. the 10d axio-dilaton), and an intermediate region with a similarly $\IZ_2$ symmetric distribution of NS5- and D5-branes. For convenience we depict the Riemann surface $\Sigma$ as a strip, parametrised by a coordinate $z$ with ${\rm Im}\, z\in [0,\pi/2]$, with the two points at infinity ${\rm Re }\, z\to\pm\infty$ corresponding to the two asymptotic AdS$_5\times\IS^5$ regions, and with NS5-brane punctures (resp. D5-brane punctures) in the lower (resp. upper) boundary\footnote{The quadrant in the ETW configurations in the previous sections can be described as a strip by simply taking $w=-e^{-z}$, see (\ref{map}) later. The point $w=0$, i.e. $z\to \infty$, corresponds to a closed off puncture, with vanishing 5-form flux, hence no actual asymptotic AdS$_5\times\IS^5$ region at that point \cite{Aharony:2011yc}.}. This strip corresponds to the disk mentioned in section \ref{sec:general-holo}, with the two boundaries of the strip corresponding to the two kinds of segment in the disk boundary, and the two points at infinity corresponding to punctures describing two asymptotic AdS$_5\times \IS^5$.

The geometry is of the kind explained in section \ref{sec:holodual}, with metric (\ref{ansatz}) and functions (\ref{the-fs}) defined in terms of $h_1,h_2$ given by (see e.g. \cite{Bachas:2011xa,Assel:2011xz} for explicit expressions of this form)
\beqa
h_1 &=& \Bigg[\, -i{\tilde\alpha} \sinh(z-{\tilde\beta})\, -\,\sum_b {\tilde\gamma}_b \,\log \left(\tanh \left( \frac{i\pi} 4- \frac{z-{\tilde \delta}_b}2\right) \right)\Bigg]\,+\, c.c.\nonumber\\
h_2 &=& \Bigg[\, \alpha \cosh(z-\beta)\, -\,\sum_a  \gamma_a \,\log \left(\tanh \left(\frac{z-\delta_a}2  \right) \right)\Bigg]\,+\, c.c.
\label{the-strip-hs}
\eeqa
These are analogous to (\ref{the-hs}), but in terms of the strip coordinate $z$ instead of the quadrant coordinate $w$, and with a slightly different asymptotic behaviour to include the second AdS$_5\times\IS^5$ region. Specifically, they expressions are related by
\begin{equation}
w=-e^{-z}\quad , \quad 
    \gamma_a\equiv 2 d_a\quad , \quad {\tilde \gamma}_b\equiv 2{\tilde d}_b \quad ,\quad k_a\equiv e^{-\delta_a}\quad ,\quad l_b\equiv e^{-{\tilde \delta}_b}
    \label{map}
\end{equation}

The configuration is quotiented by $\Omega R$, with $R$ being inherited\footnote{This description of the orientifolding ignores the backreaction of the O5-plane on the geometry. The latter could be accounted for (at least slightly far away from the O5-plane location) by introducing an additional D5-brane source in the expressions of $h_1,h_2$. In our discussion we will ignore this backreaction, but its topological content will be included in our discussion of the SymTFT in section \ref{sec:orientifold-symtft}.} from its flat space avatar $R:(x^3,x^4,x^5,x^6)\to (-x^3,-x^4,-x^5,-x^6)$. It can be expressed in terms of the $\IS^5$ in the internal space, by using its embedding in $\IR^6$ as
\begin{equation}
(x^4)^2+(x^5)^2+(x^6)^2+(x^7)^2+(x^8)^2+(x^9)^2=r^2\, .
\end{equation}
The fixed point set is hence AdS$_4\times\IS_1^2$. In the AdS$_4\times\IS_1^2\times\IS_2^2$ fibration over the strip, it acts as antipodal identification on $\IS_2^2$ together with a reflection $z\to -{\ov z}$ on the complex coordinate on the strip. The fixed point set in the strip corresponds to the segment in the imaginary axis, but the only fixed point set in the whole geometry corresponds to the $\IS_1^2$ sitting at the upper endpoint of this segment (times AdS$_4$). We thus have an O5-plane on AdS$_4\times \IS_1^2$, and located at $z=i\pi$ in the strip; this is precisely the geometry required to preserve the same supersymmetries as the D5-branes.

%%%%%%%%%%%
\begin{figure}[htb]
\begin{center}
\includegraphics[scale=.35]{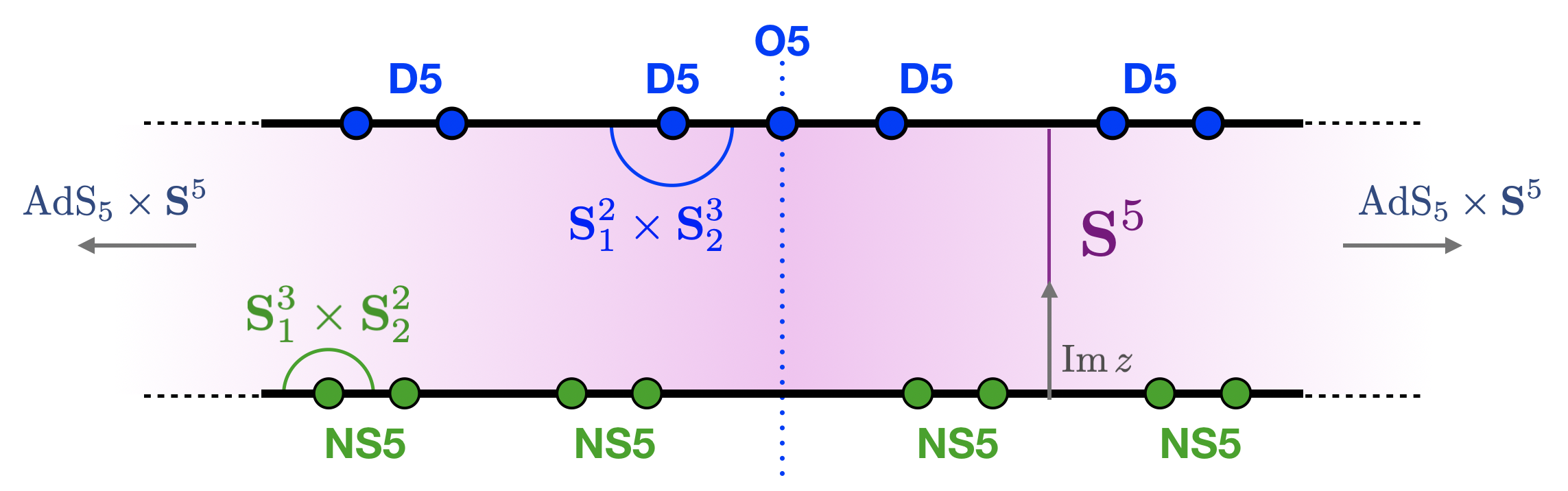}
\caption{\small Picture of the Riemann surface in the presence of and O5-plane orientifold quotient. There are two orientifold image asymptotic AdS$_5\times\IS^5$ regions, and the 5-branes are distributed also symmetrically under the reflection with respect to the blue dotted vertical line. Note however that the only fixed locus in the full geometry, i.e. the O5-plane, is the AdS$_4\times\IS^2$ on top of the blue dot. For comparison with figure \ref{fig:quadrant} we have depicted the segments corresponding to $\IS^5$ (violet vertical line) and the $\IS^2\times\IS^3$'s (blue and green arcs).}
\label{fig:orientifold}
\end{center}
\end{figure}
%%%%%%%%%%%

In order to admit this orientifold quotient, the 5-brane configurations have to be distributed on the corresponding boundaries of $\Sigma$ symmetrically under this reflection with respect to the imaginary axis, see figure \ref{fig:orientifold}. For notational convenience, let us label the 5-brane stacks with indices $a$ and $b$ taking both positive and negative values, with the $a^{th}$ and $b^{th}$ stacks being orientifold images of the $-a^{th}$ and $-b^{th}$ (In case there are 5-branes on top of the O5-plane (for instance for the ${\widetilde {O5}}^-$-plane), we simply include them by using a label $a=0$ or $b=0$). Hence, for (\ref{the-strip-hs}) to be symmetric under $z\to-{\ov z}$, we must have 
\begin{equation}
  \delta_a=-\delta_{-a}\quad , \quad   {\tilde \delta}_b=-{\tilde\delta}_{-b}\quad , \quad \gamma_a=\gamma_{-a}\quad ,\quad {\tilde\gamma}_b={\tilde\gamma}_{-b}\, ,
\end{equation}
as well as  $\beta=0$, ${\tilde\beta}=0$. Namely, the positions and multiplicities of 5-branes on the different stacks must respect the symmetry. In particular, in our ETW brane notation in previous sections, $n_{-a}=n_a$, $m_{-b}=m_b$.

This also ensures, although we will not be explicit about it, that the NSNS 2-form and RR 2- and 4-form backgrounds also respect the symmetry, implying that the assignment of 5-form flux on the asymptotic $\IS^5$'s and the $\IS^2\times\IS^3$ on the 5-brane throats are $\IZ_2$ symmetric.  Finally, noting that the NSNS 2-form is intrinsically odd under the orientifold action, and recalling (\ref{the-fluxes}) we have ${\tilde L}_{-b}=-{\tilde L}_b$; on the other hand, the RR 2-form is intrinsically even under the orientifold, but the orientation of $\IS_1^2$ flips, so recalling (\ref{the-fluxes}) we have $K_{-a}=-K_a$. In case there are 5-branes on top of the O5-plane, we simply have $K_0=L_0=0$, in agreement with the fact that there is no flux jump due to the $\IZ_2$ symmetry.

From the perspective of the configuration after the quotient, we have a configuration that away from the location of the orientifold plane has the same structure as the solutions in section \ref{sec:holodual}. The main new ingredient is thus the presence of the orientifold plane and it action on local fields and localised degrees of freedom. This will also be the only main new ingredient in our study of the SymTFT.

\subsection{The SymTFT}
\label{sec:orientifold-symtft}

\subsubsection{General structure}
\label{sec:general-orientifold-symtft}
%%%%%%%%%%%
\begin{figure}[htb]
\begin{center}
\includegraphics[scale=.5]{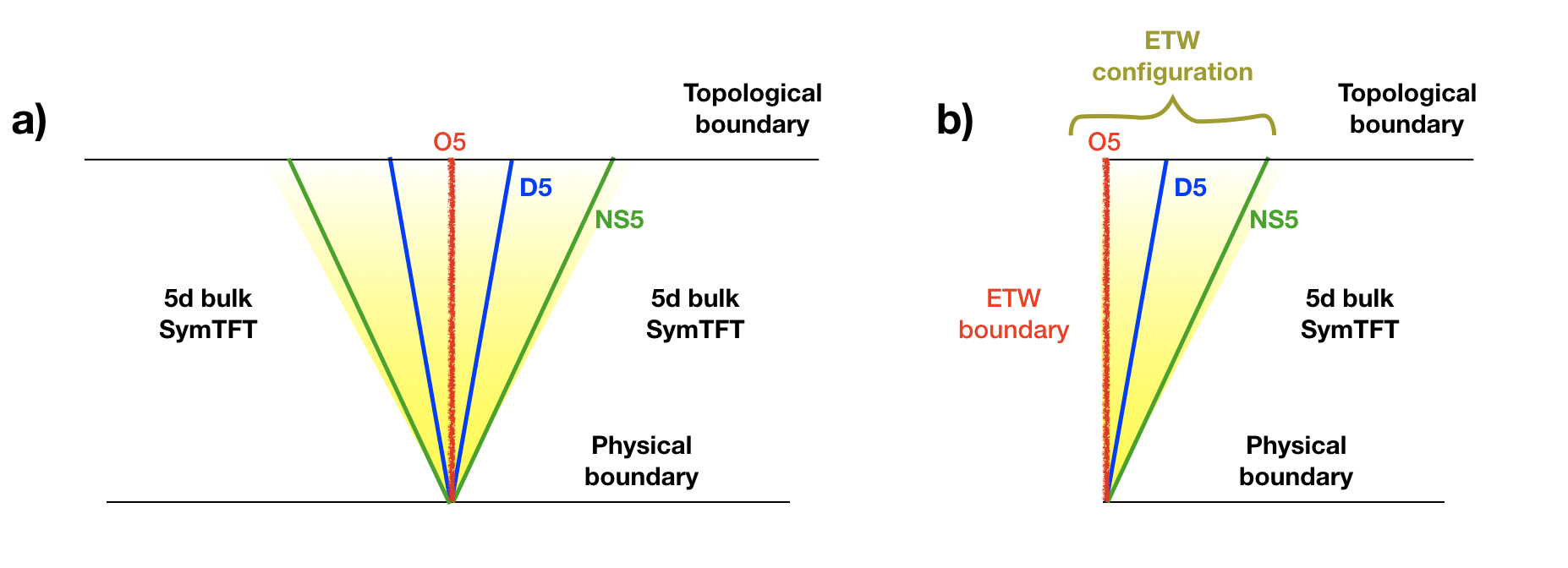}
\caption{\small a) Double cover of the SymTFT Fan of 4d $\CN=4$ $su(N)$ SYM on half-space with orientifold boundary conditions. b) The configuration in the orientifold quotient, making manifest that the O5-plane location defines an ETW boundary.}
\label{fig:orientifold-symtft}
\end{center}
\end{figure}
%%%%%%%%%%%

The most efficient way to extract the structure of the Symmetry Theory in this class of models is to describe the supergravity solutions in the previous section in Poincaré coordinates of the 5d solution. As shown in figure \ref{fig:orientifold-symtft}, in the covering space we have a full 4d Poincaré holographic boundary, with a $\IZ_2$ symmetric fan of wedges separated by 5-branes spanning AdS$_4$ slices, and the O5-plane (possibly with 5-branes on top of it) sitting in the AdS$_4$ orthogonal to the 4d holographic boundary. As in previous sections, the Symmetry Theory picture can be extracted by introducing the 4d physical boundary, and lettin the asymptotic infinity in the Poincaré radial direction to describe the topological boundary. Also, although we have described the configuration in the covering space, we can make the orientifold identification manifest, by simply folding the configuration along the vertical axis, and reach a configuration exhibiting an ETW boundary, which corresponds to the O5-plane (possibly with 5-branes).

The picture clearly shows that the structure of the Symmetry Theory is a variant of the SymTFT Fan encountered in previous sections. In fact, away from the O5-plane it is locally exactly of that kind. Namely, we have a SymTFT Fan built by putting together wedges containing branches of SymTFTs of the kind (\ref{bf-symtft}) for different values of their coefficients, joined via junction theories to SymTFT$_{\IS^2\times \IS^3}$'s corresponding to the 5-branes, with fluxes dictated by flux conservation. 

New features are however encountered in the local region around the
O5-plane, i.e. the ETW boundary. First, the 5-form flux does not go to
zero at the location of the O5-plane, we rather have a non-zero number
$n$ of flux quanta, and correspondingly the $\IS^5$ does not shrink to
zero size. Hence, spacetime is ending due to the presence of the
O5-plane; this is similar to how type IIA theory can end on O8-planes,
as in each boundary of type I$'$ theory. The second is the presence of a
new object, the O5-plane itself, whose role we discuss in the
following.

In order to understand the role of the O5-plane in the Symmetry Theory, the presence of additional 5-branes in the SymTFT Fan away from it is not relevant, so we may simply focus on the configuration in a small neighbourhood around the O5-plane. This sector of the Symmetry Theory in the covering space is therefore a 5d SymTFT$_n$ spanning the whole non-compact 4d spacetime, times the interval between the physical and topological boundaries, but cut in two halves by an O5-plane (possibly with 5-branes on top). The integer $n$ is the amount of flux remaining after the asymptotic value $N$ has been peeled of by the 5-branes away from the O5-plane.

The above picture describes the retracted version of the full Symmetry Theory. A more detailed picture is obtained by activating the full SymTree structure, considering the two orientifold images of the SymTFT$_n$ joining at a junction theory with the SymTFT$_{\IS^2\times\IS^3}$'s associated to the O5-plane (and the possible D5-branes on top of it). One may think that in the presence of an NS5-brane stuck on the O5-plane there should be an additional SymTFT$_{\IS^2\times\IS^3}$ branch in the SymTree. However since the linking number of such NS5-brane is zero, as discussed above, or equivalently the integral of the RR 2-form over the $\IS^2$ vanishes, so does the total 5-form flux over the corresponding $\IS^2\times\IS^3$. Therefore this would-be SymTFT$_{\IS^2\times\IS^3}$ is actually trivial. This is the SymTFT manifestation of the fact that the theories with a stuck NS5-brane can be Higgsed to those without it, with  no modification of their 1-form symmetry structure. 

These ingredients will be important in the discussion of the 1-form symmetries, to which we turn next.

\subsubsection{The 1-form symmetries in the SymTFT}

In this section we carry out the discussion of the 1-form symmetries in the SymTFT of the different O5-plane configurations. The discussion will parallel, and match, the discussion in section \ref{sec:orientifold-1form-sym}. For simplicity we restrict to configurations without stuck NS5-brane, since they simply add a topologically trivial sector to the SymTFT structure, as we have just explained.

\Ocase{The $O5^-$-plane configuration and 1-form symmetries}

Consider the configuration with an O5$^-$-plane, hence we need $n$ even. In the configuration with no stuck NS5-brane, we have two SymTFT$_n$ theories joining at a junction with the O5$^-$-plane. The latter may be regarded as an extra branch corresponding to a SymTFT$_{\IS^2\times\IS^3}$ with $-2$ units of 5-form flux. The relevant observation is that, because ${\rm gcd}(n,2)=2$, there is an unbroken $\IZ_2$ in the full theory. This is precisely the structure required to match the unbroken $\IZ_2$ 1-form symmetry discussed in section \ref{sec:orientifold-1form-sym}.

One may argue that the factor of 2 arises because we are working in
the covering space, and wonder about a possible intrinsic description
in the orientifold quotient. Indeed, such a description is obtained by
simply realising that the orientifold makes some of the 2-form fields
$\IZ_2$-valued. For $B_2$, this simply follows from the fact that it
is odd under the orientifold action, hence it is projected down to a
$\IZ_2$-valued gauge field. For $C_2$, this can be seen by noticing
that two stuck D1 strings can become orientifold images of each other
and move away from the orientifold plane, showing the field under
which they are charged is $\IZ_2$-valued.

\Ocase{The $O5^+$-plane configuration and 1-form symmetries: odd $n$}

Consider now the $O5^+$-plane configuration, starting with the case of an odd number $n$ of RR 5-form flux units in the bulk SymTFT. In this case the SymTree connects two SymTFT$_n$ theories with the SymTFT$_{\IS^2\times\IS^3}$ associated to the O5-plane, which contains a term of the form (\ref{bf-symtft}) with coefficient $+2$ in the covering space. Since now ${\rm gcd}(n,2)=1$, there is no left-over 1-form symmetry, in agreement with the field theory analysis in section \ref{sec:orientifold-1form-sym}.

From the perspective of the theory in the quotient, the action of the
O5-plane on the 2-form fields is as in the case of the O5$^-$-plane
above, hence it would seem that there is a $\IZ_2$ symmetry. However,
for odd $n$ this is broken by the baryonic vertex of the SymTFT$_n$
theories. Alternatively, there is a way more intrinsic to the
orientifold configuration to explain the breaking of the $\IZ_2$
symmetry. This is given by a vertex corresponding to the gravitational
realisation of the $\fso(n)$ spinor introduced in section
\ref{sec:orientifold-1form-sym}. The microscopic construction is the
near horizon version of the flat space construction mentioned there,
i.e. we consider a D7-${\ov{\rm D7}}$ pair spanning a 3d subspace
along the O5-plane volume and wrapped on the ${\bf RP}^3\times \IS^2$
around the O5-plane (with a nontrivial $\IZ_2$ Wilson exchanging the
two objects along non-trivial 1-cycles $H_1({\bf RP}^3,\IZ)=\IZ_2$, to
account for the orientifold action). This describes a spinor state in
the theory, which breaks the above $\IZ_2$ symmetry as explained in
the field theory approach.

\Ocase{The $O5^+$-plane configuration and 1-form symmetries: even $n$}

Consider now the case of an $O5^+$-plane with an even number $n$ of RR 5-form flux units in the bulk SymTFT. In this case we have a SymTree with an unbroken $\IZ_2$ symmetry because ${\rm gcd}(n,2)=2$. This is however broken by the presence of the spinor mentioned above, so no non-trivial 1-form symmetry remains, in agreement with the field theory analysis. 

\Ocase{The ${\widetilde{\rm O5}}^\pm$-plane configurations}

Let us quickly consider the ${\widetilde{\rm O5}}^\pm$-plane configurations. The ${\widetilde{\rm O5}}^-$-plane corresponds to the O5$^-$-plane with an extra D5-brane on top, hence its charge is $+1$ in the covering space. This means that, even if $n$ is even, there is no $\IZ_2$ symmetry preserved in the SymTree because ${\rm gcd}(n,1)=1$. This agrees with the field theory description. Equivalently, from the perspective of the orientifold quotient theory, the projection on 2-form fields is as in the O5$^-$-plane, so one might think that there is a $\IZ_2$ symmetry. However, the electric 1-form symmetry is broken by the presence of explicit flavours in the fundamental due to the stuck D5-brane, in agreement with the above result. Either way, there is no non-trivial 1-form symmetry in this case.

For the case of the ${\widetilde{\rm O5}}^+$-plane, we have a situation similar to the O5$^+$-plane, and we will not discuss it further.

\section{Conclusions}

In this work we have disclosed the structure of the 5d SymTFT of 4d
$\cN=4$ $\fsu(N)$ SYM on a space with boundary coupled to a
Gaiotto-Witten BCFT$_3$, by extracting the key topological information
from the gravitational dual of configurations of D3-branes ending on a
system of NS5- and D5-branes. The Symmetry Theory turns out to display
an extremely rich structure, consisting of a fan of SymTFTs
(\ref{bf-symtft}) with different levels, coupled via $\fu(1)$ junction
theories to the SymTFTs of the 5-branes, in a SymTree-like
structure. The SymTFT Fan realises 0-form flavour symmetries and their
enhancement, and allows for the identification of unbroken 1-form
symmetries of the SYM theory with boundaries.

\medskip

We have performed a systematic characterisation of the physical
effects undergone by different topological operators when moved across
different SymTFTs of the Symmetry Theory, by using the holographic
string theory realisation, truncated at the topological level. Subtle
field theory effects, such as the variation of the order of the
discrete symmetries, are very simply realized in terms of familiar
brane dynamics, such as Hanany-Witten brane creation effects and
Freed-Witten consistency conditions.

\medskip

We have further discussed the introduction of 7-branes to provide
boundary conditions for the 5-brane theories, and studied their role
as duality interfaces of the physical gauge theory, implementing
$SL(2,\IZ)$ duality transformation on topological operators moved
across the 7-brane branch cuts.

\medskip

We have carried out a similar analysis for orientifold boundary conditions. By extracting the topological information of configurations including the different kinds of O5-planes in the gravitational duals,  we have shown that the corresponding Symmetry Theories are obtained by enriching the SymTFT Fan with the local physics of the O5-plane neighbourhood. We have used string theory techniques to characterized the structure of 1-form symmetries for different kinds of orientifold boundary conditions. 

\medskip

Our work opens several interesting directions, some of which are:

\begin{itemize}

\item {We have devoted this work to uncovering the structure of the Symmetry Theory for 4d ${\cal N}=4$ $\fsu(N)$ SYM with Gaiotto-Witten BCFT$_3$ boundary conditions. It would be interesting to perform a fully systematic study of the SymTFT regarding the symmetry and operator structure of the Gaiotto-Witten BCFT, in particular matching the existing QFT body of knowledge for this class of theories}.

\item The use of 5-brane configurations to define boundary conditions allows to construct holographic dual pairs for boundaries of 4d theories with less supersymmetry. We expect our techniques to be helpful in extracting the corresponding SymTFTs for such constructions. For instance, as discussed in \cite{Huertas:2023syg} the holographic duals of $\cN=3$ S-fold \cite{Garcia-Etxebarria:2015wns} or $\cN=2$ orbifold theories with boundaries can be obtained by considering quotients of the parent ETW configurations with symmetric sets of 5-branes. It is straightforward to apply our techniques to formulate the SymTFT Fans for these configurations, and it would be interesting to explore possible new features arising from the quotients.

\item It is possible to construct 5-brane configurations with smaller supersymmetry but still providing BCFT$_3$ boundary conditions for the 4d $\fsu(N)$ SYM theories, for instance by using rotated configurations of NS5- and D5-branes preserving 4 supersymmetries \cite{Elitzur:1997fh,Barbon:1997zu,Giveon:1998sr}. Such BCFT$_3$ 5-brane configurations were considered in \cite{Hashimoto:2014vpa,Hashimoto:2014nwa} and share many features of those in \cite{Gaiotto:2008sa,Gaiotto:2008ak}. Hence, although they do not have a known supergravity dual, our techniques may well allow for the construction of their corresponding SymTFTs, hopefully uncovering new classes of SymTFT Fans, thus enlarging the classification of symmetry structures for supersymmetric boundary conditions of 4d $\fsu(N)$ SYM theories.

\item Several features of the Symmetry Theories we have considered seem related to the fact that we are dealing with conformal boundary conditions for a conformal field theory. In particular, the fan-like structure of junction theories in the holographic dual is  directly related to conformal invariance. This suggests that the Symmetry Theories of general conformal field theories with conformal boundary conditions are given in terms of a suitable SymTFT Fan combining wedges of bulk 5d SymTFTs and flavour SymTFTs into a local SymTree structure. It would be interesting to explore a general formulation of this SymTFT Fans for other classes of CFTs.

\item The characterisation of the topological sector of supergravity backgrounds with ETW branes is also relevant from the perspective of the Cobordism Conjecture \cite{McNamara:2019rup}, as it provides a direct link between the topological sector in the bulk and that on the ETW branes. This provides a powerful tool in the characterisation of cobordism defects in general theories of Quantum Gravity. Our work is an important first step in explicitly developing this program in the context of holographic dual pairs, and we expect it to seed further developments relevant to swampland studies.

We hope to come back to these and other interesting questions in the future.

\end{itemize}

\acknowledgments

We thank Roberta Angius, Andrés Collinucci, Matilda Delgado, Miguel
Montero and Xingyang Yu for helpful discussions. We also thank the anonymous JHEP referee for suggestions to improve the readability of the paper and make its implications more manifest. I.G.E. and
J.H. thank Harvard University for hospitality during part of the
development of this work. I.G.E. is partially supported by STFC grants
ST/T000708/1 and ST/X000591/1 and by the Simons Foundation
collaboration grant 888990 on Global Categorical Symmetries. The work
by J.H. and A.U is supported through the grants CEX2020-001007-S and
PID2021-123017NB-I00, funded by MCIN/AEI/10.13039/ 501100011033 and by
ERDF A way of making Europe. The work by J. H. is also supported by
the FPU grant FPU20/01495 from the Spanish Ministry of Education and
Universities.

\paragraph{Data access statement.} There is no additional research
data associated with this work.

\newpage

\appendix

\section{Review of the SymTree construction}
\label{app:symtree}

In this section we review the SymTree construction in \cite{Baume:2023kkf}, a generalized 5d Symmetry Theory which includes non-topological ingredients, and therefore goes beyond the SymTFT paradigm to encode the symmetry structure of certain theories. A variant of this construction, the SymTFT Fan, will play a prominent role in the main text. 

\subsection{Introduction}
\label{app:intro}

A SymTree \cite{Baume:2023kkf} is the Symmetry Theory (generalizing the concept of SymTFT) for a QFT with multiple ``decoupled'' sectors. These sectors are usually coupled by operator mixing terms, but these terms can vanish in a specific limit of mass scales or parameters of the theory, $\langle \mathcal{O}_1 \mathcal{O}_2 \rangle_{\mathrm{conn}} \rightarrow 0$, hence decoupling the sectors. Each sector is itself a non-trivial interacting relative QFT.

Although the various sectors have independent local dynamics, their global form can still be non-trivially coupled topologically due to their coupling in the UV theory. For instance, in a multi-sector model, the global form of the gauge group in the parent theory can impose non-trivial constraints on the spectrum of Wilson lines in the daughter ones.

A clear example of this phenomenon is a UV $\mathfrak{su}(N+M)$ gauge theory that undergoes adjoint Higgsing to an $\mathfrak{su}(N)\times\mathfrak{su}(M)\times\mathfrak{u}(1)$ gauge theory in the IR. For instance, whereas defining Wilson line operators in the parent $\mathfrak{su}(N+M)$ theory poses no difficulty, constructing Wilson lines for the isolated $\mathfrak{su}(N)$ or $\mathfrak{su}(M)$ gauge theory sectors introduces immediate subtleties, as the parent and daughter theories have seemingly different electric 1-form symmetries. The SymTree construction helps to construct Wilson lines on which the action of all 1-form symmetries is consistent, as we review in \ref{app:adjoint}.

\begin{figure}
    \centering
    \includegraphics[width=0.5\linewidth]{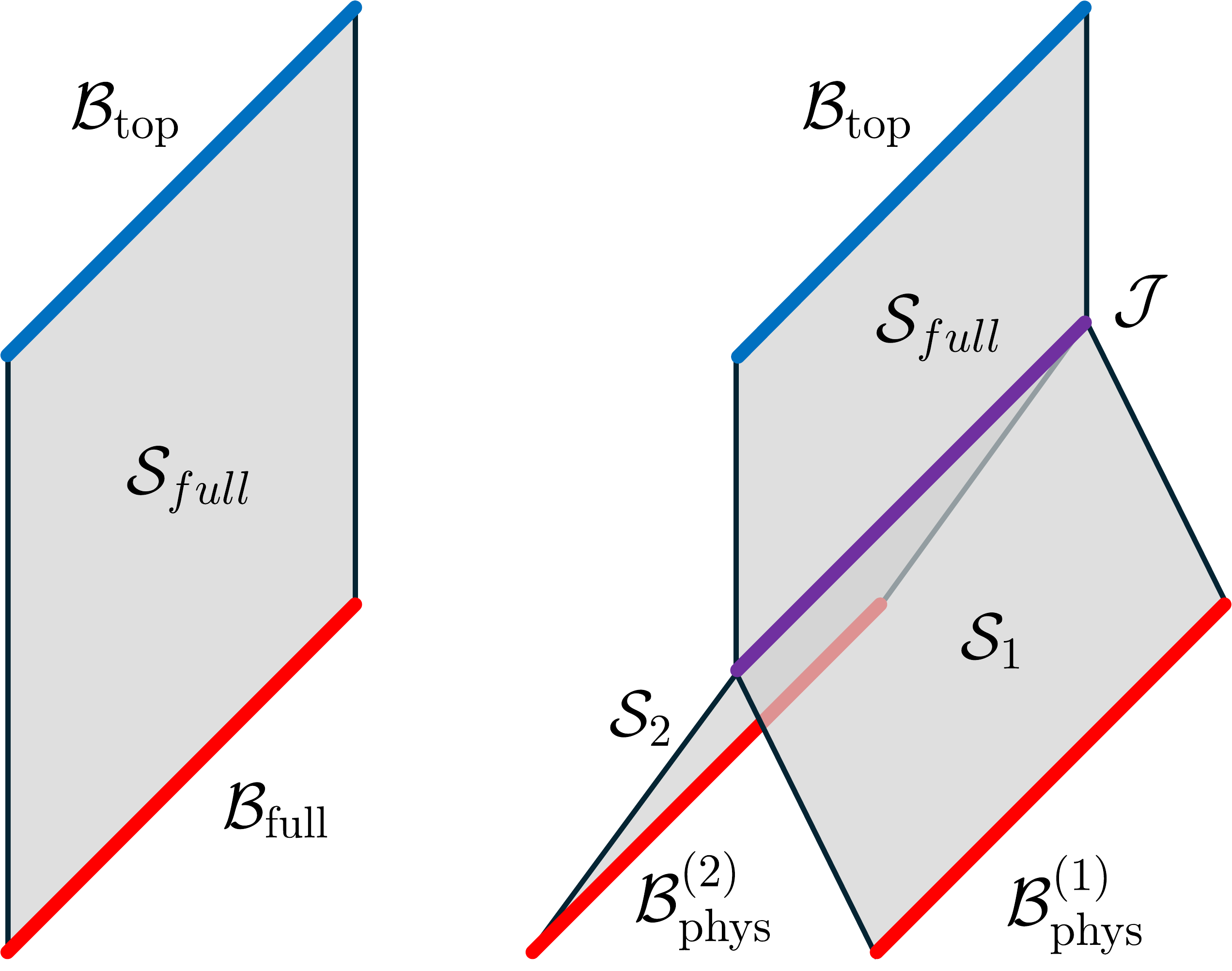}
    \caption{At the left, a standard SymTFT, with the physical boundary conditions $\mathcal{B}_{\text{full}}$ and the topological boundary conditions $\mathcal{B}_{\text{top}}$. At the right, a SymTree, the SymTFT of a multi-sector QFT. There are now two independent physical boundary conditions, $\mathcal{B}_{\text{phys}}^{(1)}$ and $\mathcal{B}_{\text{phys}}^{(2)}$, each with its own SymTFT $\mathcal{S}_1$ and $\mathcal{S}_2$, glued along $\mathcal{J}$.}
    \label{fig:SymTree1}
\end{figure}

In a SymTree, the SymTFTs for the different individual sectors (the branches) have the daughter theories at their physical boundaries, but, at the would-be topological boundaries, the SymTFTs are glued together at a {\em junction theory}, and onto the would-be physical boundary of a new SymTFT (the trunk) that encodes the symmetries of the parent theory, which has its own topological boundary, see figure \ref{fig:SymTree1}. In general there is no guarantee that the theory residing at the junction will be topological, and in all known examples it is not. In short, a SymTree consists of branches that merge at junctions, where each branch is associated with a SymTFT, and each junction specifies the gluing conditions between these bulk TFTs. The junctions themselves are not necessarily topological and can often support additional degrees of freedom.

Clearly, the above picture can be iterated, if the trunk SymTFT is actually just a branch of an even larger SymTree. This would describe the symmetry structure of a parent theory, which splits into a set of decoupled daughter theories, each of which subsequently splits into a set of grand-daughter theories. For this brief recap, we will stick to the simplest picture of a trunk theory splitting into several branches.

\subsection{Formal structure and retraction}
\label{app:formal-retraction}

These ideas can be formalized as follow. A $d$-dimensional multi-sector QFT is described by a set of relative theories $\mathcal{T}_i$ each at the physical boundary $\mathcal{B}_{\mathrm{phys}}^{(i)}$ of a corresponding $(d+1)$-dimensional SymTFT $\mathcal{S}_i$. Their would-be topological boundaries, $\mathcal{B}_{\mathrm{top}}^{(i)}$, are actually neither decoupled nor topological, as they couple through an additional junction theory $\mathcal{J}$, which interacts with the $\mathcal{T}_i$ only through purely topological couplings. The junction theory $\mathcal{J}$ is relative, in the sense that it is a boundary of the SymTFT $\mathcal{S}_{\rm full}$, which has a further boundary ${\cal B}_{top}$ with topological boundary conditions, see figure \ref{fig:SymTree1}.

The action of the system has the following structure, for the case of a 2-sector QFT for simplicity. Denoting by $S_1$ and $S_2$ the actions of the SymTFTs ${\cal S}_i$, and $S_{\mathcal{J}_{12}}$ that of the junction theory, we have
\begin{equation}
    S_{\rm full}=S_1+S_2+S_{\mathcal{J}_{12}}+S_{\rm full}
    \label{action-tree}
\end{equation}
where $S_{\rm full}$ describes the action of the trunk SymTFT. Hence, the original sectors $\mathcal{T}_1$ and $\mathcal{T}_2$ now interact via topological terms, as well as with an intermediate gluing theory $\mathcal{J}_{12}$. We will later discuss concrete realizations of this structure.

The generalization to theories with additional decoupled sectors should be clear. In general, there may be several ways to combine the decoupled physical theories into a network of junctions, leading to different treelike structures. But for each choice of tree, there is a notion of a $(d+1)$-dimensional bulk. 

For any SymTree we may recover a standard SymTFT description by pushing all of the junctions into the physical theories ${\mathcal T}_i$, resulting in a single slab filled by $\mathcal{S}_{\mathrm{full}}$ with a topological boundary and a physical boundary. This operation is dubbed \emph{retracting} the SymTree branches, and leads to a physical boundary $\mathcal{B}_{\textnormal{phys}}^{(\textnormal{retract})}$ combining all junctions and physical boundary conditions of the multi-sector QFT. In the 2-sector example, the physical boundary condition $\mathcal{B}_{\textnormal{phys}}^{(\textnormal{retract})}$ after retracting is obtained by dimensionally reducing $\mathcal{S}_1\otimes \mathcal{S}_2$ along the interval with boundary conditions $\mathcal{J}$ on the junction end, and $\mathcal{B}^{(1)}_{\mathrm{phys}}\otimes \mathcal{B}^{(2)}_{\mathrm{phys}}$ on the physical end.

\begin{figure}
    \centering
    \includegraphics[width=0.5\linewidth]{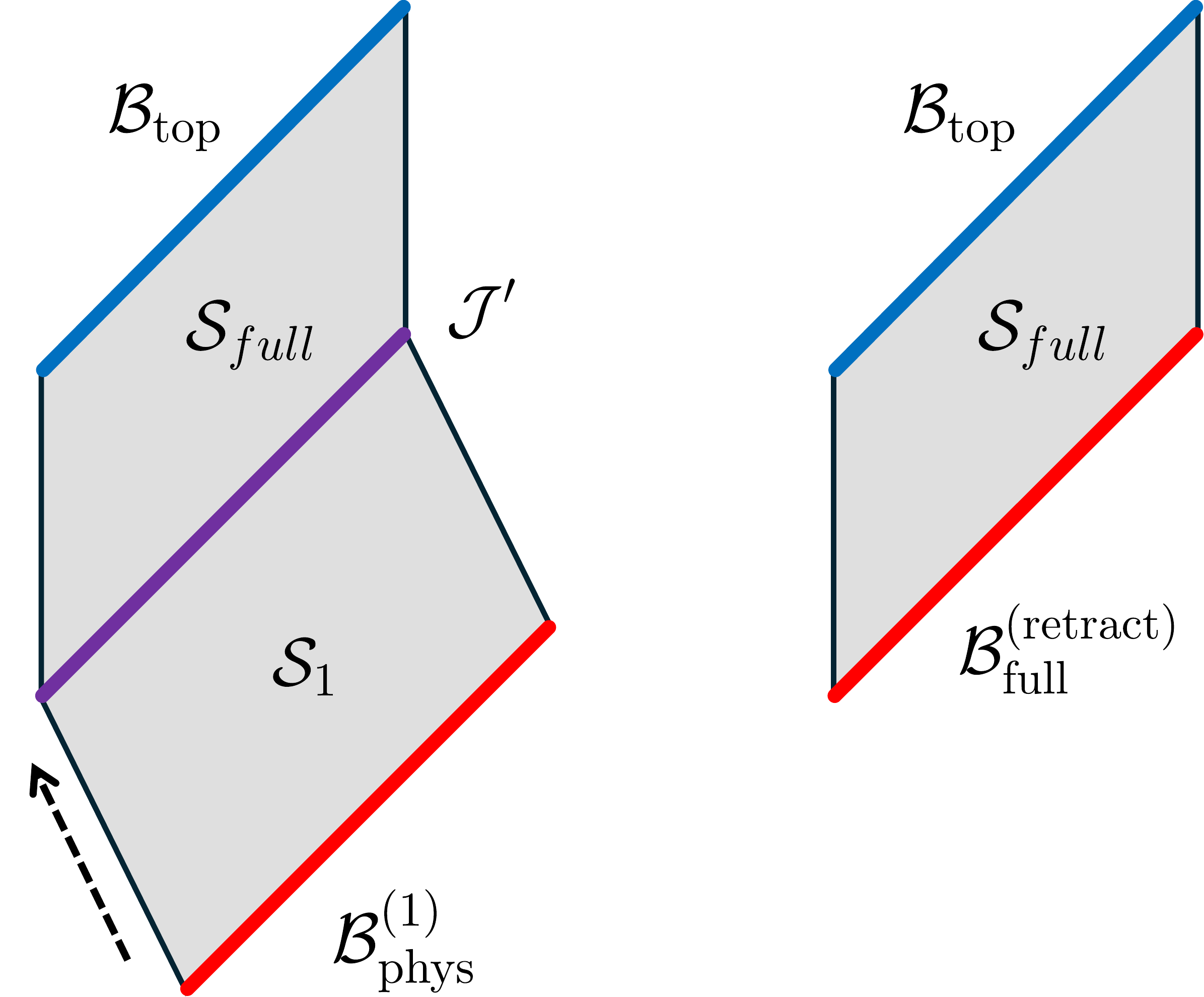}
    \caption{Procedure of retraction of a branch of a SymTree. Once all the branches are retracted on their junctions, we are left with the boundary condition $\mathcal{B}_{\text{full}}^{\text{(retract)}}$.}
    \label{fig:SymTree2}
\end{figure}

The retraction of SymTree branches can be carried out individually for each branch,  see figure \ref{fig:SymTree2}. 

\subsection{Crossing of operators}
\label{app:crossing}

\begin{figure}[htb]
\begin{center}
\includegraphics[scale=.45]{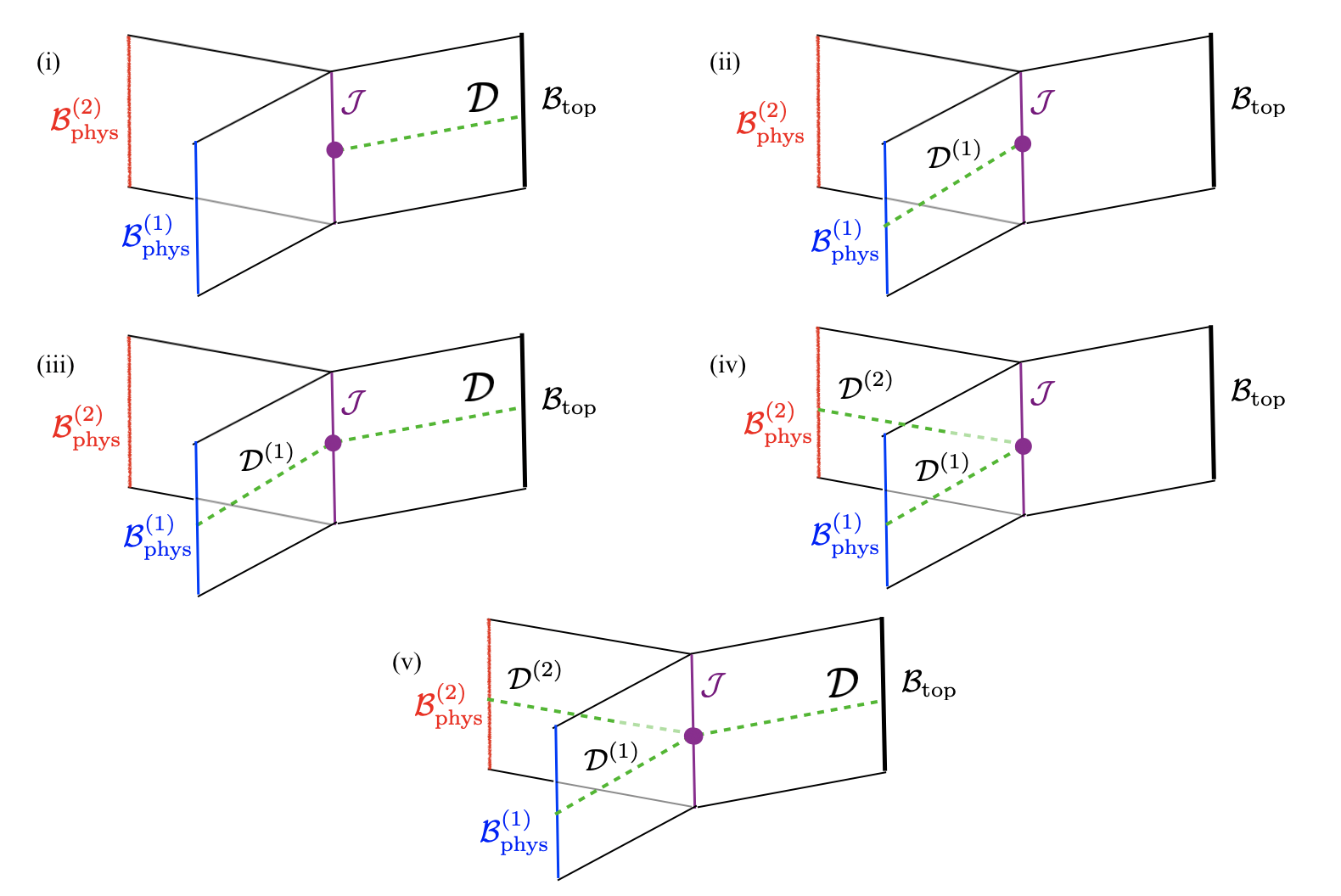}
\caption{\small Sketch of various possible topological defect operators in a SymTree. They necessarily intersect the junctions, and violet dots indicate the possible dressing of topological operators by some operator of the junction theory. The figure is essentially figure 7 in \cite{Baume:2023kkf}, modulo a left-right orientation flip, to fit better our later applications}
\label{fig:symtree-defects}
\end{center}
\end{figure}

An interesting phenomenon in SymTrees is the crossing of topological operators across junctions. In standard SymTFTs charged defects correspond to topological operators stretching from the topological boundary to the physical boundary. In the SymTrees described above, they necessarily cross the junction theory. Hence the charged defects are in general a combination of charged defects of the trunk SymTFT ${\mathcal S}_{\rm full}$ and one or several of the branch theories ${\mathcal S}_i$, plus some possible dressing with an operator of the junction theory. There are several possibilities to combine these objects in different ways, and in figure \ref{fig:symtree-defects} we provide a few illustrative (but not necessarily exhaustive) list for a 2-sector theory. 

%%%%%%%%%%%
%%%%%%%%%%%

\begin{figure}[htb]
\begin{center}
\includegraphics[scale=.45]{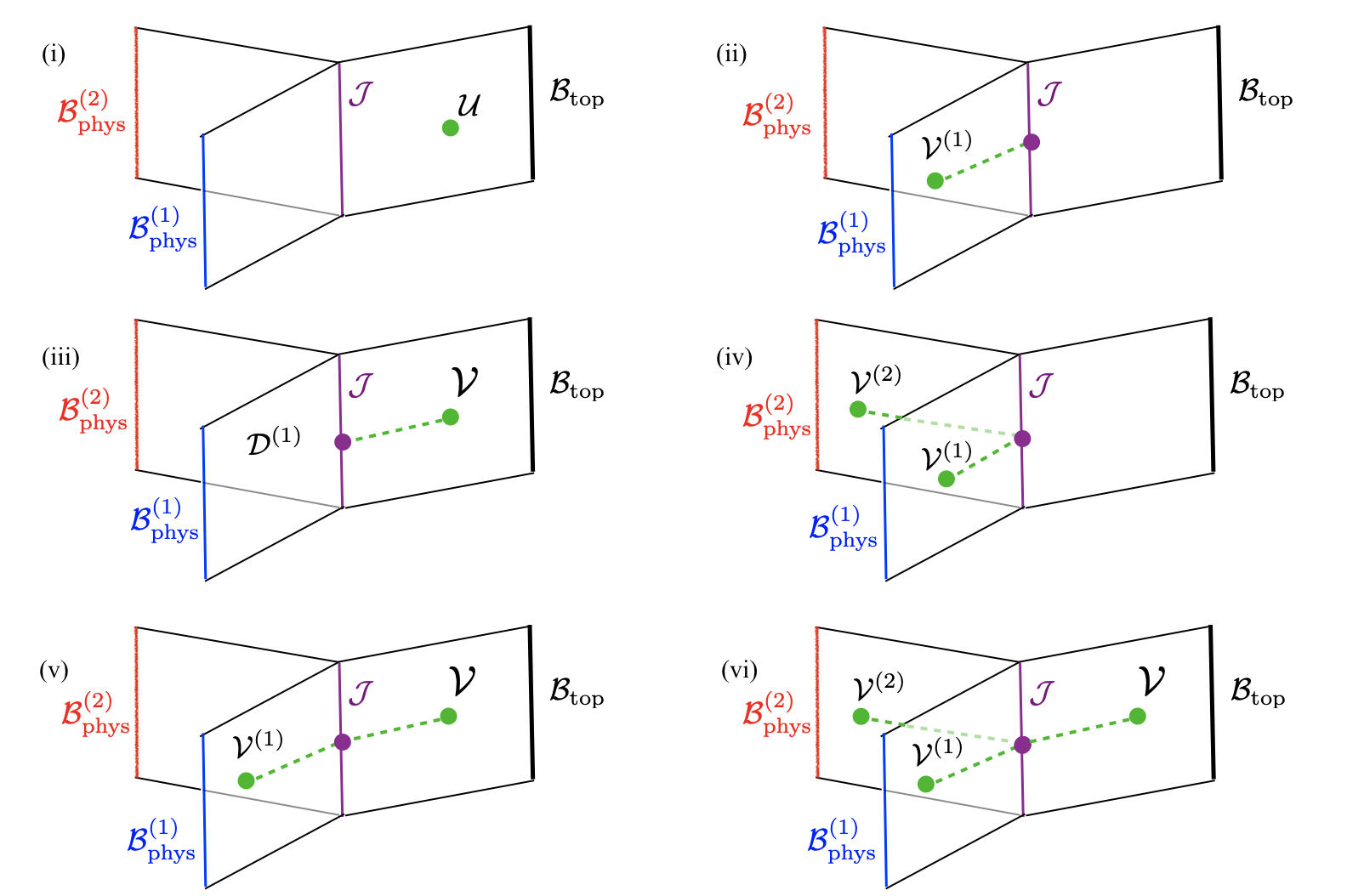}
\caption{\small Sketch of various possible topological symmetry generators in a SymTree. They are related by crossings across the junctions, and violet dots indicate the possible dressing of topological operators by some operator of the junction theory. The figure is essentially figure 9 in \cite{Baume:2023kkf}, modulo a left-right orientation flip, to fit better our later applications}
\label{fig:symtree-generators}
\end{center}
\end{figure}

Regarding symmetry generators, in standard SymTFTs they correspond to topological operators parallel to the physical and topological boundaries. In a SymTree they do not intersect the junction theories, but motion of such operators across junction leads to non-trivial relations between symmetry generators of the trunk ${\cal S}_{\rm full}$ and the brane ${\cal S}_i$ theories, in particular involving the appearance of operators of the junction theories.There are several possibilities to combine these objects in different ways, and in figure \ref{fig:symtree-generators} we provide a few illustrative (but not necessarily exhaustive) list in a 2-sector theory. 

%%%%%%%%%%%
%%%%%%%%%%%

Some of these are illustrated in the example of adjoint Higgsing in the next section

\subsection{Example: Adjoint Higgsing}
\label{app:adjoint}

A prototypical setup in which the SymTree construction arises is in adjoint Higgsing. We will exploit intuitions from this example in the 4d case, so we express it in such terms, although the arguments hold in general dimensions. We moreover follow the discussion of the theory in \cite{Baume:2023kkf} realized on a set of $N$ D3-branes in flat space, separated into several stacks of $N_i$ D3-branes, $i=1,\ldots, k$,  with $\sum_i N_i=N$. The QFT description corresponds to an adjoint Higgsing $\fsu(N)\to \fs[\fu(N_1)\oplus\ldots \oplus \fu(N_k)]$. For simplicity we will mostly write explicit expressions for the 2-stack case. The adjoint vevs provide the necessarily scales for the physical decoupling of the different sectors explained in section \ref{app:intro}.

The Symmetry Theory of this system is a SymTree where the trunk theory ${\mathcal S}_{\rm full}$ is the 5d SymTFT of 4d $\fsu(N)$ SYM, and the branch theories ${\mathcal S}_i$ (assuming a single junction at which the trunk theory splits into $k$ branches) are given by the 5d SymTFTs of 4d $\fsu(N_i)$ SYM. These theories are obtained from the dimensional reduction of the 10d type IIB string theory on $\IS^5$'s (morally, surrounding the D3-branes) with $N$, or $N_i$, units of RR 5-form flux, respectively. The deformation of the $\IS^5$ surrounding all the $N$ D3-branes into the sum of all the $\IS^5$'s surrounding each stack of $N_i$ D3-branes, as its size shrinks below a  critical radius $r_*$, allows for the computation of the coupling of these theories. Focusing on the $k=2$ 2-stack case, this results in an action of the from (\ref{action-tree}), with, in the language of (\ref{bf-symtft})
\beqa
S^{(i)}_{5d}=\frac {N_i}{2\pi}\int_{\IX_4\times (0,r_*)} B_2\wedge dC_2\quad , \quad S_{\rm full}=\frac {N}{2\pi}\int_{\IX_4\times (r_*,\infty)} B_2\wedge dC_2
\eeqa
The junction theory $S_{\mathcal{J}_{12}}$ is just 4d ${\cal N}=1$ $U(1)$ theory, coupled to $B_2$ and $C_2$ as background fields for the electric and magnetic 1-forms symmetries of the $U(1)$ Maxwell theory. For instance for the bulk 2-form field $B_2$ the junction conditions are
\beqa
\frac{N_1}g B_2^{(1)}=\frac{N_2}g B_2^{(2)}=\frac Ng B^{(\rm full)}
\label{juction-condi2}
\eeqa
with $g={\rm gcd}(N_1,N_2)$. These junction conditions were obtained in \cite{Baume:2023kkf} from the holographic dual realization, by computing the dimensional reduction of the 10d action on the geometry of an $\IS^5$ splitting into two $\IS^5$'s using the Mayer-Vietoris sequence. It would be interesting to find a derivation of the above conditions from a more direct field theoretic computation (for instance, by extending the SymTFT branches with additional Maxwell fields, which are eventually gapped in the interior of the SymTFT but stay as non-topological edge modes \cite{GarciaEtxebarria:2024fuk}, in our case localized on the junction theory).

Let us now describe how certain operators of the theory behave and reproduce some of the pictures in figures \ref{fig:symtree-defects} and \ref{fig:symtree-generators}. In particular, let us consider the global form of the gauge group and of its unbroken subgroup after Higgsing to be $SU(N)\to S[U(N_1)\times U(N_2)]$, for a 2-sector splitting, and let us focus on how Wilson line operators of the $\fsu(N)$ theory and and of the $\fsu(N_i)$ theories are related. For this purpose it is useful to formally introduce the $U(N_i)$ gauge connections
\beqa
{\cal A}_i=A_i + \frac{a_i}{N_i}{\bf 1}_{N_i}
\eeqa
where ${\bf 1}_{N_i}$ is the $N_i\times N_i$ identity matrix. For the combined connection to fall in $SU(N)$ rather than in $U(N)$, we need that $\sum_{i=1}^k a_iN_i=0$. Let us now try to construct a Wilson line, for simplicity in the fundamental representation of the $U(N_i)$ theory. This sounds problematic because the $SU(N)$ theory has an electric $\IZ_N$ symmetry, whereas the $SU(N_i)$ gauge theory has electric defects acted on by $\IZ_{N_i}$. This is solved by dressing the naive $SU(N_i)$ Wilson line with a $U(1)_i\subset U(N_i)$ Wilson line as
\beqa
%{\cal W}_{{\bf R}_i,q_i}={\cal W}_{{\bf R}_i}^{\rm naive} \exp \left( -iq({\bf R}_i)\int a_i\right)
{\cal W}_i={\cal W}_i^{\rm naive} \exp \left( -i\frac{1}{N_i}\int a_i\right)
\label{dressed-wls-tree}
\eeqa
To combine the operators of the different theories, one simply needs to use the fact that the $U(1)_i$ are actually not independent, and relate to the $U(1)$ in the junction theory, which is the one arising when the Higgsing is written as $SU(N)\to SU(N_1)\times SU(N_2)\times U(1)/\IZ_L$, with $L={\rm lcm}(N_1,N_2)$. This leads to the junction conditions
\beqa
a_1=\frac{N_2}g a\quad ,\quad a_2=-\frac{N_1}g a
\eeqa
This is the precise way in which the $U(1)$ in the junction theory allows sectors with discrete symmetries of different order to talk to each other in a way consistent with the symmetries. Note in particular that the above condition is compatible with (\ref{juction-condi2}), as should be the case since the $U(1)$ backgrounds are a trivialization of the 2-form field.

Using the dressed Wilson line operators (\ref{dressed-wls-tree}) one can form combinations transforming consistently under the $\IZ_N$ of the full theory. We simply build
\beqa
{\mathbb W}_{\fund}=\otimes_{i=1}^k {\cal W}_i
\label{dressed-wl-sum}
\eeqa
This operator transforms appropriately under the $\IZ_N$, while maintaining the fact that the individual naive Wilson lines ${\cal W}_i^{\rm naive}$ in (\ref{dressed-wls-tree}) transform under the corresponding $\IZ_{N_i}$. This operator can then be combined with the naive Wilson line operator of the $SU(N)$ theory in the fundamental representation. The operators for general representations are simply obtained by fusion of the previous one. It is clear that different choices of representations lead to defect operators of the kind displayed in some of the figures \ref{fig:symtree-defects}. Generically they have the structure of figure (v), with non-trivial representations in all the SymTree branches; but for instance they can turn into figure (iii) when one gets the trivial representation in $SU(N_2)$, but non-trivial ones in $SU(N_1)$ and $SU(N)$, or figure (i) when they are trivial in both daughter theories. On the other hand, there are no defects realizing figures (ii) or (iv).

Clearly, a similar discussion can be carried out for topological operators corresponding to symmetry generators. Taking a line operator of this kind in the theory ${\cal S}_{\rm full}$ in the fundamental representation, it transforms in a combination similar to (\ref{dressed-wl-sum}) of symmetry generators of the $SU(N_i)$ theories, realizing some of the pictures of figure \ref{fig:symtree-generators}. The generalization of the behaviour of charge defect operators and symmetry generators to theories with general global structures, and hence both Wilson and 't Hooft topological line operators is then straightforward.

This analysis is useful in the discussion of similar processes in SymTFT Fans in the main text.

\bibliographystyle{JHEP}
\bibliography{refs}

\end{document}